\documentclass[11pt,a4paper]{article}
\usepackage[colorlinks=true, linkcolor=black!50!blue, urlcolor=blue, citecolor=blue, anchorcolor=blue]{hyperref}
\usepackage[font=small,labelfont=bf,margin=0mm,labelsep=period,tableposition=top]{caption}
\usepackage[a4paper,top=3cm,bottom=2.5cm,left=2.5cm,right=2.5cm,bindingoffset=0mm]{geometry}

\usepackage{graphicx,placeins}
\usepackage{float}
\usepackage{afterpage}
\usepackage{epsfig,cite}
\usepackage{amssymb}
\usepackage{amsmath}
\usepackage{dsfont}
\usepackage{multirow}
\usepackage{url}
\usepackage{xcolor,colortbl}
\usepackage{float}
\usepackage{afterpage}
\usepackage{url}
\usepackage{hyperref}
\usepackage{booktabs}
\usepackage{mathrsfs}

\usepackage{tikz}
\usepackage{tikz-3dplot}
\usepackage[compat=1.0.0]{tikz-feynman}

\usepackage{enumitem}
\usepackage{hyperref}
\usepackage{cite}

\usepackage{pifont}

\usetikzlibrary{shapes, arrows}
\usetikzlibrary{decorations.pathreplacing}
\usetikzlibrary{positioning, calc}
\tikzstyle{fitted} = [rectangle, minimum width=5cm, minimum height=1cm, text centered, draw=black, fill=red!30]
\tikzstyle{operations} = [rectangle, rounded corners, minimum width=2cm,text centered, draw=black, fill=red!30]
\tikzstyle{roundtext} = [rectangle, rounded corners, minimum width=2cm, minimum height=0.8cm, text centered, draw=black, fill=red!30]
\tikzstyle{n3py} = [rectangle, rounded corners, minimum width=3cm, minimum height=1cm, text centered, draw=black, fill=green!30]
\tikzstyle{myarrow} = [thick,->,>=stealth]
\tikzstyle{line} =[draw, -latex']
\tikzstyle{decision} = [diamond, draw, fill=red!20, text width=7.5em, text centered,  inner sep=0pt, minimum height=2em, aspect=4]
\tikzstyle{cloud} = [draw, ellipse,fill=green!20, minimum height=2em]
\tikzstyle{inout} = [rectangle, draw, fill=green!20, text width=9.5em, text centered, rounded corners, minimum height=2em, minimum width=10em]
\tikzstyle{block}=[rectangle, draw, fill=blue!20, text width=9.5em, 
                   text centered, rounded corners, minimum height=2em, 
                   minimum width=10em]

\definecolor{darkgreen}{rgb}{0.0, 0.5, 0.13}

\newcommand{\la}{\left\langle}
\newcommand{\ra}{\right\rangle}
\newcommand{\lc}{\left[}
\newcommand{\rc}{\right]}
\newcommand{\lp}{\left(}
\newcommand{\rp}{\right)}

\def\gsim{\mathrel{\rlap{\lower4pt\hbox{\hskip1pt$\sim$}}
    \raise1pt\hbox{$>$}}}         %greater than or approx. symbol
\def\lsim{\mathrel{\rlap{\lower4pt\hbox{\hskip1pt$\sim$}}
    \raise1pt\hbox{$<$}}}         %less than or approx. symbol

\newcommand{\draft}[1]{}

\def\beq{\begin{equation}}
\def\eeq{\end{equation}}

\def\lapprox{\lower .7ex\hbox{$\;\stackrel{\textstyle <}{\sim}\;$}}
\def\gapprox{\lower .7ex\hbox{$\;\stackrel{\textstyle >}{\sim}\;$}}

\def\d{{\rm d}}

\numberwithin{equation}{section}
\numberwithin{figure}{section}
\numberwithin{table}{section}

\usepackage{tabularx}
\newcolumntype{C}[1]{>{\centering\arraybackslash}p{#1}}

\usepackage{amsmath}
\usepackage{amsfonts}
\usepackage{amssymb}
\usepackage{dsfont}
\usepackage{pifont}
\usepackage{booktabs}
\usepackage{bbold}
\usepackage{graphicx}
\usepackage{epstopdf}
\usepackage{epsfig}
\usepackage{framed}
\usepackage{makeidx}
\usepackage{siunitx}
\usepackage[capitalise]{cleveref}
\usepackage{hyperref}
\usepackage{placeins}
\usepackage[font=small,labelfont=bf]{caption}
\newcommand{\aem}{a_{em}}

\newcommand{\MSbar}{\overline{\rm MS}}

\begin{document}
\newgeometry{top=1.5cm,bottom=1.5cm,left=1.5cm,right=1.5cm,bindingoffset=0mm}

\vspace{-2.0cm}
\begin{flushright}
%Nikhef-2023-aaa\\
TIF-UNIMI-2023-17\\
Edinburgh 2023/19\\
CERN-TH-2023-159\\
\end{flushright}
\vspace{0.3cm}

\begin{center}
  {\Large \bf Photons in the proton: implications for the LHC}
  \vspace{1.1cm}

  {\bf The NNPDF Collaboration}: \\[0.1cm]
  Richard D. Ball$^1$,
Andrea Barontini$^2$,
Alessandro Candido$^{2,3}$,
Stefano Carrazza$^2$,
Juan Cruz-Martinez$^3$,\\[0.1cm]
Luigi Del Debbio$^1$,
Stefano Forte$^2$,
Tommaso Giani$^{4,5}$,
Felix Hekhorn$^{2,6,7}$,
Zahari Kassabov$^8$,\\[0.1cm]
Niccol\`o Laurenti,$^2$
Giacomo Magni$^{4,5}$,
Emanuele R. Nocera$^9$,
Tanjona R. Rabemananjara$^{4,5}$,
Juan Rojo$^{4,5}$,\\[0.1cm]
Christopher Schwan$^{10}$,
Roy Stegeman$^1$, and
Maria Ubiali$^8$

 \vspace{0.7cm}
 
 {\it \small

 ~$^1$The Higgs Centre for Theoretical Physics, University of Edinburgh,\\
   JCMB, KB, Mayfield Rd, Edinburgh EH9 3JZ, Scotland\\[0.1cm]
 ~$^2$Tif Lab, Dipartimento di Fisica, Universit\`a di Milano and\\
   INFN, Sezione di Milano, Via Celoria 16, I-20133 Milano, Italy\\[0.1cm]
   ~$^3$CERN, Theoretical Physics Department, CH-1211 Geneva 23, Switzerland\\[0.1cm]
    ~$^4$Department of Physics and Astronomy, Vrije Universiteit, NL-1081 HV Amsterdam\\[0.1cm]
   ~$^5$Nikhef Theory Group, Science Park 105, 1098 XG Amsterdam, The Netherlands\\[0.1cm]
~$^6$University of Jyvaskyla, Department of Physics, P.O.\ Box 35, FI-40014 University of Jyvaskyla, Finland\\[0.1cm]
~$^7$Helsinki Institute of Physics, P.O.\ Box 64, FI-00014 University of Helsinki, Finland\\[0.1cm]
   ~$^8$ DAMTP, University of Cambridge, Wilberforce Road, Cambridge, CB3 0WA, United Kingdom\\[0.1cm]
     ~$^9$ Dipartimento di Fisica, Universit\`a degli Studi di Torino and\\
   INFN, Sezione di Torino, Via Pietro Giuria 1, I-10125 Torino, Italy\\[0.1cm]
   ~$^{10}$Universit\"at W\"urzburg, Institut f\"ur Theoretische Physik und Astrophysik, 97074 W\"urzburg, Germany\\[0.1cm]

   }

 \vspace{0.7cm}

{\bf \large Abstract}

\end{center}
We construct a set of parton distribution functions (PDFs), based on the recent
NNPDF4.0 PDF set, that also include a photon PDF. The photon PDF is constructed
using the LuxQED formalism, while QED evolution accounting for
$\mathcal{O}\lp \alpha\rp$, $\mathcal{O}\lp \alpha \alpha_s\rp$, and
$\mathcal{O}\lp \alpha^2\rp$ corrections is implemented and benchmarked
by means of the {\sc\small EKO} code. We investigate the impact of QED effects
on NNPDF4.0, and compare our results both to our previous NNPDF3.1QED PDF set
and to other recent PDF sets  that include the  photon. 
We assess the impact of photon-initiated processes and electroweak corrections
on a variety of representative LHC processes, and find that they can
reach the 5\% level in vector boson pair
production at large invariant mass. 

\clearpage

% Table of contents
\tableofcontents

% Introduction
\section{Introduction}
\label{sec:intro}

High precision physics at the LHC, especially in its high-luminosity
era~\cite{CidVidal:2018eel,Azzi:2019yne,Cepeda:2019klc}, will demand the
inclusion of electroweak corrections in the computation of theoretical
predictions for hard processes~\cite{Heinrich:2020ybq}. This requires an extension of the set of
proton parton distribution functions (PDFs). In particular a photon PDF has to
be provided, and evolution equations need to be supplemented with QED splitting
functions. The photon PDF enables the inclusion of photon-initiated processes,
which typically are enhanced in the high-mass and large transverse-momentum
tails of the distributions. In principle, at high-enough scales, proton
PDFs should also include PDFs for leptons~\cite{Bertone:2015lqa}, gauge
bosons~\cite{Fornal:2018znf} and indeed for the full set of 52
standard model fields~\cite{Bauer:2017isx}, or even new hypothetical particles such
as dark photons~\cite{McCullough:2022hzr}. Even at the LHC lepton
PDFs are needed in searches for leptoquarks~\cite{Buonocore:2020erb}
or exotic resonances that couple to leptons~\cite{Buonocore:2021bsf}.
However, in practice, the main
requirement of current precision physics at the LHC is the availability of
PDF sets that also include a photon PDF, that mixes upon combined
QED$\times$QCD evolution with the standard quark and gluon PDFs.
We will henceforth refer to such PDF sets as ``QED PDFs''.

Initial attempts at the construction of QED PDF sets relied on
models for the photon PDF at the initial evolution
scale~\cite{Martin:2004dh,Schmidt:2015zda}.  
A first data-driven determination of QED PDFs, based on a 
fit using the NNPDF2.3 methodology~\cite{Ball:2012cx,Ball:2013hta},
resulted in a photon PDF with large uncertainties. The fact that
determining the  photon PDF from  the data yields a result affected by large uncertainties
was more recently confirmed in determinations based on fitting to the
data PDFs with a fixed
functional forms 
within the {\sc\small xFitter} methodology~\cite{Giuli:2017oii} applied
to high-mass ATLAS Drell--Yan distributions~\cite{Aad:2016zzw}.

A breakthrough in the determination of QED PDFs was achieved in 2016 in
Refs.~\cite{Manohar:2016nzj,Manohar:2017eqh} (see also related results
in Refs.~\cite{Harland-Lang:2016kog,Harland-Lang:2016lhw}), where it
was shown that the photon PDF can be computed perturbatively in QED,
given as input the proton structure functions at all scales, from the
elastic ($Q^2\to 0$) to the deep-inelastic ($Q^2\to\infty$)
regimes --- the so-called LuxQED method.
Since then, this LuxQED framework has been the basis of all QED PDF
sets~\cite{Bertone:2017bme,Harland-Lang:2019pla,Cridge:2021pxm,
  Xie:2021ajm,Xie:2023qbn}. 

While the effect of the inclusion of the photon PDF on other PDFs is small,
it is not negligible within current uncertainties.
For instance, the photon typically carries a fraction of the proton
momentum which is about two orders of magnitude smaller than that of
the gluon, so the corresponding depletion of the gluon momentum
fraction is relevant at the percent level, which is comparable to the
size of the current uncertainty on the gluon PDF.
Precision calculations of LHC processes with electroweak effects
thus require a consistent global QED PDF determination.

With this motivation, we construct a QED PDF set based on the recent NNPDF4.0
PDF determination~\cite{Ball:2021leu,NNPDF:2021uiq}, including
QED corrections to parton evolution up to
$\mathcal{O}\lp \alpha \alpha_s\rp$ and $\mathcal{O}\lp \alpha^2\rp$.
We follow closely the methodology developed for the
NNPDF3.1QED PDFs in Ref.~\cite{Bertone:2017bme},
based on an iterative procedure. Namely, we evaluate the photon PDF at some
fixed scale using the LuxQED formula and the structure function computed from
an existing PDF set, now NNPDF4.0. We evolve it together with the other 
PDFs to the initial parametrization scale $Q_0=1.65$~GeV. We then
re-determine the quark and gluon PDFs  while also including this
photon PDF as a boundary condition to the QCD$\times$QED evolution. We compute
the photon PDF again using LuxQED, and we iterate until convergence.

All results presented in this paper are obtained using a new 
theory implementation, now adopted by default in the NNPDF public code~\cite{NNPDF:2021uiq}.
In particular, the {\sc\small APFEL}~\cite{Bertone:2013vaa} and
{\sc APFELgrid}~\cite{Bertone:2016lga} codes have been replaced by a suite of
newly developed tools including the evolution
equations solver {\sc\small EKO}~\cite{Candido:2022tld},
the DIS structure functions calculator {\sc\small YADISM}~\cite{yadism},
and the interpolator of hard-scattering cross-sections
{\sc\small PineAPPL}~\cite{christopher_schwan_2023_7995675,Carrazza:2020gss}.
Taken together, they provide a pipeline for the  efficient
automatization of the
computation of theory predictions~\cite{Barontini:2023vmr}. An integral component of this theory
pipeline is the use of interpolation grids in the format provided by
{\sc\small PineAPPL}~\cite{christopher_schwan_2023_7995675,Carrazza:2020gss},
which can be interfaced to Monte Carlo generators. Among these is 
{\sc\small mg5\_aMC@NLO}~\cite{Frederix:2018nkq}, which automates NLO
QCD+EW calculations for a wide range of LHC processes where knowledge of QED
PDFs is crucial.

The outline of this paper is as follows. First, Sect.~\ref{sec:theory} reviews
the theoretical framework underlying the NNPDF4.0QED determination and in
particular the implementation of QED evolution in {\sc\small EKO}.
Then the NNPDF4.0QED PDFs are presented in Sect.~\ref{sec:results}, where
they are compared to the previous NNPDF3.1QED PDF set, and to other
recent  QED PDF sets. Implications for LHC phenomenology are studied in
Sect.~\ref{sec:pheno} by means of the {\sc\small PineAPPL} interface to
{\sc\small mg5\_aMC@NLO}. Conclusions and an outline of future developments are
finally presented in Sect.~\ref{sec:summary}.
Because this paper is, as mentioned, the first to make use of a new
theory pipeline, and specifically its implementation of combined
QCD$\times$QED evolution, a set of benchmarks is collected in two appendices.
Specifically, in \cref{app:pineline} we benchmark NNPDF4.0 PDFs
determined using the old and new pipeline and show that they are
indistinguishable; in \cref{app:exatrn} the new implementation of joint QCD
and QED evolution of PDFs in the {\sc\small EKO} code is discussed and
benchmarked against the previous implementation in the {\sc\small APFEL} code.

% Theory
\section{Evolution and determination of QED PDFs}
\label{sec:theory}

In this section, we discuss the structure of combined QED$\times$QCD evolution
and we briefly review the methodology used to determine QED PDFs.
This methodology is based on the LuxQED
formalism~\cite{Manohar:2016nzj,Manohar:2017eqh}, and was developed for the
construction of the NNPDF3.1QED PDF set~\cite{Bertone:2017bme}. 

\subsection{QCD$\times$QED evolution equations and basis choice}
\label{sec:QED-DGLAP}

The scale dependence of PDFs is  determined by evolution equations of
the form
\begin{equation}
  \label{eq:DGLAPN}
  \mu^2 \frac{d f_i(N,\mu^2)}{d \mu^2}
  =
  - \sum_{j} \gamma_{ij}\left(N,a_s(\mu^2), \aem(\mu^2) \right) f_j(N,\mu^2)\,,
\end{equation}
where $f_i(N,\mu^2)=\int_0^1 dx\,x^{N-1}f_i(x,\mu^2)$ is the Mellin
transform of the $i$-th PDF. The anomalous dimensions $\gamma_{ij}(N,a_s, \aem)$
are determined as a simultaneous perturbative expansion in the strong coupling
$a_s=\alpha_s/(4 \pi)$ and in the electromagnetic coupling
$\aem=\alpha/(4 \pi)$:
\begin{align}
  \label{eq:gammaij}
  \gamma_{ij}(N,a_s, \aem)
  &=
  \sum_{\substack{n,m=0 \\ (n,m) \neq (0,0)}}^\infty a_s^n \aem^m \gamma_{ij}^{(n,m)}(N)\\
  &= a_s \gamma_{ij}^{(1,0)}(N)
  + a_s^2 \gamma_{ij}^{(2,0)}(N)
  + a_s^3 \gamma_{ij}^{(3,0)}(N) \nonumber\\
  &\hspace{10pt}
  + \aem \gamma_{ij}^{(0,1)}(N)
  + \aem^2 \gamma_{ij}^{(0,2)}(N)
  + a_s \aem \gamma_{ij}^{(1,1)}(N) + \dots
  \nonumber\,.
\end{align}
In this work we include pure QCD corrections up to NNLO, $\gamma^{(1,0)}$,
$\gamma^{(2,0)}$, and $\gamma^{(3,0)}$~\cite{Vogt:2004mw, Moch:2004pa};
the pure NLO QED corrections $\gamma^{(0,1)}$ and
$\gamma^{(0,2)}$~\cite{deFlorian:2016gvk}; and the leading mixed correction
$\gamma^{(1,1)}$~\cite{deFlorian:2015ujt}.

The scale dependence of the strong and electromagnetic
couplings is in turn determined by coupled renormalization group
equations  of the form 
\begin{align}
  \mu^2 \frac{d a_s }{d \mu^2}
  & =\beta_{_{\rm QCD}} (a_s, \aem)
  =-a_s^2\left( \beta_{_{\rm QCD}}^{(2,0)}
  + a_s \beta_{_{\rm QCD}}^{(3,0)}
  + \aem \beta_{_{\rm QCD}}^{(2,1)}
  + a_s^2 \beta_{_{\rm QCD}}^{(4,0)}\dots \right)\,,
  \label{RGE:as}\\
  \mu^2 \frac{d \aem }{d \mu^2}
  & =\beta_{_{\rm QED}} (a_{s}, \aem)
  = -\aem^2 \left( \beta_{_{\rm QED}}^{(0,2)}
  + \aem \beta_{_{\rm QED}}^{(0,3)}
  + a_s \beta_{_{\rm QED}}^{(1,2)}
  + \dots \right)\,,
  \label{RGE:aem}
\end{align}
in terms of the coefficients of the corresponding QCD and QED beta functions.
Consistently with the treatment of evolution equations, we include
pure QCD contributions up to NNLO, namely $\beta_{_{\rm QCD}}^{(2,0)}$,
$\beta_{_{\rm QCD}}^{(3,0)}$, and $\beta_{_{\rm QCD}}^{(4,0)}$; pure
QED up to NLO, namely $\beta_{_{\rm QED}}^{(0,2)}$ and
$\beta_{_{\rm QED}}^{(0,3)}$; and the leading mixed terms
$\beta_{_{\rm QCD}}^{(2,1)}$ and $\beta_{_{\rm QED}}^{(1,2)}$~\cite{Surguladze:1996hx}.
We adopt  the $\overline{\rm {MS}}$ scheme. 
Schemes in which the electroweak coupling does not run, such as the $G_\mu$
scheme, are commonly used in the computation of electroweak corrections,
but $\overline{\rm { MS}}$ is more convenient  when considering combined QCD
and QED corrections~\cite{Fanchiotti:1992tu,Denner:2019vbn}.
Equations~(\ref{eq:DGLAPN}--\ref{RGE:aem}) must then be simultaneously solved
with a common scale $\mu$.

The solution is most efficiently obtained in a maximally decoupled basis in
quark flavor space. This requires adopting a suitable combination of quark and
antiquark flavors such that the sum over $j$ in Eq.~(\ref{eq:DGLAPN}) contains
the smallest possible number of entries. In the case of QCD-only evolution,
this is achieved in the so-called evolution basis, in which one separates off
the singlet combination $\Sigma = \sum_i (q_i + \bar q_i)$, which mixes with the
gluon, and then one constructs nonsinglet combinations of individual C-even
(sea-like) and C-odd (valence-like) $q^\pm_i=q_i\pm\bar q_i$ quark and antiquark
flavors, each of which evolves independently.

Because the photon couples differently to up-like $u_k = \{u,c,t\}$ and
down-like $d_k=\{d,s,b\}$ quarks, a different basis choice is necessary when
also including  QED. To this end, given $n_f$  active quark flavors, we split
them into $n_u$ up-like and $n_d$ down-like flavors, such that $n_f=n_u + n_d$,
and define the four combinations
\begin{equation}
  \Sigma_u=\sum_{k=1}^{n_u}u_k^+,
  \quad
  \Sigma_d=\sum_{k=1}^{n_d}d_k^+,
  \quad
  V_u=\sum_{k=1}^{n_u}u_k^-,
  \quad
  V_d=\sum_{k=1}^{n_d}d_k^-.
\end{equation}
Also, we separate off the QCD contributions to the anomalous
dimensions, by rewriting the perturbative expansion of the anomalous dimensions,
\cref{eq:gammaij}, as
\begin{equation}
  \gamma_{ij}(N,a_s,\aem)
  =
  \gamma_{ij}(N,a_s)+\tilde{\gamma}_{ij}(N,a_s,\aem)\,,
\end{equation}
where $\gamma_{ij}(a_s)$ contains the pure QCD contributions and
$\tilde{\gamma}_{ij}$ contains both the pure QED and the mixed QCD$\times$QED
corrections.

The maximally decoupled evolution equations are then constructed as follows.
The nonsinglet combinations
\begin{align}
  T_3^d  &= d^+ - s^+\,,
  &V_3^d &= d^- - s^-\,,
  &T_3^u &=u^+ - c^+\,,
  & V_3^u &=u^- - c^-\,,
  \label{eq:TV3ud} \\
  T_8^d &= d^+ + s^+ - 2b^+\,,
  &V_8^d &= d^- + s^- - 2b^-\,,
  &T_8^u &= u^+ + c^+ - 2t^+\,,
  &V_8^u &= u^- + c^- - 2t^-\,,
  \label{eq:TV8ud}
\end{align}
evolve independently according to nonsinglet evolution equations of the form
\begin{align}
  \label{eq:non-singlet}
  \mu^2\frac{d}{d\mu^2}T^{u/d}_{3/8}
  & =
  - (\gamma_{\rm ns,+} +\tilde{\gamma}^{\rm ns,+}_{u/d}) T^{u/d}_{3/8}\,,
  \\
  \mu^2\frac{d}{d\mu^2}V^{u/d}_{3/8}
  & = - (\gamma_{\rm ns,-} +\tilde{\gamma}^{\rm ns,-}_{u/d} )V^{u/d}_{3/8}\,.
\end{align}
The valence sum and difference combinations,
defined as
\begin{equation}
  V  = V_u + V_d\,,
  \quad
  V_\Delta  = \frac{n_d}{n_u}V_u - V_d\,,
  \label{eq:VVDelta}
\end{equation}
satisfy coupled evolution equations 
\begin{equation}
  \label{eq:valence}
  \mu^2\frac{d}{d\mu^2}
  \begin{pmatrix}
     V \\
     V_\Delta
  \end{pmatrix}
  =-
  \begin{pmatrix}
    \gamma^{{\rm ns},V}+\langle \tilde{\gamma}^{{\rm ns},-}_{q} \rangle
    &
    \nu_u\tilde{\gamma}^{{\rm ns},-}_{\Delta q}\\
    \nu_d\tilde{\gamma}^{{\rm ns},-}_{\Delta q}
    &
    \gamma^{{\rm ns},-}+\{ \tilde{\gamma}^{{\rm ns},-}_{q} \}
  \end{pmatrix}
  \begin{pmatrix}
    V \\
    V_\Delta
  \end{pmatrix}\,,
\end{equation}
in terms of the linear combinations of anomalous dimensions
\begin{equation}
  \label{eq:AD:combination}
  \nu_{u/d} = \frac{n_{u/d}}{n_f}\,,
  \quad
  \langle \tilde{\gamma}^{{\rm ns},\pm}_{q} \rangle
  =
  \nu_u \tilde{\gamma}^{{\rm ns},\pm}_{u}+\nu_d \tilde{\gamma}^{{\rm ns},\pm}_{d}\,,
  \quad
  \{ \tilde{\gamma}^{{\rm ns},\pm}_{q} \}
  =
  \nu_d \tilde{\gamma}^{{\rm ns},\pm}_{u}+\nu_u \tilde{\gamma}^{{\rm ns},\pm}_{d}\,,
  \quad
  \tilde{\gamma}^{{\rm ns},\pm}_{\Delta q}
  =
  \tilde{\gamma}^{{\rm ns},\pm}_{u} - \tilde{\gamma}^{{\rm ns},\pm}_{d}\,.
\end{equation}

Finally, the gluon and photon satisfy coupled evolution equations
together with the quark singlet sum and difference combinations,
defined as
\begin{equation}
  \Sigma
  = \Sigma_u + \Sigma_d, \quad \Sigma_\Delta
  = \frac{n_d}{n_u}\Sigma_u - \Sigma_d\,.
  \label{eq:ggSSDelta}
\end{equation}
These coupled evolution equations read
\begin{equation}
  \mu^2\frac{d}{d\mu^2}
  \begin{pmatrix}
    g \\
    \gamma \\
    \Sigma \\
    \Sigma_\Delta
  \end{pmatrix}
  =
  -{\mathbf\Gamma}
  \begin{pmatrix}
    g \\
    \gamma \\
    \Sigma \\
    \Sigma_\Delta
  \end{pmatrix}\,,
\end{equation}
with ${\mathbf\Gamma}$ a $4\times 4$ anomalous dimension matrix of the form
\begin{equation}
  {\mathbf\Gamma}
  =
  \begin{pmatrix}
    \gamma_{gg}+\tilde{\gamma}_{gg}
    & \tilde{\gamma}_{g\gamma}
    & \gamma_{gq} + \langle\tilde{\gamma}_{gq}\rangle
    & \nu_u\tilde{\gamma}_{g \Delta q} \\
    \tilde{\gamma}_{\gamma g}
    & \tilde{\gamma}_{\gamma \gamma}
    & \langle \tilde{\gamma}_{\gamma q} \rangle
    & \nu_u \tilde{\gamma}_{\gamma \Delta q} \\
    2n_f (\gamma_{qg} +\langle  \tilde{\gamma}_{qg} \rangle )
    & 2 n_f  \langle \tilde{\gamma}_{q \gamma} \rangle
    & \gamma_{qq} +  \langle \tilde{\gamma}^{{\rm ns},+}_{q} \rangle
    + \langle e^2_q\rangle^2\tilde{\gamma}_{{\rm ps}}
    & \nu_u\tilde{\gamma}^{{\rm ns},+}_{\Delta q}
    + \nu_ue^2_{\Delta q}\langle e^2_q\rangle \tilde{\gamma}_{{\rm ps}}\\
    2n_f \nu_d \tilde{\gamma}_{\Delta qg}
    & 2n_f \nu_d \tilde{\gamma}_{\Delta q\gamma}
    & \nu_d\tilde{\gamma}^{{\rm ns},+}_{\Delta q}
    + \nu_d e^2_{\Delta q}\langle e^2_q\rangle \tilde{\gamma}_{{\rm ps}}
    & \gamma^{{\rm ns},+} + \{ \tilde{\gamma}^{{\rm ns},+}_{q} \}
    + \nu_u \nu_d (e^2_{\Delta q})^2\tilde{\gamma}_{{\rm ps}}
  \end{pmatrix}\,,
  \nonumber
\end{equation}
where $\tilde{\gamma}^{{\rm ps}}_{qq'}=e^2_q e^2_{q'}
\tilde{\gamma}_{{\rm ps}}$~\cite{deFlorian:2016gvk}, and the combinations
$\langle \tilde{\gamma}_{gq} \rangle$, $\langle \tilde{\gamma}_{qg} \rangle$,
$\tilde{\gamma}_{g \Delta q}$, $\tilde{\gamma}_{\Delta q g}$,
$\langle \tilde{\gamma}_{\gamma q} \rangle$,
$\langle \tilde{\gamma}_{q \gamma} \rangle$,
$\tilde{\gamma}_{\gamma \Delta q}$, and $\tilde{\gamma}_{\Delta q \gamma}$
are constructed analogously to those listed in Eq.~\eqref{eq:AD:combination}.

The basis defined by \cref{eq:TV3ud,eq:TV8ud,eq:VVDelta,eq:ggSSDelta}
is denoted as the unified evolution basis. Compared to the basis used
in {\sc\small APFEL}~\cite{Bertone:2013vaa} to solve the
QCD$\times$QED evolution equations, our definitions of $\Sigma_\Delta$ and
$V_\Delta$ differ due to the prefactors $n_d/n_u$ that make the basis fully
orthogonal. In the presence of scale-independent intrinsic heavy quarks
({\it e.g.} a charm-quark PDF in a three-flavor scheme~\cite{Ball:2015dpa}),
a further decomposition needs to be applied~\cite{Candido:2022tld}.

The unified  flavor basis has been implemented in the {\sc\small EKO}
code~\cite{Candido:2022tld}, which, as discussed in the introduction,
is now used to solve evolution equations as part of the new theory
pipeline. The basic ingredient of the EKO code is the construction of
evolution kernel operators $E(Q^2 \leftarrow Q_0^2)$ (EKOs) such that
\begin{equation}\label{eq:basiceko}
  f(Q^2)= E(Q^2 \leftarrow Q_0^2)   f(Q_0^2),
\end{equation}
where $f(Q^2)$ is a vector whose components are all PDF flavors,
including all active quark and antiquark flavors in a suitable basis,
the photon and the gluon.
Formally, the EKOs are given by
\begin{equation}
  \label{eq:exeko}
  E(Q^2 \leftarrow Q_0^2)
  =
  \mathcal P \exp\left(-\int_{Q_0^2}^{Q^2} \frac{\d\mu^2}{\mu^2}\,
  \gamma\left(a_s(\mu^2), \aem(\mu^2)\right)\right) \, ,
\end{equation}
where $\gamma$ is the full matrix of anomalous dimensions, and $\mathcal P$
denotes path ordering. In the unified evolution basis the evolution kernel is a
block diagonal matrix, with individual diagonal entries for the nonsinglet
combinations Eqs.~(\ref{eq:TV3ud},\ref{eq:TV8ud}) (without path-ordering),
a $2\times2$ block for the valence combinations
Eq.~(\ref{eq:VVDelta}), and a $4\times4$ block in the singlet sector.

A variety of implementations of the solution of the evolution
equations, each of those corresponding to a determination of the EKO,
and which differ by higher-order terms, are available in the EKO code.
These are discussed in Ref.~\cite{Candido:2022tld}. At NNLO in QCD, the
solutions are essentially indistinguishable in the data region. The
solution adopted here is the iterated-exact (EXA) of
Ref.~\cite{Candido:2022tld}, while the truncated (TRN) solution was
adopted for the NNPDF3.1QED~\cite{Bertone:2017bme} and
NNPDF4.0~\cite{Ball:2021leu} PDF sets, using the
{\sc\small APFEL}~\cite{Bertone:2013vaa} implementation. These different
implementations (at NNLO) are benchmarked and shown to be equivalent in
\cref{app:exatrn}, where we also discuss the motivations for this choice.

\subsection{Construction of  the photon PDF}
\label{subsec:methodology}

As mentioned in Sect.~\ref{sec:intro}, the photon PDF is determined using
the LuxQED~\cite{Manohar:2016nzj,Manohar:2017eqh} formalism, implemented in a
PDF fit using the same methodology as the one used for the NNPDF3.1QED PDF
set~\cite{Bertone:2017bme}, but now with the new theory pipeline.
The LuxQED result amounts to proving that the photon PDF is perturbatively
determined in QED by knowledge of the proton inclusive structure functions
$F_2$ and $F_L$:
\begin{equation}
  \label{eq:LUX:QED}
  \begin{split}
    & x \gamma(x, \mu^2)
    =
    \frac{2}{\aem (\mu^2)} \int\limits_x^1 \frac{dz}{z}
    \Biggl\{ \int_{m_p^2x^2 \over (1-z)}^{\mu^2 \over (1-z)} \frac{dQ^2}{Q^2}
    \aem^2(Q^2) \Biggl[ -z^2 F_L(x/z, Q^2) \\
    & + \left( z P_{\gamma q}(z) + \frac{2 x^2 m_p^2}{Q^2} \right)
    F_2(x/z, Q^2)\Biggr] - \aem^2(\mu^2) z^2 F_2(x/z, \mu^2)\Biggr\}\,,
  \end{split}
\end{equation}
where $m_p$ is the mass of the proton and $P_{\gamma q}$ is the
photon-quark splitting function. Equation~(\ref{eq:LUX:QED}) holds
in the $\MSbar$ scheme, including terms of order $\aem$ and
$\aem^2\ln\mu^2/m_p^2$, times the accuracy of the QCD
determination of the structure functions.

The integration
over the scale $Q^2$ and  Bjorken-$x$ dependence of the structure
functions $F_i(x,Q^2)$ in Eq.~(\ref{eq:LUX:QED}) includes four
different regions and 
corresponding contributions to the structure functions: an 
elastic contribution at  $x=1$, a  resonance contribution when $x$ is
large and close to $x=1$, an inelastic non-perturbative contribution at low $Q^2$ and
intermediate $x$ and an inelastic perturbative region for
intermediate $x$ and large $Q^2$. The first three contributions must be
determined by fits to lepton-proton scattering
data, but the latter contribution may be determined
by expressing the structure functions through perturbative QCD
factorization in terms of
quark and gluon PDFs. In a global PDF determination, one may choose to
determine the photon PDF using the LuxQED
formula Eq.~(\ref{eq:LUX:QED}) at a single chosen scale
$\mu^2=Q_\gamma^2$, and then evolve jointly the photon and all other
PDFs through the QCD$\times$QED evolution equations discussed in
Sect.~\ref{sec:QED-DGLAP}. An alternative option is to determine the
photon PDF using Eq.~(\ref{eq:LUX:QED}) at all scales. The two choices
are equivalent because the LuxQED photon Eq.~(\ref{eq:LUX:QED})
satisfies the joint QED$\times$QCD evolution equations to the accuracy
of the LuxQED formula itself, though they
differ by higher-order corrections and also due to the fact that the
LuxQED photon is partly determined from a parametrization of data.
The former choice was made 
for the construction of our previous NNPDF3.1QED
set~\cite{Bertone:2017bme}, which we follow here, also in the choice
of parametrization of the structure functions in the non-perturbative
region. The same choice  was  made in
Refs.~\cite{Bertone:2017bme,Harland-Lang:2019pla,Cridge:2021pxm,Xie:2021ajm,Xie:2023qbn},
though in Refs.~\cite{Xie:2021ajm,Xie:2023qbn} a PDF set in which the
photon PDF is determined using Eq.~(\ref{eq:LUX:QED}) at all scales
was also presented.

%-----------------------------------------------------------------------------
\begin{figure}[!t]
  \centering
  \includegraphics[width=.49\textwidth]{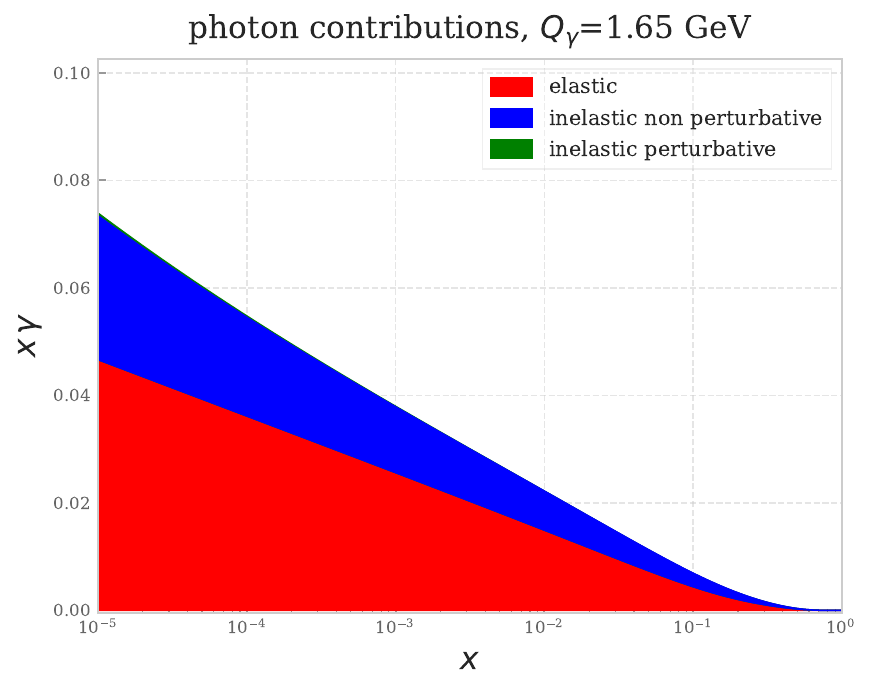}
  \includegraphics[width=.49\textwidth]{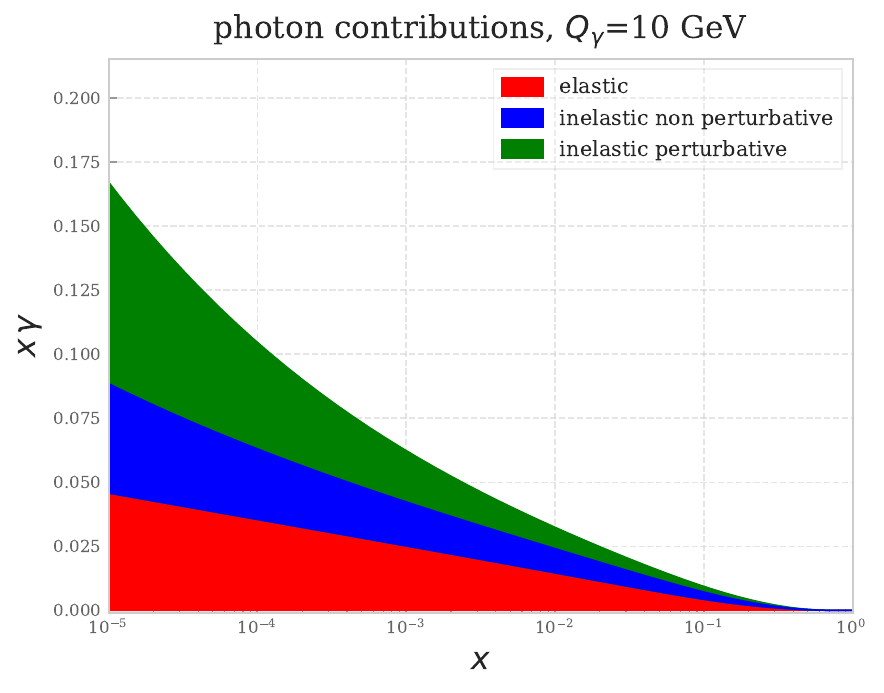}
  \includegraphics[width=.49\textwidth]{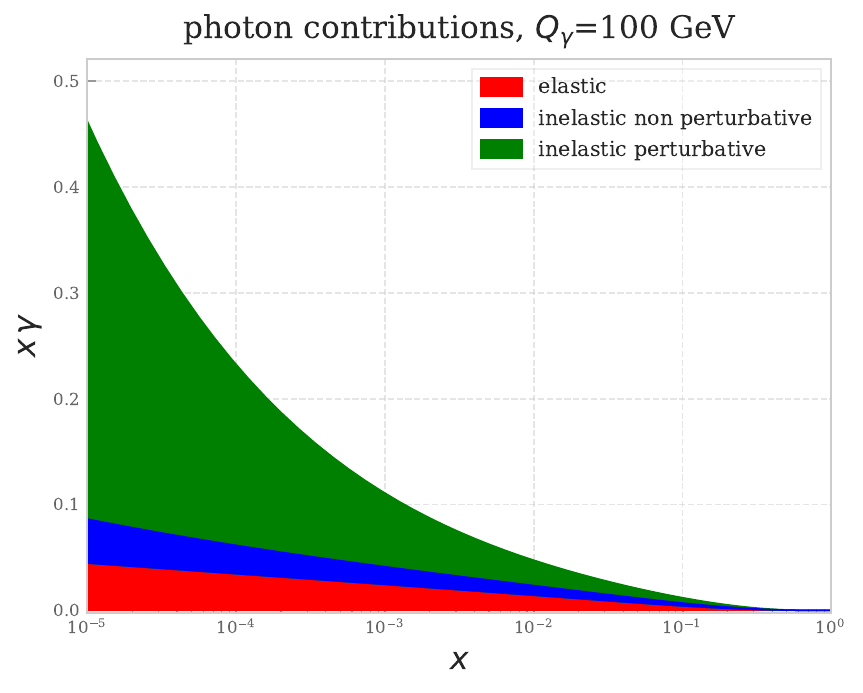}
  \includegraphics[width=.49\textwidth]{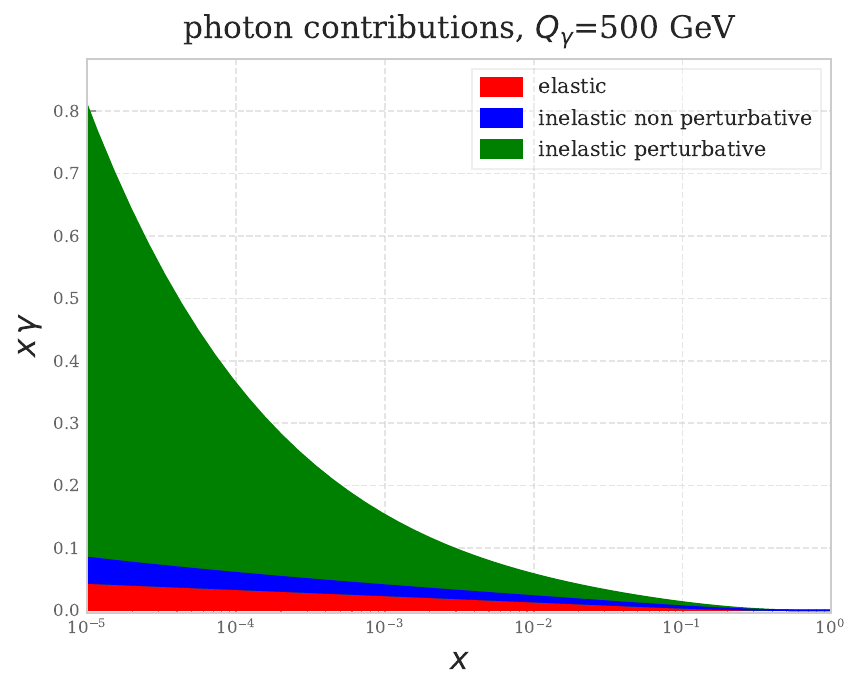}\\
  
  \caption{\small Breakdown of the photon PDF $\gamma(x,\mu^2)$  into
    the contributions  coming from
  different contributions to the proton structure functions
  $F_i(x,Q^2)$ that determine it according to the
    LuxQED formula Eq.~(\ref{eq:LUX:QED}). The result is shown as a
  function of $x$ for four
  different choices of the scale $\mu=Q_\gamma$: $Q_\gamma=1.65$~GeV
  (top left), 10~GeV (top right), 100~GeV (bottom left) and 500~GeV  (bottom right).
}
  \label{fig:luxcomponents} 
\end{figure}
%-------------------------------------------------------------------------------

The relative size of the elastic, inelastic
non-perturbative (including resonances) and elastic perturbative
contributions to $\gamma(x,\mu^2)$ is displayed   in
Fig.~\ref{fig:luxcomponents} as a function of $x$ for four 
different choices of the scale $\mu=Q_\gamma$ at which the LuxQED
formula Eq.~(\ref{eq:LUX:QED}) is used to determine the photon (see
also Fig.~18 in Ref.~\cite{Manohar:2017eqh}).
If the LuxQED formula is used at the scale $Q_\gamma=1.65$~GeV at which the
NNPDF4.0 PDFs are parametrized, the photon is entirely determined  by
the elastic and non-perturbative contributions, but as the scale is
increased, an
increasingly large contribution comes from the perturbative
region. By the time a value $Q_\gamma\gtrsim100$~GeV  is reached, the
photon at small $x\lsim10^{-3}$ is almost entirely determined by the
perturbative contribution, while the elastic and non-perturbative
contributions only remain dominant at large $x\gsim0.1$ where however
the photon PDF is tiny.
Choosing a large value of $Q_\gamma$ has the dual advantage
that the LuxQED result is more accurate at high scale, because it
includes contributions of order $\aem$ and
$\aem^2\ln\mu^2/m_p^2$, and also that one is then mostly relying on
the accurate perturbative determination of the structure functions,
that exploits global information on quarks and gluon PDFs and not just
the lepton-proton scattering  data. As in  Ref.~\cite{Bertone:2017bme}, we
choose $\mu=Q_\gamma=100$~GeV, as also advocated in
Ref.~\cite{Manohar:2017eqh}.\footnote{Note that in
Ref.~\cite{Xie:2021equ}, Table 1, the value at which the photon is
evaluated in Ref.~\cite{Manohar:2017eqh} (called $\mu_{\rm eval}$, see Sect.~10.1
of that Ref.) is incorrectly reported as 
$Q=10$~GeV, instead of the correct $Q=100$~GeV.} We will discuss the
dependence of our results on the choice of scale  $Q_\gamma$ in 
Sect.~\ref{sec:results}.

Equation~(\ref{eq:LUX:QED}) must be viewed as a constraint on the set of
photon, quark and gluon PDFs that are simultaneously determined. 
Because of the small impact of the photon PDF on the other PDFs, it was
suggested in Refs.~\cite{Manohar:2017eqh,Bertone:2017bme} that the constraint can be implemented
iteratively. Namely, we first determine the photon PDF using
Eq.~(\ref{eq:LUX:QED}) by means of structure functions that are determined from
an existing PDF set at a given scale $\mu=Q_\gamma$. This photon PDF is then
evolved, by solving joint QCD$\times$QED evolution equations with the
unchanged given set, to a
chosen PDF parametrization scale
where it is taken as fixed. All other
PDFs are then re-determined, with the constraint that the  momentum sum
rule now also includes a contribution from the given (fixed) photon, {\it i.e.}
\begin{equation}
  \label{eq:momsr}
  \int_0^1 dx\, \lp  x\Sigma(x,Q^2) + xg(x,Q^2) + x\gamma(x,Q^2) \rp =1\,.
\end{equation}
The photon PDF at a scale $\mu=Q_\gamma$ is then determined again from
Eq.~(\ref{eq:LUX:QED}), in which structure functions are obtained from the
new fit. The procedure is iterated until convergence,
which was achieved in two iterations in the NNPDF3.1QED
determination~\cite{Bertone:2017bme}.

Here we follow the same procedure, starting with a re-determination of
the NNPDF4.0 PDF set with the new pipeline, which is compared and
shown to be equivalent to the published NNPDF4.0 in
\cref{app:pineline}.

% Results
\section{The NNPDF4.0QED parton distributions}
\label{sec:results}

We present here the NNPDF4.0QED PDFs: we
first summarize our procedure, then discuss the effect on fit quality
of the inclusion of the photon PDF, examine the photon PDF itself,
also in comparison to other determinations, and finally study the photon
momentum fraction.

\subsection{Construction of the NNPDF4.0QED parton set}
\label{sec:construction_NNPDF40QED}

As mentioned in  Sect.~\ref{sec:theory}
all the methodological aspects and settings of the PDF
determination are the same as used for the underlying pure QCD NNPDF4.0
PDF~\cite{Ball:2021leu}, but now 
using a new theory 
pipeline. Even if in principle
theory predictions should be independent of implementation details,
in practice differences may arise, {\it e.g.} due to issues of numerical
accuracy. Also, in the process of transitioning to this new pipeline, a few
minor bugs in data implementation were uncovered and fixed. Finally,
the new pipeline includes a new implementation of heavy-quark mass
effect in deep-inelastic structure functions that differs by
subleading terms from the previous one. 
Benchmarks showing the equivalence of the old and new
pipeline are briefly presented  in Appendix~\ref{app:pineline}. Because of this
equivalence, NNPDF4.0 PDF replicas produced with the new pipeline should be
considered equivalent to the published ones, and indeed for phenomenological
applications (specifically those presented in Sect.~\ref{sec:pheno}) we will
compare the results obtained using NNPDF4.0QED PDFs to pure QCD results
obtained using the published NNPDF4.0 replicas.
Nevertheless, in this section only, for all comparisons
between pure QCD and QCD$\times$QED, we will use NNPDF4.0 pure QCD
replicas generated using the new pipeline, in order to avoid even
small confounding effects. Note however that, as discussed in the end of
Sect.~\ref{sec:QED-DGLAP}, the QED PDFs are based on the EXA solution
of the evolution equations, that differs by subleading terms from the
TRN solution used in the pure QCD fit. The effect of this difference
is assessed in Appendix~\ref{app:exatrn} and is very small at
NNLO. However, all comparisons between
pure QCD and QED PDFs also include the effect of this change.

The NNPDF4.0QED PDFs are determined at NLO and NNLO by supplementing
with a
photon PDF the pure QCD PDF set, according to the
methodology outlined in the previous section. However, all theory
predictions are obtained as in the pure QCD determination: hence in
particular no photon-induced contributions are included, and thus the
only effect of the inclusion of a photon PDF is through its mixing
with other PDFs. This is justified because
the NNPDF4.0 dataset was constructed including
cuts that remove all datapoints for which the effect of electroweak
corrections is larger than the experimental uncertainties (see
Ref.~\cite{Ball:2021leu}, Sect.~4.1). The inclusion of the electroweak
corrections, which will allow relaxing these cuts,  is left for future work. 
As in Ref.~\cite{Bertone:2017bme}, the final PDF set is obtained
after two iterations, with a third iteration providing a check of
convergence.

\subsection{Fit quality}
\label{sec:fit_quality}

Table~\ref{tab:chi2_qed_global} displays the statistical estimators obtained
using a set of 100 NNPDF4.0QED NLO and NNLO PDF replicas, compared to their
QCD-only counterparts, generated using the
new theory pipeline. Specifically, we show the $\chi^2$, for both the full
dataset and
for datasets grouped by process; for the full dataset we also show the average
over replicas of the training and validation figures of merit
$\la E_{\rm tr}\ra_{\rm rep}$ and $\la E_{\rm val}\ra_{\rm rep}$, and the average
$\chi^2$ over replicas $\la \chi^2\ra_{\rm rep}$, all as defined in Table~9 of
Ref.~\cite{Ball:2010de}. Note that $\chi^2$ and $\la \chi^2\ra_{\rm rep}$ are
computed using the experimental covariance matrix, while, as in all NNPDF
determinations, the figure of merit used for minimization is computed using the
$t_0$ covariance matrix~\cite{Ball:2009qv}.

%-----------------------------------------------------------------------
\begin{table}[!t]
  \centering
  \footnotesize
  \renewcommand{\arraystretch}{1.30}
  \begin{tabularx}{\textwidth}{Xlcccc}
  \toprule
  & \multirow{2}{*}{Dataset}
  & \multicolumn{2}{c}{NNPDF4.0 NLO}
  & \multicolumn{2}{c}{NNPDF4.0 NNLO} \\
  &
  & QCD$\times$QED
  & QCD
  & QCD$\times$QED
  & QCD \\
  \midrule
  $\chi^2$ &  \multirow{4}{*}{Global}
  & 1.31 & 1.26
  & 1.17 & 1.17  \\
  $\la E_{\rm tr}\ra_{\rm rep}$
  &
  & 2.47$\pm$0.07 & 2.41$\pm$0.06
  & 2.27$\pm$0.06 & 2.28$\pm$0.05 \\
  $\la E_{\rm val}\ra_{\rm rep}$
  &
  & 2.66$\pm$0.11 & 2.57$\pm$0.10
  & 2.39$\pm$0.10 & 2.37$\pm$0.11  \\
  $\la \chi^2\ra_{\rm rep}$
  &
  & 1.337$\pm$0.016 & 1.286$\pm$0.017
  & 1.192$\pm$0.014 & 1.195$\pm$0.015  \\
  \midrule
  \multirow{8}{*}{$\chi^2$}
  & DIS neutral-current
  & 1.38 & 1.31
  & 1.22 & 1.23 \\
  & DIS charged-current
  & 0.94 & 0.92
  & 0.90 & 0.90 \\
  & Drell--Yan (inclusive and with one jet)
  & 1.56 & 1.56 
  & 1.30 & 1.31 \\
  & Top-quark pair production
  & 2.31 & 1.98
  & 1.31 & 1.24 \\
  & Single-top production
  & 0.38 & 0.36
  & 0.39 & 0.36 \\
  & Inclusive jet production
  & 0.83 & 0.85
  & 0.93 & 0.96  \\
  & Dijet production
  & 1.56 & 1.55
  & 1.94 & 2.03  \\
  & Direct photon production
  & 0.64 & 0.58
  & 0.74 & 0.75 \\
  \bottomrule
\end{tabularx}
\vspace{0.2cm}
\caption{\small Statistical estimators for NNPDF4.0QED NLO and NNLO,
  compared to NNPDF4.0 pure QCD.
  From top to bottom: total $\chi^2$ per number of data points, average
  over replicas of the training and validation figures of merit
  $\la E_{\rm tr}\ra_{\rm rep}$ and $\la E_{\rm val}\ra_{\rm rep}$,
  average $\chi^2$ over replicas $\la \chi^2\ra_{\rm rep}$,
  $\chi^2$ for datasets grouped by process. The total number of data
  points is 4424 (4616) at NLO (NNLO).
  \label{tab:chi2_qed_global}}
\end{table}
%--------------------------------------------------------------------------

It is clear from  Table~\ref{tab:chi2_qed_global} that all estimators
corresponding are essentially unchanged by the inclusion of
QED corrections, with slightly larger differences seen at NLO than at
NNLO: the impact of QED effects on fit quality is negligible,
both globally and for individual processes. Specifically, the training
and validation figures of merit in the pure QCD and QCD$\times$QED
determinations differ by less than one sigma.
As mentioned, these differences also include the effect of
switching from the TRN solution of evolution equations in the pure QCD
fit to the EXA solution. The impact of this change is yet smaller than
that of the QED corrections, but it slightly reduces it.

\subsection{The photon PDF}
\label{sec:gpdf}

%-----------------------------------------------------------------------------
\begin{figure}[!t]
  \centering
  \includegraphics[width=.49\textwidth]{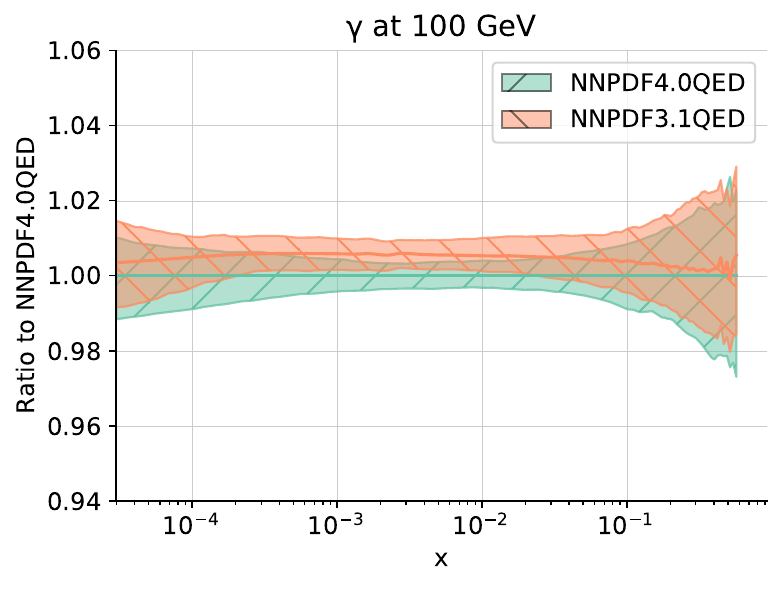}
  \includegraphics[width=.49\textwidth]{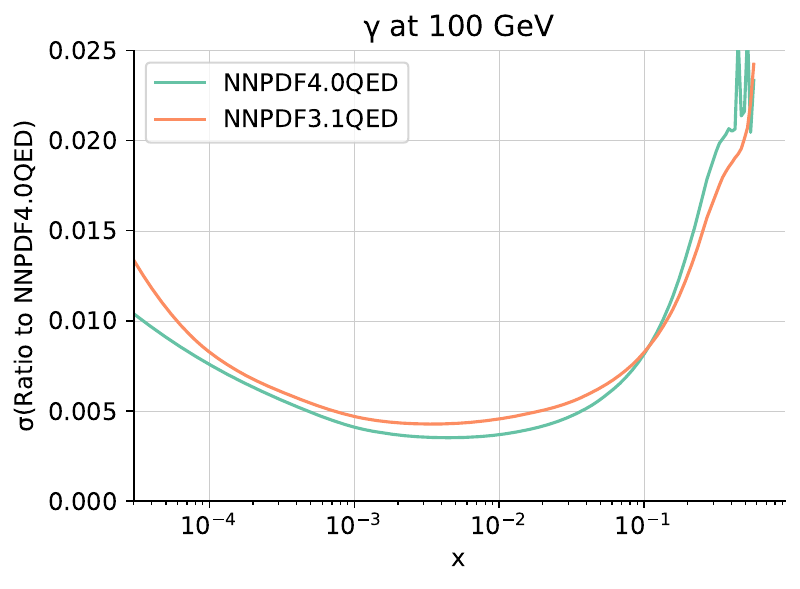}\\
  \includegraphics[width=.49\textwidth]{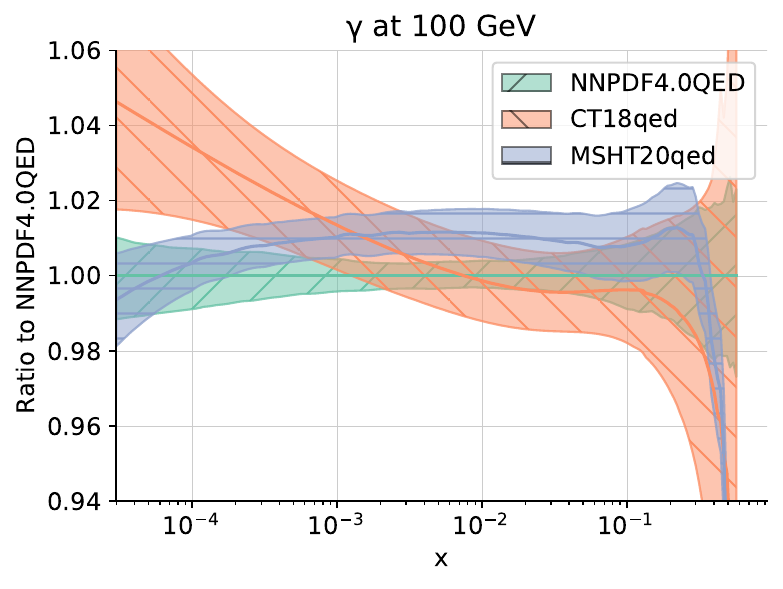}
  \includegraphics[width=.49\textwidth]{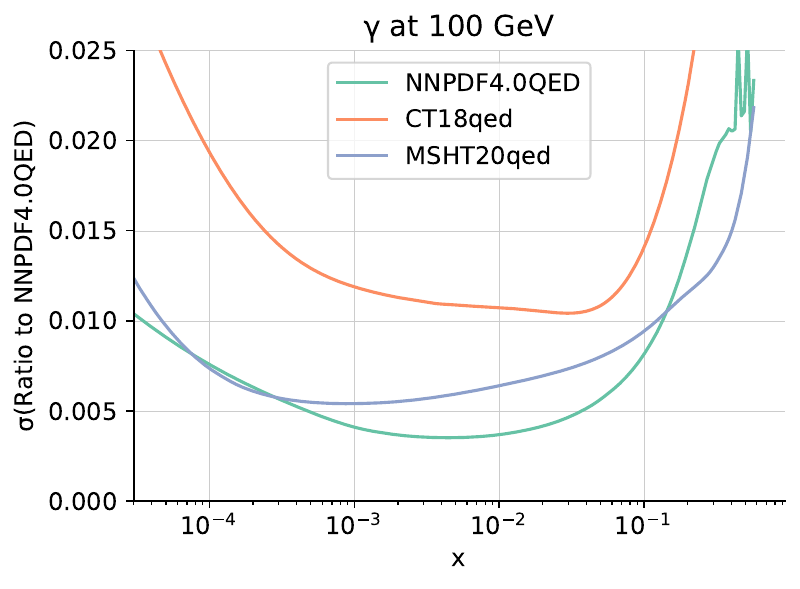}\\
  
  \caption{\small Top left: the photon PDF at $Q=100$~GeV in NNPDF4.0QED NNLO
    compared to its NNPDF3.1QED counterpart as a ratio to the central value of
    the former. Top right: the relative PDF uncertainties on the photon PDF in
    these two determinations. Bottom: same as top panels now comparing
    NNPDF4.0QED with MSHT20QED~\cite{Cridge:2021pxm} and CT18QED~\cite{Xie:2021ajm} (all NNLO). Bands correspond to 1$\sigma$
    uncertainties.
}
  \label{fig:PDFQED-q100gev-ratios} 
\end{figure}
%-------------------------------------------------------------------------------
In Fig.~\ref{fig:PDFQED-q100gev-ratios}, the NNLO NNPDF4.0QED photon PDF at
$Q=100$~GeV is compared to its counterpart in our previous
NNPDF3.1QED~\cite{Bertone:2017bme} set, and in the recent QED
PDF sets MSHT20QED~\cite{Cridge:2021pxm},
and CT18QED~\cite{Xie:2023qbn}.\footnote{The MSHT group has recently also
released the MSHT20qed\_an3lo PDF set, in which a photon PDF set is
added to PDFs treated with approximate N$^3$LO QCD theory~\cite{Cridge:2023ryv}.}
Here and elsewhere in this paper all
uncertainties correspond to 1$\sigma$.
Results agree at the percent level, despite the fact that quark and
gluon PDFs in these sets can display much larger differences.
This is a consequence of the fact that in all these PDF sets the photon PDF is
determined with the LuxQED formalism in terms of the proton structure function,
and that the latter, in turn, is well constrained by experimental data
both at high and low scale. In fact, we have checked that the dominant
contribution to the difference between the NNPDF3.1QED and NNPDF4.0QED
photon PDFs seen in Fig.~\ref{fig:PDFQED-q100gev-ratios} is due to the
difference between the TRN and EXA solutions of evolution equations
(see Section~\ref{sec:construction_NNPDF40QED} and
Appendix~\ref{app:exatrn}), i.e. to higher-order corrections,
with the residual difference being due to the change in the PDFs used
in order to compute the structure function. The fact that the 3.1 and
4.0 PDFs are compatible within uncertainties is thus a consequence of
the fact that the uncertainties due to higher-order  corrections and to
PDFs are
correctly accounted for by the LuxQED construction, while
the NNPDF4.0 PDFs
are backward-compatible with the NNPDF3.1 PDFs~\cite{Ball:2021leu}. 

The uncertainty on the photon PDF is completely dominated by
theoretical uncertainties on the LuxQED procedure, which
include~\cite{Manohar:2017eqh} missing higher order corrections, uncertainties on the experimentally measured low-scale
structure function and so on. In our uncertainty determination  we follow
Ref.~\cite{Manohar:2017eqh}; MSHT also mostly follows this reference,
with an extra higher-twist contribution due to the low choice of scale
$Q_\gamma$,~\cite{Harland-Lang:2019pla} and indeed finds a very similar
uncertainty. A somewhat more conservative uncertainty estimate is provided by
CT18QED~\cite{Xie:2023qbn}, which also adopts a somewhat different
determination of  the elastic
contribution.

%-------------------------------------------------------------------------------
\begin{figure}[!t]
  \centering
  \includegraphics[width=.49\textwidth]{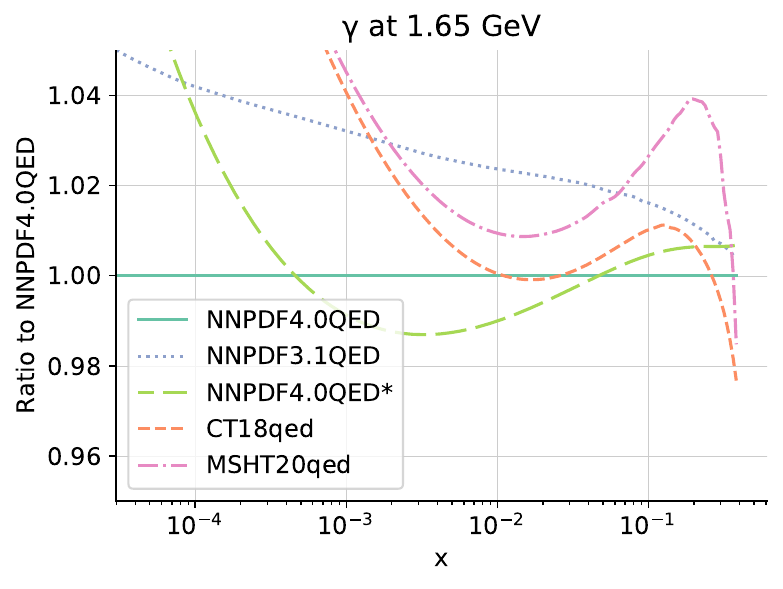}
  \includegraphics[width=.49\textwidth]{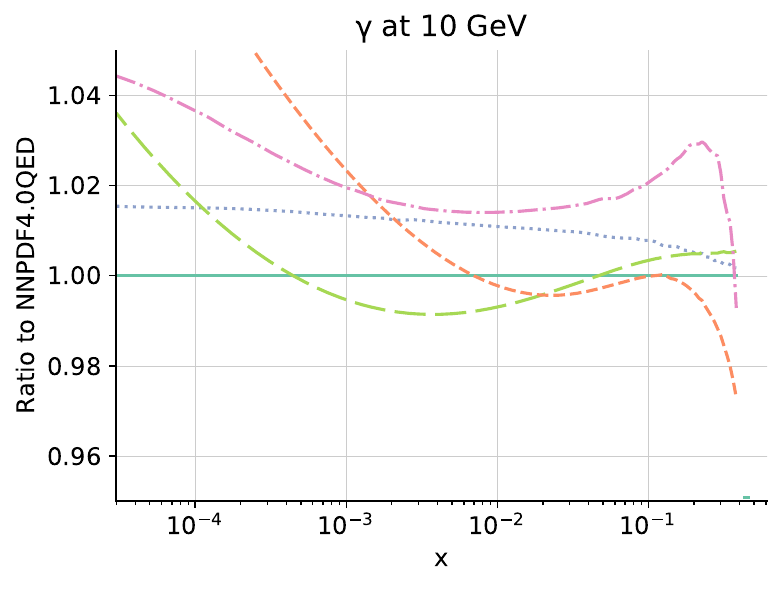}\\
  \includegraphics[width=.49\textwidth]{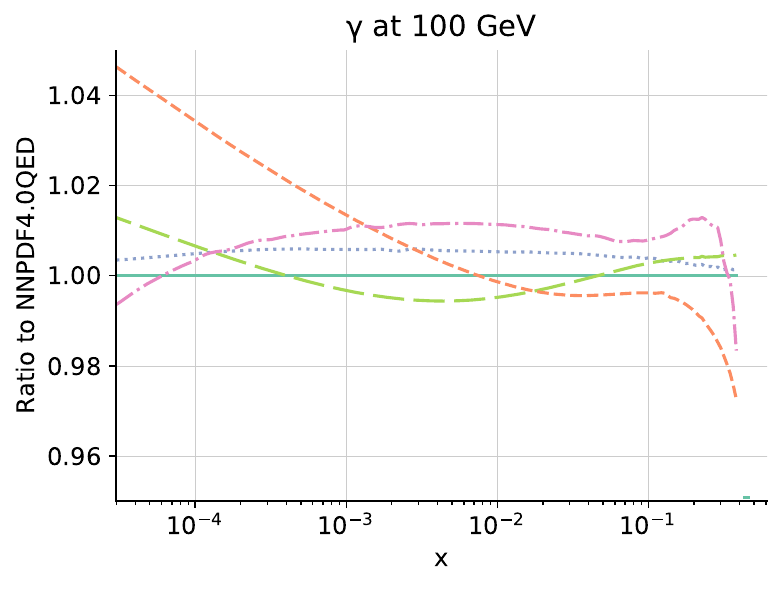}
  \includegraphics[width=.49\textwidth]{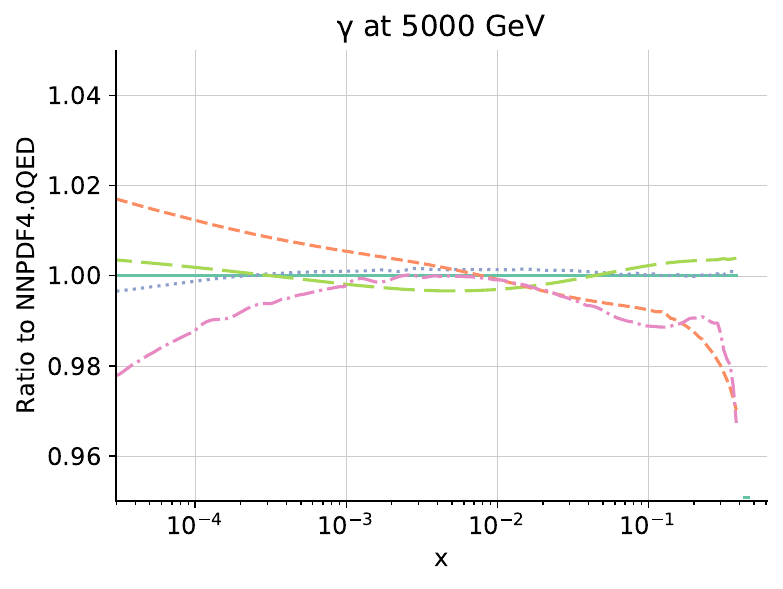}\\  
  \caption{\small Scale dependence of the central photon PDF in the
    NNPDF4.0QED, NNPDF3.1QED, MSHT20QED, and CT18QED PDF sets.
    The NNPDF4.0QED result in which the photon is determined at $Q=10$~GeV
    (denoted as NNPDF4.0QED*) is also shown.}
  \label{fig:QEDphoton-qdep-ratios} 
\end{figure}
%-------------------------------------------------------------------------------

In Fig.~\ref{fig:QEDphoton-qdep-ratios} the central value of the photon PDF
in all these sets is compared at different scales, using the native scale
dependence of each PDF set. Uncertainties are not shown in order not to clutter
the plot. Note that differences in the scale
dependence of the PDFs compared in the Figure also
arise due to somewhat different
treatments of the QCD$\times$QED evolution equations.  Specifically,
the NNPDF3.1QED PDF set adopts a numerical implementation of the TRN
solution (see  Appendix~\ref{app:exatrn}). However, one would expect
differences to be mostly driven by the mixing with the quark and gluon PDFs.
The fact that differences grow at low scales suggests that this is
indeed the case. Even so, all photon PDFs agree within about 3\% for
all $x\gtrsim10^{-3}$,
even at the lowest scale $Q=1.65$~GeV.

In Fig.~\ref{fig:QEDphoton-qdep-ratios} we also show the central
photon PDF which is found by repeating the NNPDF4.0QED determination 
with the photon PDF determined at a scale $Q_\gamma=10$~GeV instead of
the default 
$Q_\gamma=100$~GeV. If a low value of $Q_\gamma$ is adopted, the upper
limit of integration in $Q^2$, \cref{eq:LUX:QED}, is accordingly
lower. As discussed in Sect.~\ref{subsec:methodology} in such a case a
sizable contribution to the LuxQED  
formula comes from the low-scale region in which the structure
function is determined from a fit to the data, and ${\cal O}\left(\frac{m_p}{Q}\right)$
corrections to the LuxQED formula may then become relevant.  We see that the
shift in central photon PDF which is found by making this choice instead of the
default one is at  most of the order of the uncertainty on the photon
PDF. Again, this shows that the  uncertainty on  the LuxQED procedure
is correctly estimated.

%-------------------------------------------------------------------------------
\begin{figure}[!t]
  \centering
  \includegraphics[width=.49\textwidth]{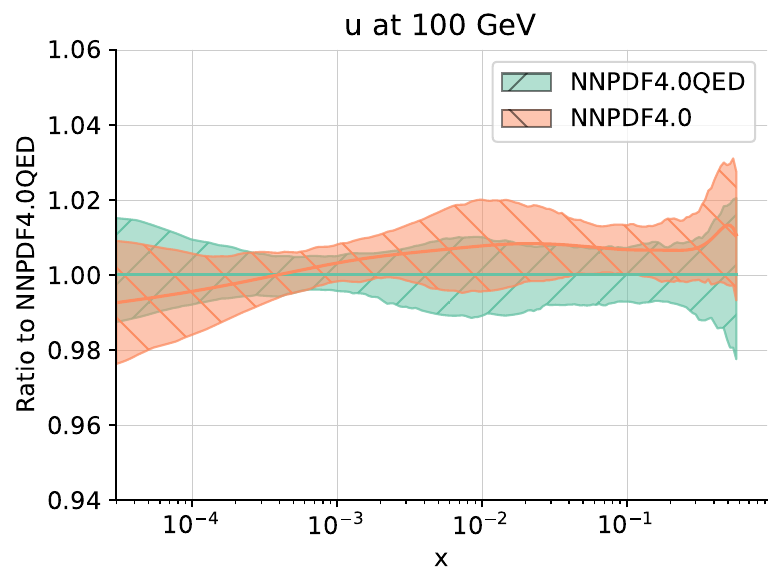}
  \includegraphics[width=.49\textwidth]{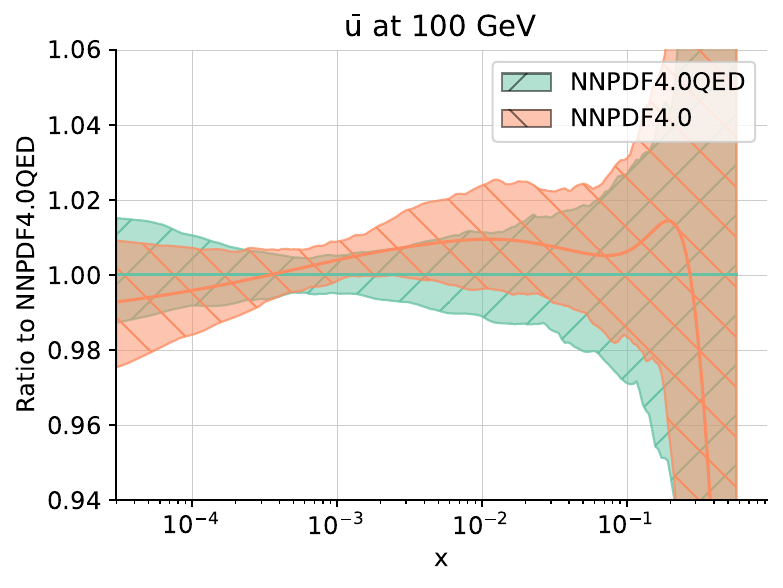}\\
  \includegraphics[width=.49\textwidth]{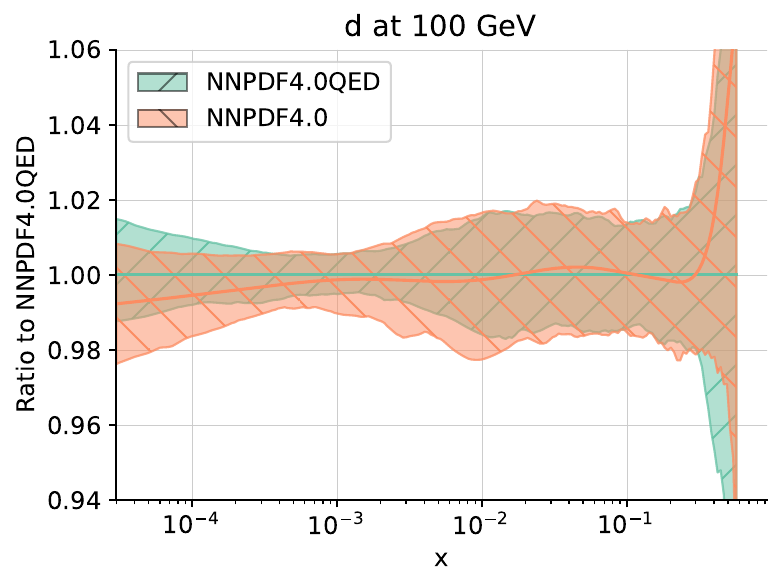}
  \includegraphics[width=.49\textwidth]{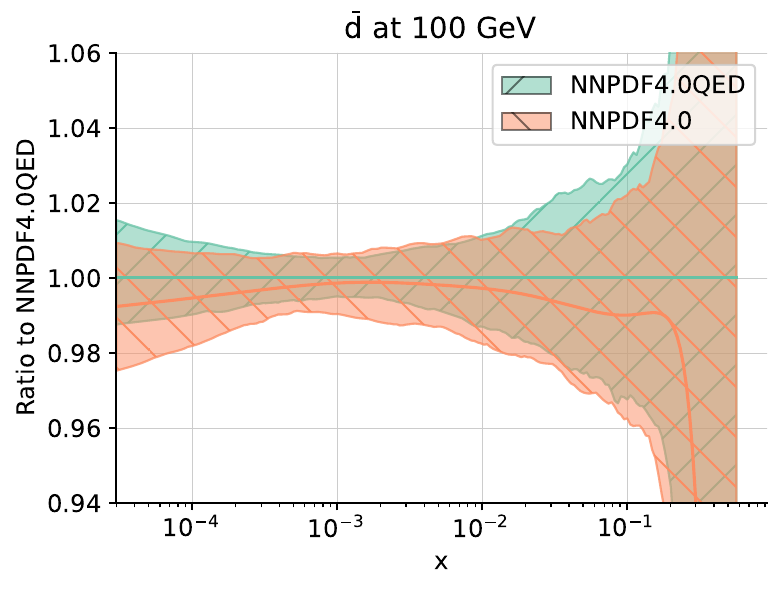}\\
  \includegraphics[width=.49\textwidth]{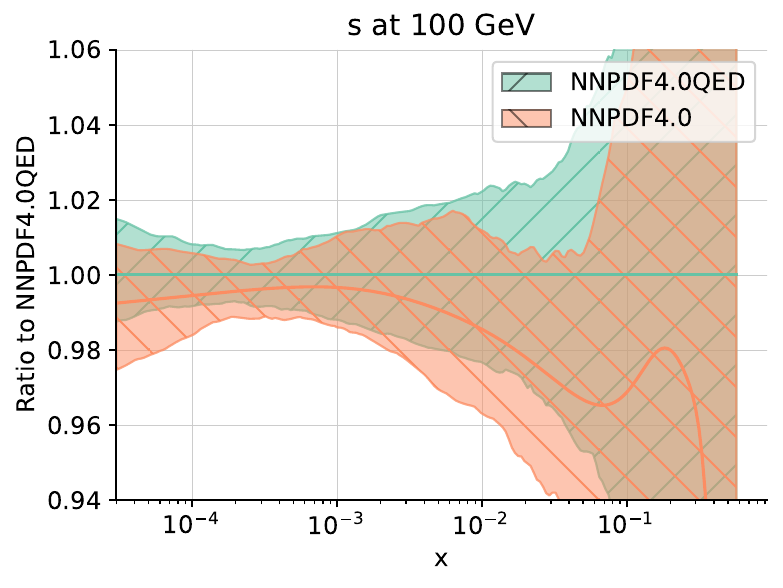}
  \includegraphics[width=.49\textwidth]{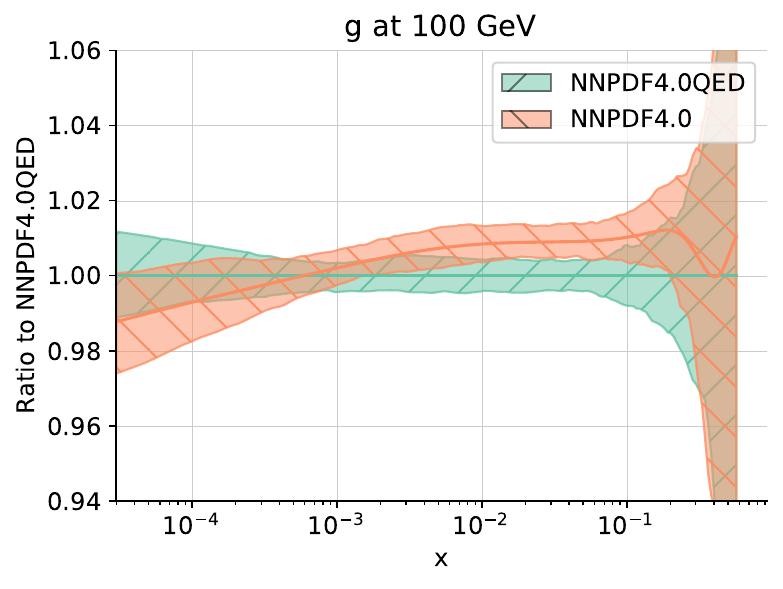}\\
  \caption{\small Comparison of PDFs
    in the NNPDF4.0QED  and the  NNPDF4.0 (pure QCD) sets, shown as a
    ratio to the  former
    at $Q=100$ GeV. Bands correspond to 1$\sigma$ uncertainties.
    From left to right and from top to bottom, the up, anti-up, down,
    anti-down, strange, and gluon PDFs are shown.}
  \label{fig:QED-q100gev-ratios} 
\end{figure}
%-------------------------------------------------------------------------------

We finally assess the impact of the inclusion of  a photon PDF on the other PDFs,
by comparing the NNLO NNPDF4.0QED and NNPDF4.0 (pure QCD) PDFs.
In Fig.~\ref{fig:QED-q100gev-ratios} we present this comparison at
$Q=100$~GeV. It is clear that the impact of the inclusion of the
photon is moderate, with the NNPDF4.0 (pure QCD) and NNPDF4.0QED PDFs
generally differing by less than one sigma and always in agreement
within uncertainties. The largest effects are seen in the
gluon PDF, which is suppressed at the percent level due to the momentum
fraction transferred from the gluon to the photon.

\subsection{The photon momentum fraction}
\label{sec:momentum_fraction}

\begin{figure}[!t]
  \centering
  \includegraphics[width=.49\textwidth]{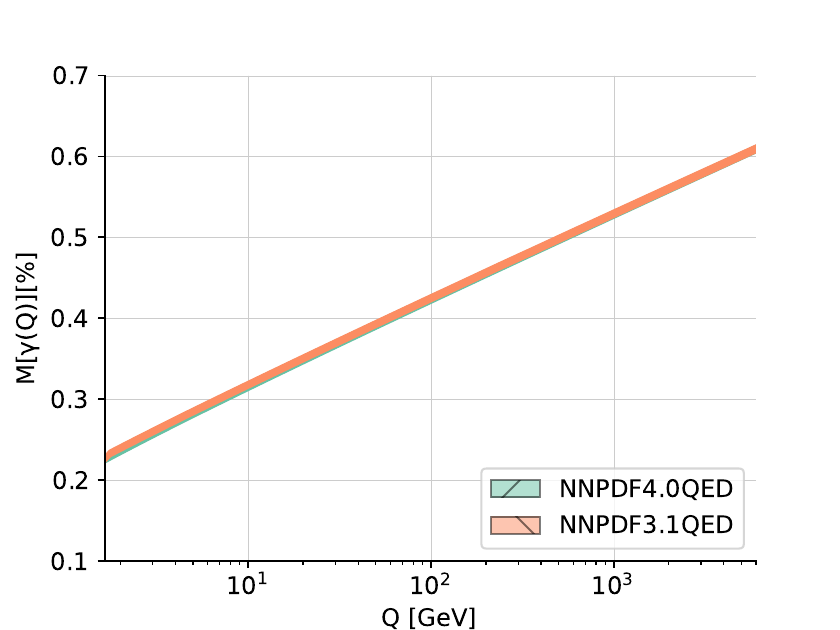}
  \includegraphics[width=.49\textwidth]{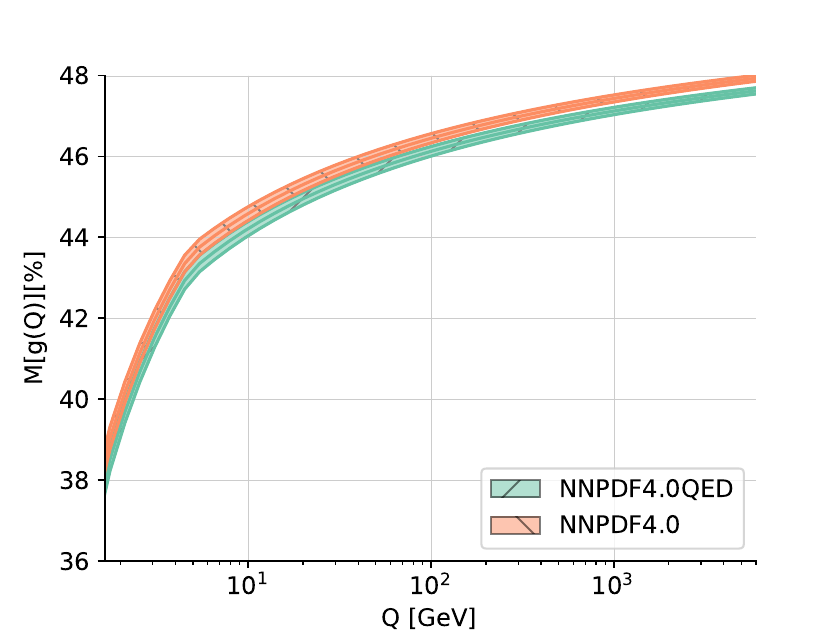}
  \caption{\small  Left: the percentage momentum fraction
    carried by the photon PDF $\gamma(x,Q^2)$ in NNPDF4.0QED
    and in NNPDF3.1QED as a function of the scale $Q$, where the
    bands indicate 1$\sigma$ uncertainties.
    Right: same for the momentum fraction carried by the gluon
    PDF in NNPDF4.0QED and in NNPDF4.0 (pure QCD).}
  \label{fig:momentum_fraction} 
\end{figure}

The main impact of the photon on other PDFs is through its contribution to the
momentum sum rule, Eq.~(\ref{eq:momsr}). We quantify this
contribution by evaluating the photon momentum fraction
\begin{equation}
  \label{eq:mom_fraction}
  M\lc \gamma(Q)\rc \equiv \int_0^1 dx\, x\gamma(x,Q)\,.
\end{equation}
In Fig.~\ref{fig:momentum_fraction} the momentum fractions carried by the
photon and by the gluon PDFs in the NNPDF4.0QED set are shown (in percentage)
as a function of scale. For the photon the result is compared to that of
NNPDF3.1QED, and for the gluon to that of NNPDF4.0 (pure QCD).
The photon momentum fraction in the NNPDF3.1QED and NNPDF4.0QED PDF sets is
essentially the same: the photon carries around 0.2\% of the proton momentum
at a low (Q$\sim$1 GeV) scale, growing logarithmically with $Q$ 
up to around 0.6\% at the multi-TeV scale. The momentum fraction
carried by the gluon is reduced by a comparable amount upon inclusion
of the photon.

%-------------------------------------------------------------------------------
%-------------------------------------------------------------------------------

% Phenomenology
\section{Implications for LHC phenomenology}
\label{sec:pheno}

We now study the phenomenological implications of the NNPDF4.0QED
PDF set. First we compare parton luminosities to
those computed using other QED PDF sets. Then we assess the impact of QED
corrections on selected processes, by comparing to the pure QCD case
calculations that include photon-induced contributions, and also by directly
comparing results obtained using NNPDF4.0 (pure QCD) and NNPDF4.0QED PDFs.

\subsection{Luminosities}
\label{sec:lumis}

%-------------------------------------------------------------------------------
\begin{figure}[!t]
  \centering
  \includegraphics[width=.49\textwidth]{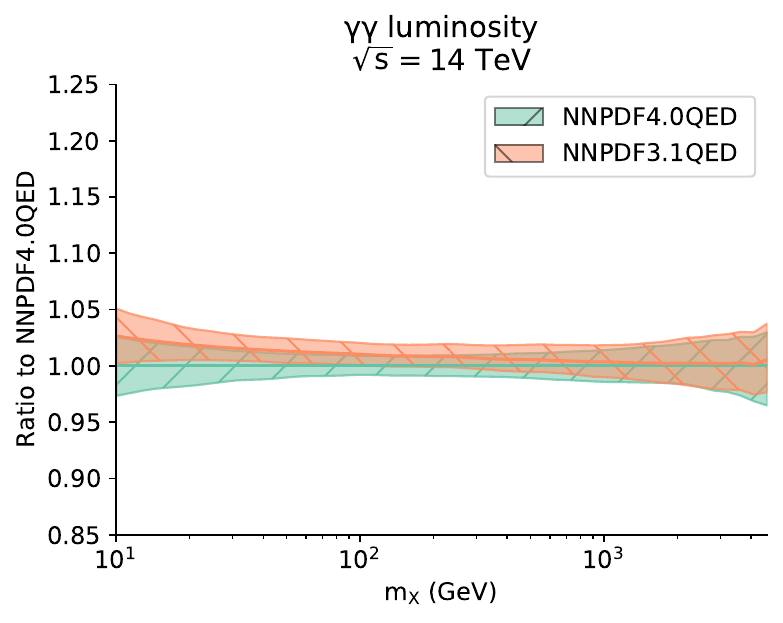}
  \includegraphics[width=.49\textwidth]{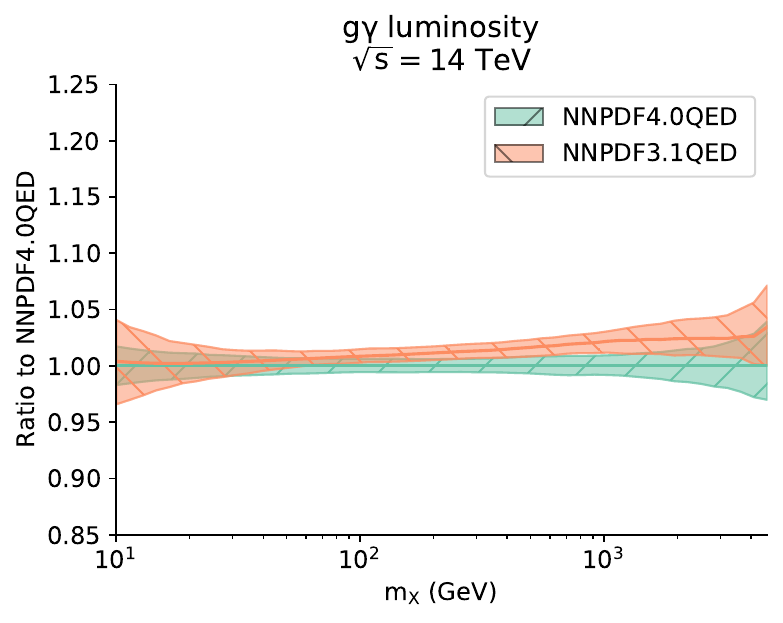}\\
  \includegraphics[width=.49\textwidth]{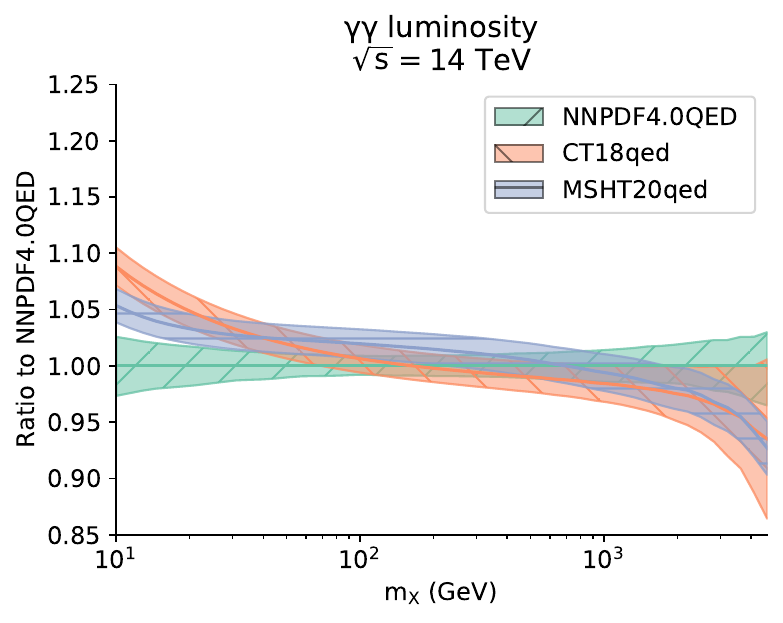}
  \includegraphics[width=.49\textwidth]{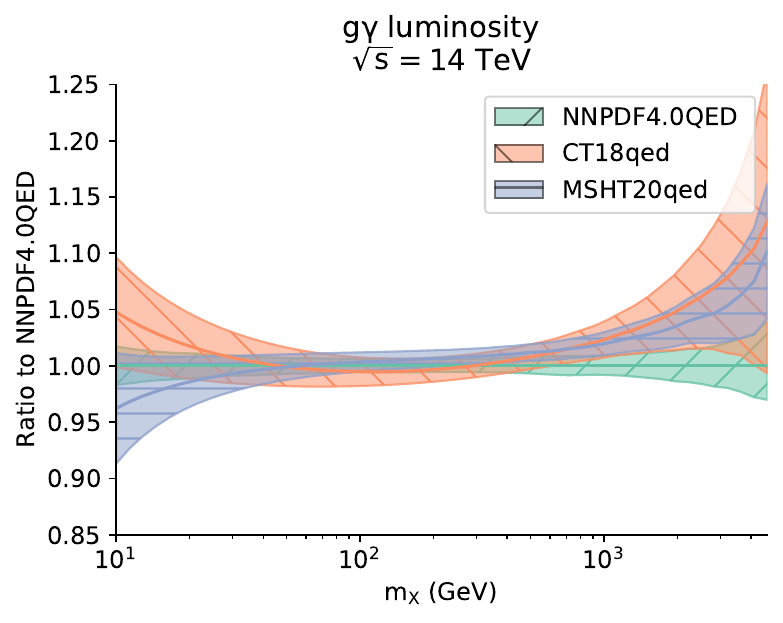}\\
  \caption{\small The photon-induced contributions to the luminosity
    at the LHC with 
    $\sqrt{s}=14$~TeV as a function of the invariant mass $m_X$
    for NNPDF4.0QED NNLO compared to its NNPDF3.1QED
    counterpart (top) and compared to MSHT20QED 
and CT18QED,  all shown as a ratio to NNPDF4.0QED. }
  \label{fig:lumis_0} 
\end{figure}
%-------------------------------------------------------------------------------

The phenomenological effect on the parton luminosity of the inclusion
of a photon PDF is both direct, through the presence of photon-induced
partonic channels, and indirect, through the effect of the photon on
other PDFs, mostly through the depletion of the gluon that is
necessary in order to preserve the momentum sum rule, as discussed in
Sect.~\ref{sec:gpdf}-\ref{sec:momentum_fraction}. The photon-induced
contributions for NNPDF4.0QED NNLO are compared in
Fig.~\ref{fig:lumis_0} to their counterpart in 
NNPDF3.1QED (top) and 
in  MSHT20QED
and CT18QED (bottom). The level of agreement is high and directly follows from
that seen between the respective photon PDFs in
Figs.~\ref{fig:PDFQED-q100gev-ratios}

%-------------------------------------------------------------------------------
\begin{figure}[!t]
  \centering
  \includegraphics[width=.49\textwidth]{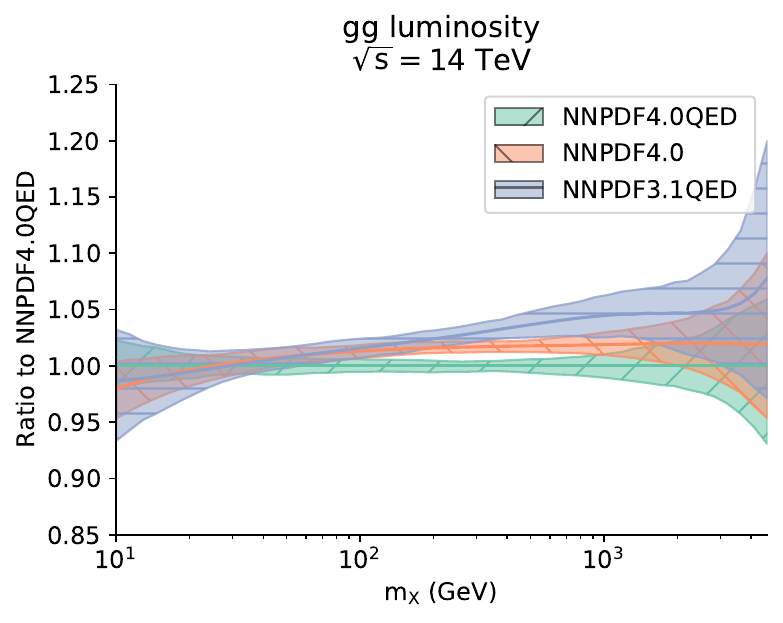}
  \includegraphics[width=.49\textwidth]{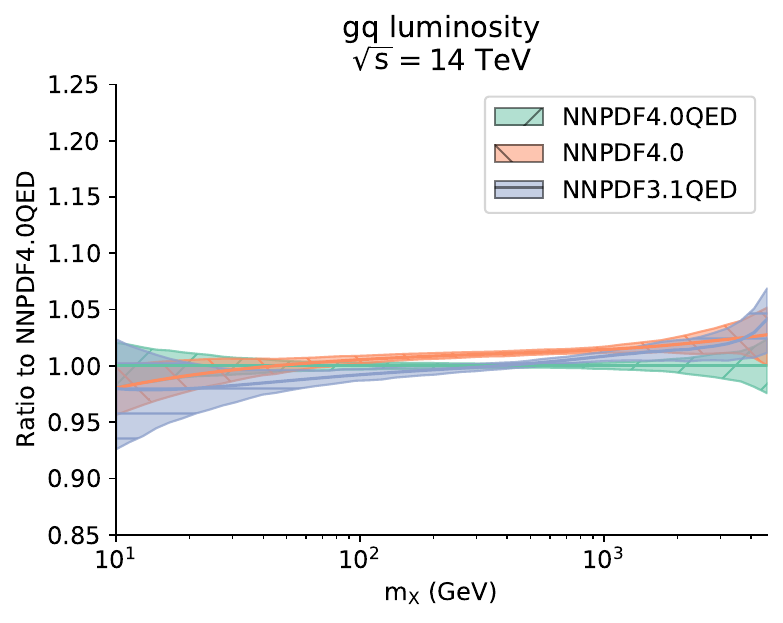}\\
  \includegraphics[width=.49\textwidth]{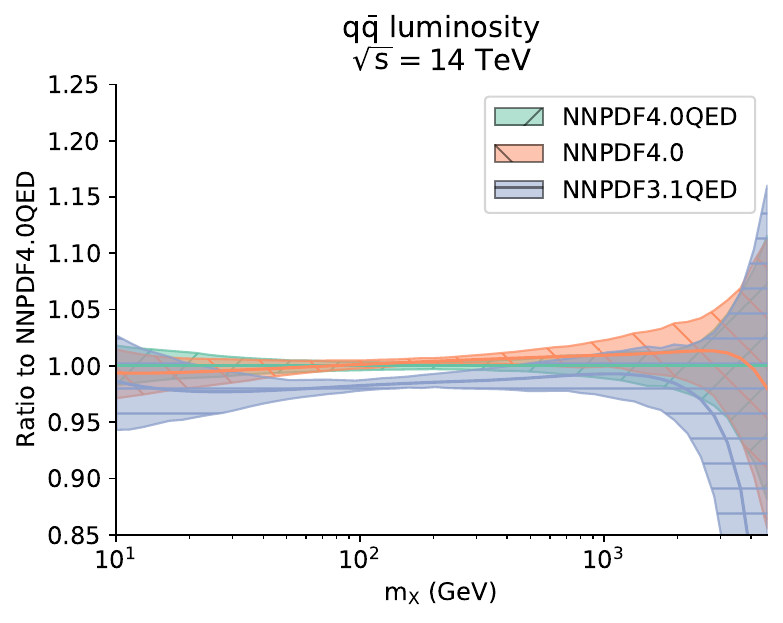}
  \includegraphics[width=.49\textwidth]{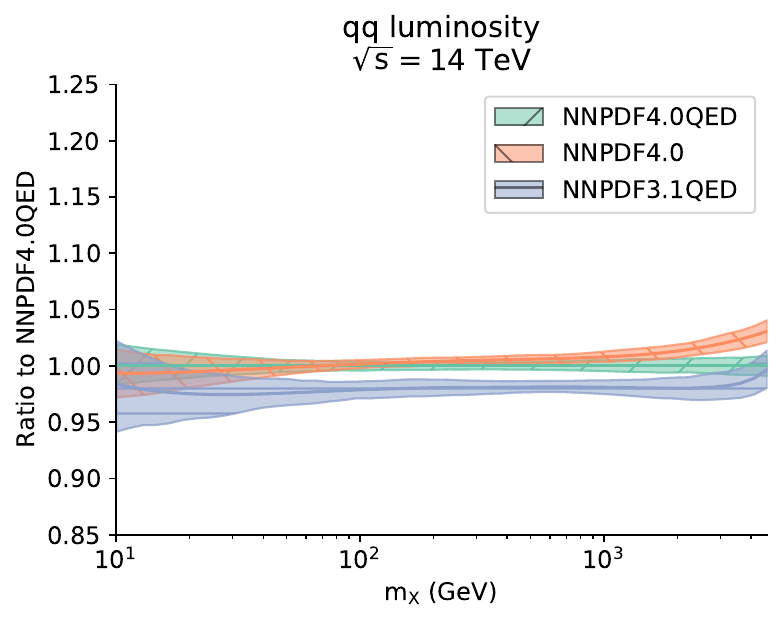}\\  
  \caption{\small Comparison of parton
  luminosities in the NNPDF4.0QED and the NNPDF4.0 (pure QCD) sets,
  shown as a ratio to the former for the LHC with 
    $\sqrt{s}=14$~TeV. From left to right and from top to bottom the
gluon-gluon, gluon-quark, quark-antiquark and quark-quark luminosities
are shown.}
  \label{fig:lumis_1} 
\end{figure}
%-------------------------------------------------------------------------------
The luminosities for all other
parton channels are shown in Fig.~\ref{fig:lumis_1}, where we compare
NNPDF4.0QED both to the previous set NNPDF3.1QED and to the pure QCD
NNPDF4.0. Because, as shown in Sect.~\ref{sec:results}, the effect of
the inclusion of 
the photon on the other PDFs is moderate, the comparison between the
two QED sets is very similar to the comparison between the pure QCD
NNPDF4.0 and NNPDF3.1 shown in Fig.~9.1 of Ref.~\cite{Ball:2021leu},
where it was shown that  NNPDF4.0 are backward compatible,
i.e.\ generally agree with NNPDF3.1 within uncertainties. The effect of
the inclusion of QED corrections is, as expected, mostly seen in the
gluon--gluon channel, where it leads to a suppression of  a few
percent in the $m_X\sim100$~GeV region in  order to account for the
transfer of a small amount of momentum to the photon.
The comparison to other PDF sets is dominated by the differences
in the quark and gluon PDFs, which are rather more significant than
the difference in the photon PDF, and thus very similar to the
corresponding comparison of pure QCD luminosities shown in Fig.~9.3 of Ref.~\cite{Ball:2021leu}.

%-------------------------------------------------------------------------------
\begin{figure}[!t]
  \centering
  \includegraphics[width=.49\textwidth]{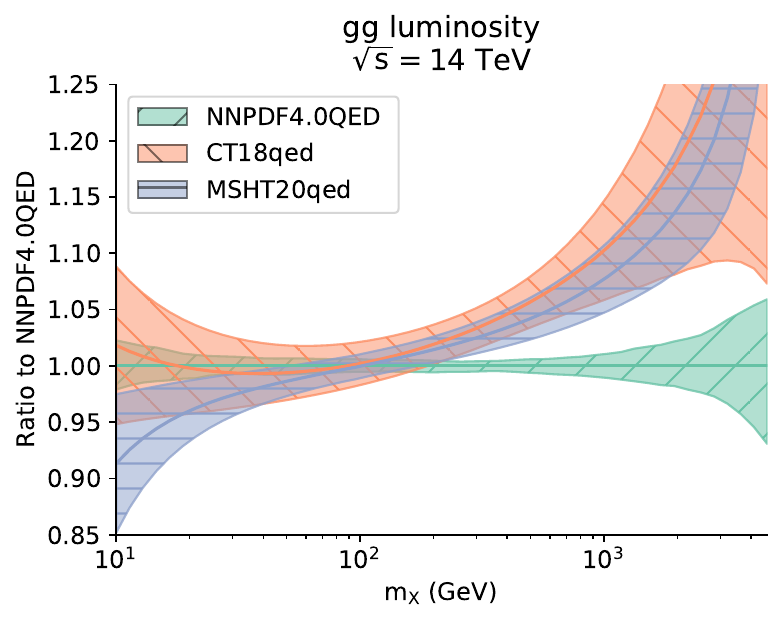}
  \includegraphics[width=.49\textwidth]{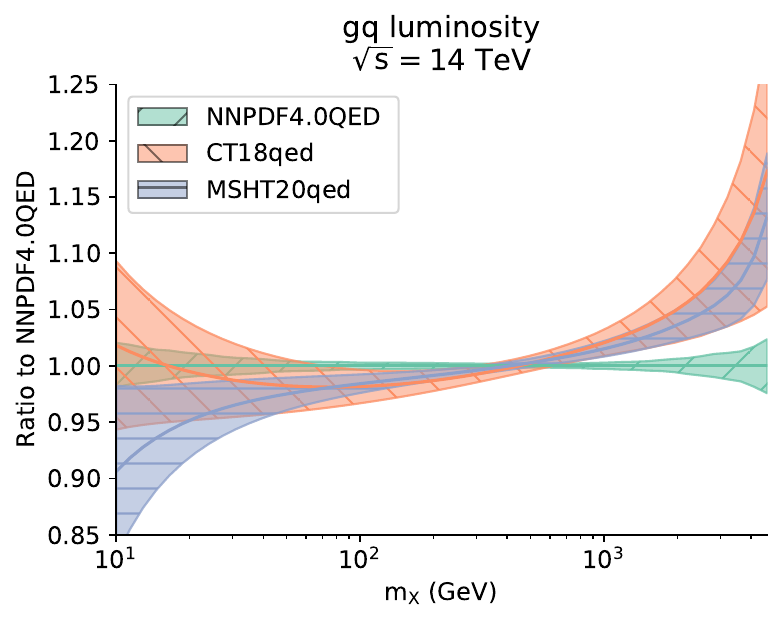}\\
  \includegraphics[width=.49\textwidth]{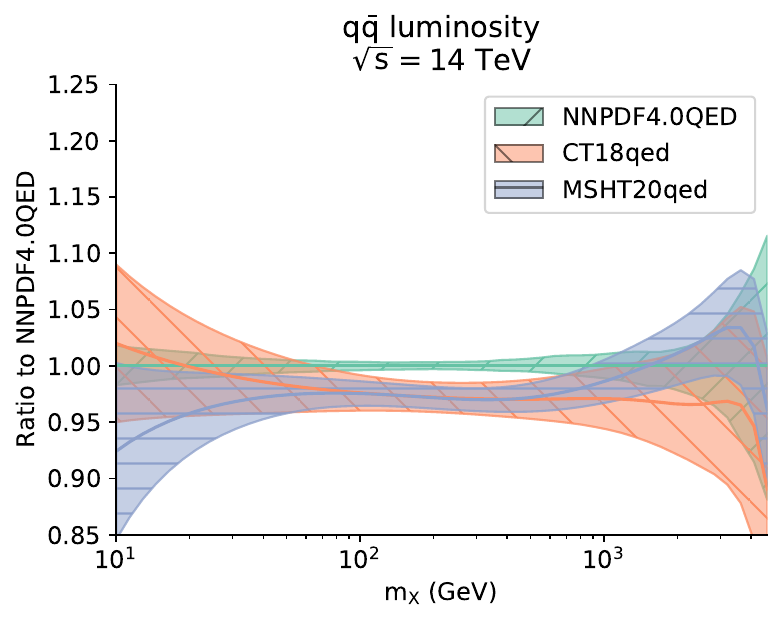}
  \includegraphics[width=.49\textwidth]{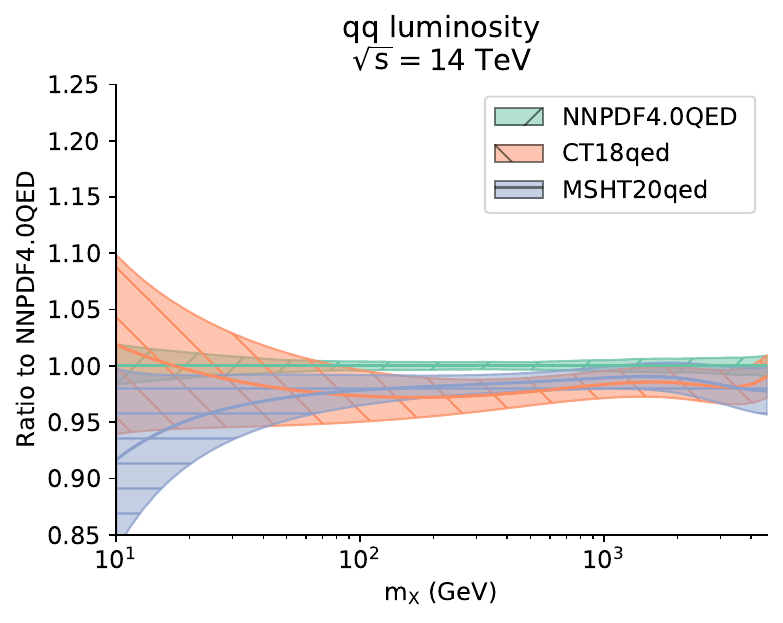}\\ 
  \caption{\small Same as Fig.~\ref{fig:lumis_1}, now comparing NNPDF4.0QED to
    MSHT20~\cite{Cridge:2021pxm} and CT18qed~\cite{Xie:2023qbn}.}
  \label{fig:lumis_2} 
\end{figure}
%-------------------------------------------------------------------------------

\subsection{Physics processes}
\label{sec:processes}

We now study the impact of the photon PDF on a few representative processes:
Drell--Yan production (neutral- and charged-current), Higgs production in
gluon-gluon fusion, in vector boson fusion, and in associated production with weak
bosons, diboson production, and top-quark pair production, all at the LHC with
center-of-mass energy $\sqrt{s}=14$ TeV. We have computed theory predictions
for these processes exploiting the
{\sc\small PineAPPL}~\cite{christopher_schwan_2023_7995675,Carrazza:2020gss} 
interface to the automated QCD and EW calculations provided by
{\sc\small mg5\_aMC@NLO}~\cite{Frederix:2018nkq}. {\sc\small PineAPPL} produces
interpolation grids, accurate to NLO in both the strong and electroweak
couplings, that are independent of PDFs. They therefore
make it easy to vary the input PDF set, since the same grid can be used for all
PDF sets considered. As we are interested in assessing differences between PDF
sets, rather than in doing precision phenomenology, we do not include NNLO
QCD corrections, and we only show PDF
uncertainties. For a detailed list of parameters and cuts used in the
calculation of these grids, see Sect.~9 of Ref.~\cite{Ball:2021leu}.

For all processes, we display predictions in figures below with a standardized
format, as follows. We show the absolute distributions (top panels) and the
ratio to the central value obtained using NNPDF4.0QED PDFs and including
photon-induced channels, which we call NNPDF4.0QED (bottom panels). In the left
plots we compare NNPDF4.0QED to: NNPDF4.0QED but with no photon-initiated
channels; NNPDF4.0 pure QCD;  NNPDF3.1QED. In the right plots we compare
NNPDF4.0QED to MSHT20QED and CT18QED. The left plots allow for  assessing
the overall size of the QED corrections (by comparison of the QED and pure QCD
results), and disentangling the size of the photon-initiated contributions
(by comparison of predictions with the photon-initiated channels switched
on and off) and the impact of the changes in the quark and gluon PDFs due to
QED effects (by comparison of NNPDF4.0 with NNPDF4.0QED with the
photon-initiated channels switched off).

%-------------------------------------------------------------------------------
\begin{figure}[!t]
  \centering
  \includegraphics[width=.49\textwidth]{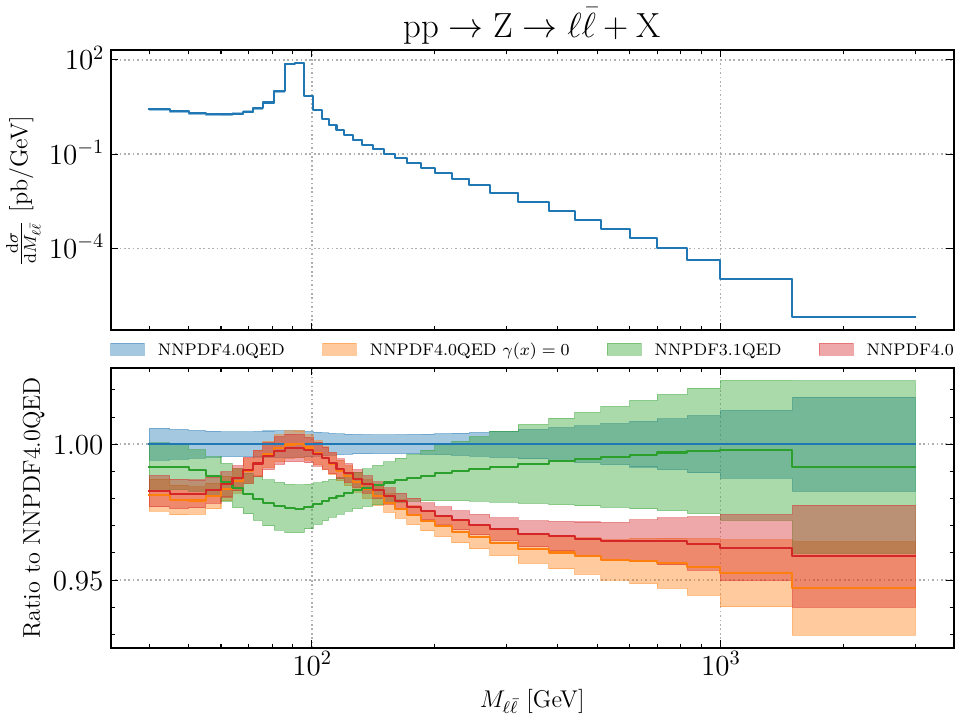}
  \includegraphics[width=.49\textwidth]{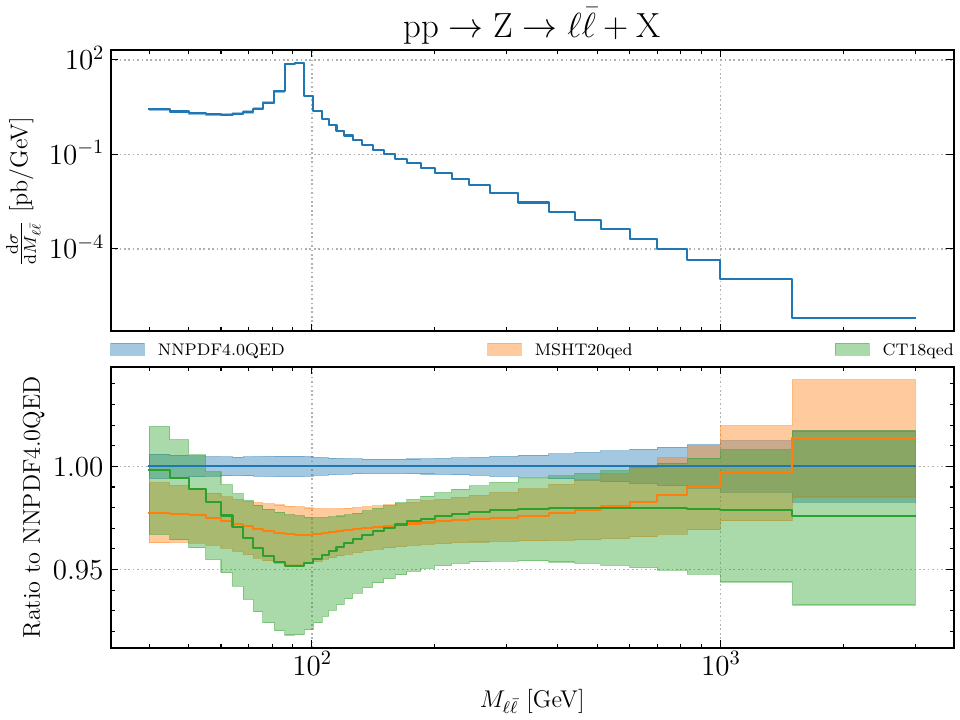}
  \includegraphics[width=.49\textwidth]{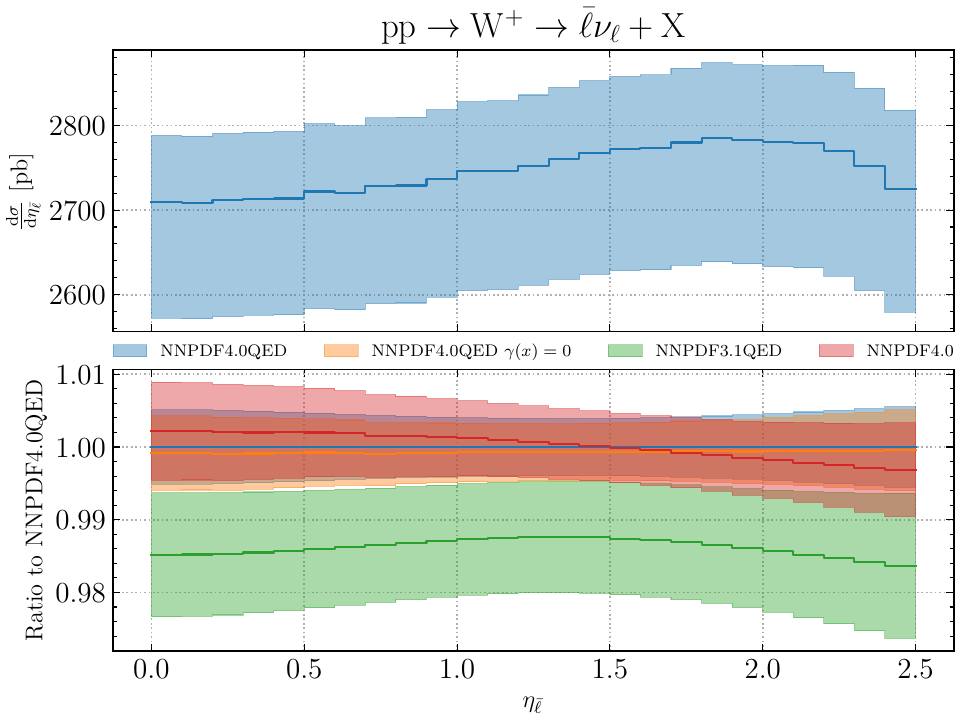}
  \includegraphics[width=.49\textwidth]{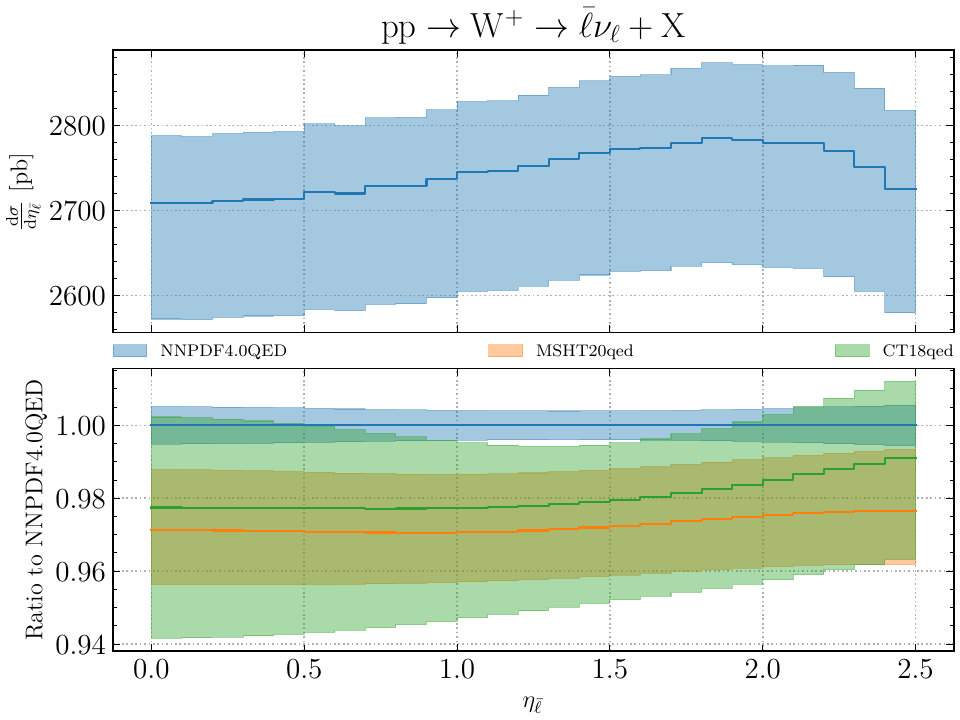}
  \includegraphics[width=.49\textwidth]{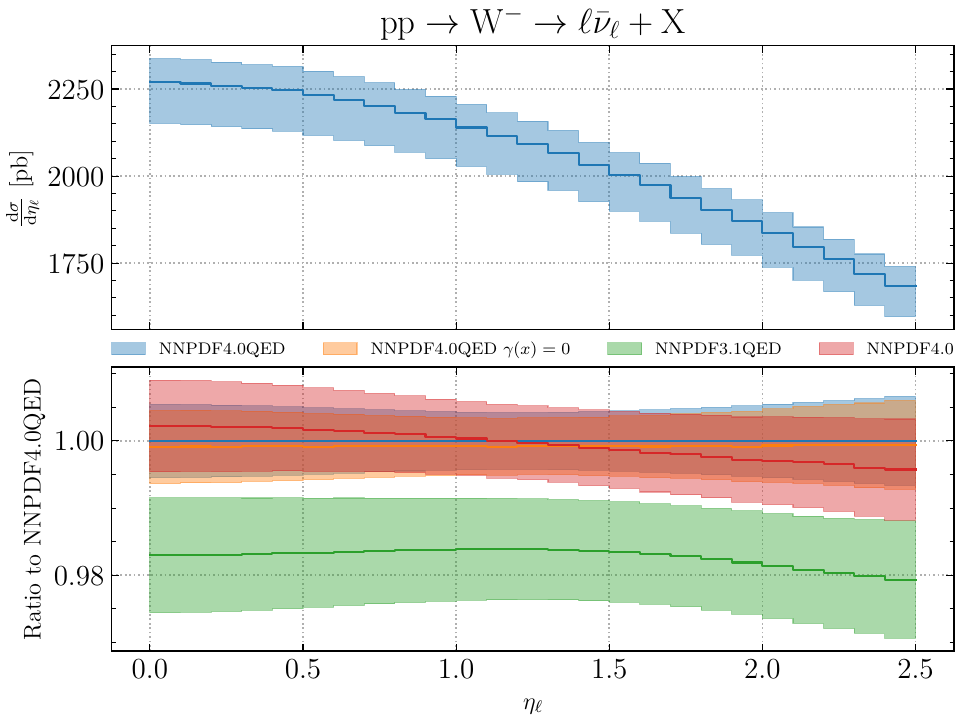}
  \includegraphics[width=.49\textwidth]{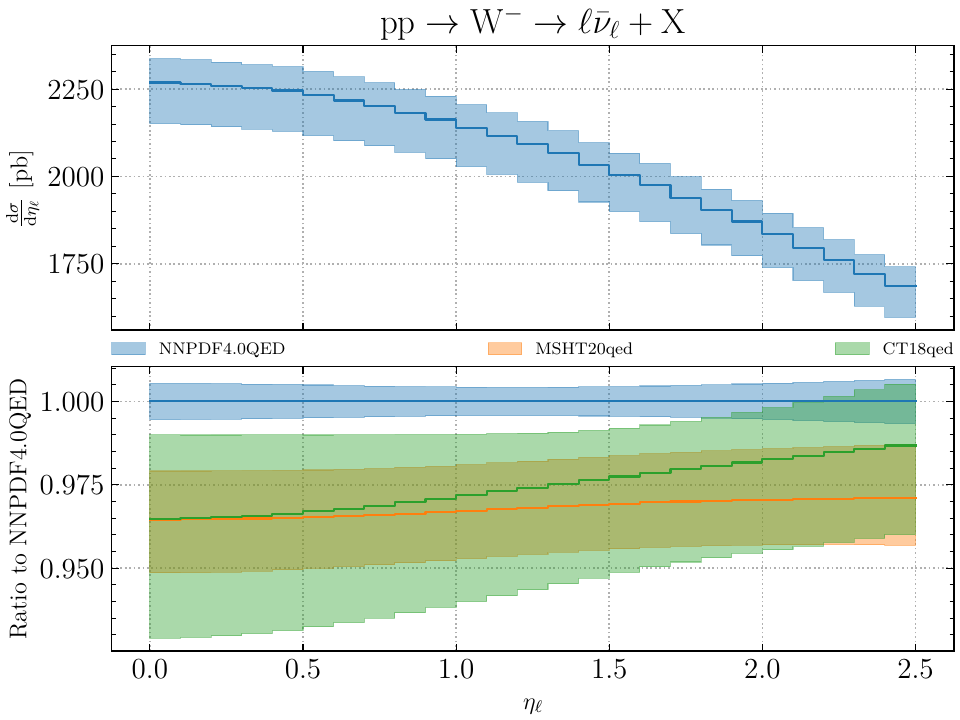}
  \caption{\small Predictions for inclusive Drell--Yan production
    at the LHC with center-of-mass energy $\sqrt{s}=14$ TeV,
    computed at NLO accuracy in the QCD and electroweak couplings. All
    uncertainties shown are PDF uncertainties only.
    From top to bottom: neutral-current dilepton
    production  
    as a function of the dilepton invariant mass
    $m_{\ell\bar{\ell}}$;
    $W^+$ production as a function of the antilepton
    pseudo-rapidity $\eta_{\bar{l}}$;
    $W^-$ production as a function of the antilepton
    pseudo-rapidity $\eta_{l}$.
    For each process, we display the absolute distributions (top panels)
    and the ratio to the central value obtained with NNPDF4.0QED
    PDFs, including photon-initiated channels (bottom panels).
    In the left panels the full NNPDF4.0 result is compared to
    NNPDF4.0 (QCD only), NNPDF3.1QED, NNPDF4.0QED (no photon-initiated);
    in the right panels it is compared to
    MSHT20QED~\cite{Cridge:2021pxm} and
    CT18QED~\cite{Xie:2023qbn}. Note that the experimentally
    measurable quantity is shown, so for dilepton production the
    $t$- and $u$-channel photon-induced contribution is also included. }
  \label{fig:NNPDF_DY_14TEV_40_PHENO} 
\end{figure}
%-------------------------------------------------------------------------------

\begin{figure}[!t]
  \centering
  \includegraphics[width=.49\textwidth]{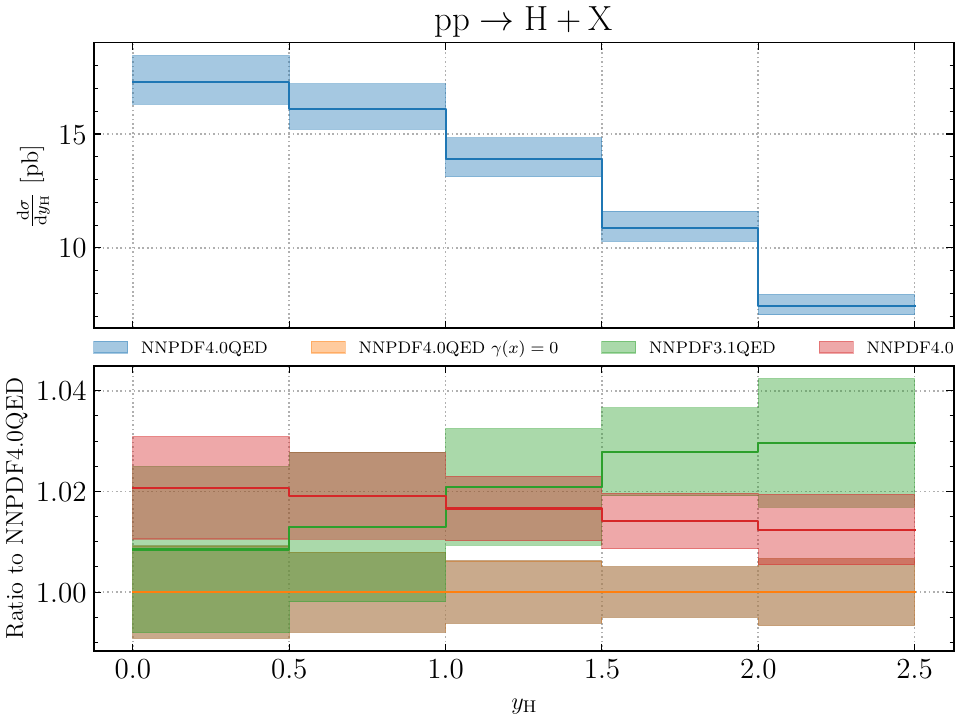}
  \includegraphics[width=.49\textwidth]{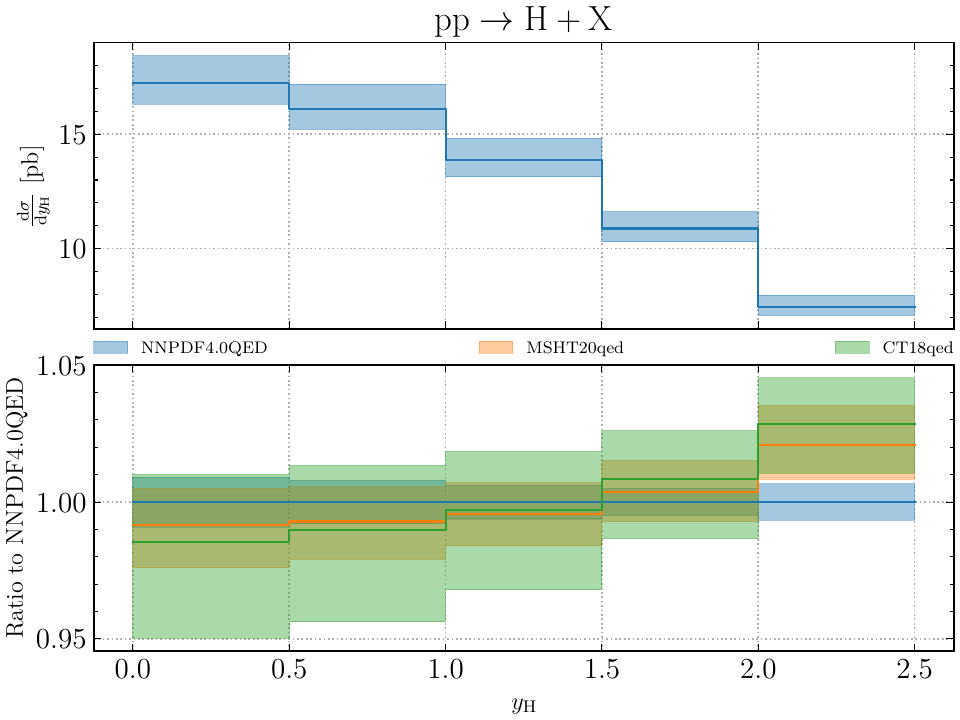}
  \includegraphics[width=.49\textwidth]{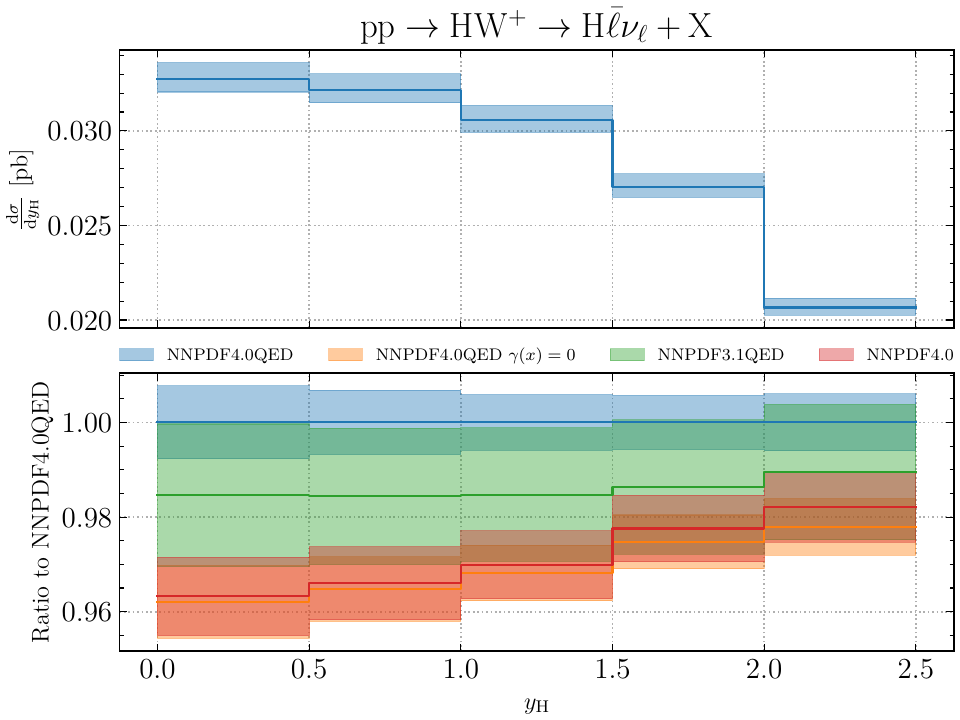}
  \includegraphics[width=.49\textwidth]{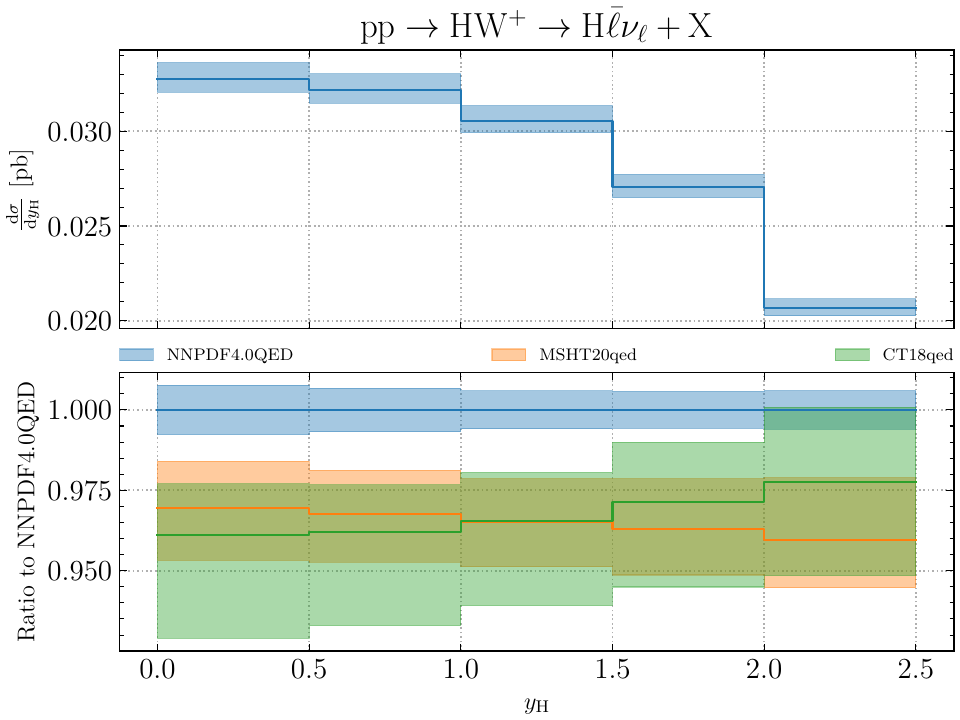}
   \includegraphics[width=.49\textwidth]{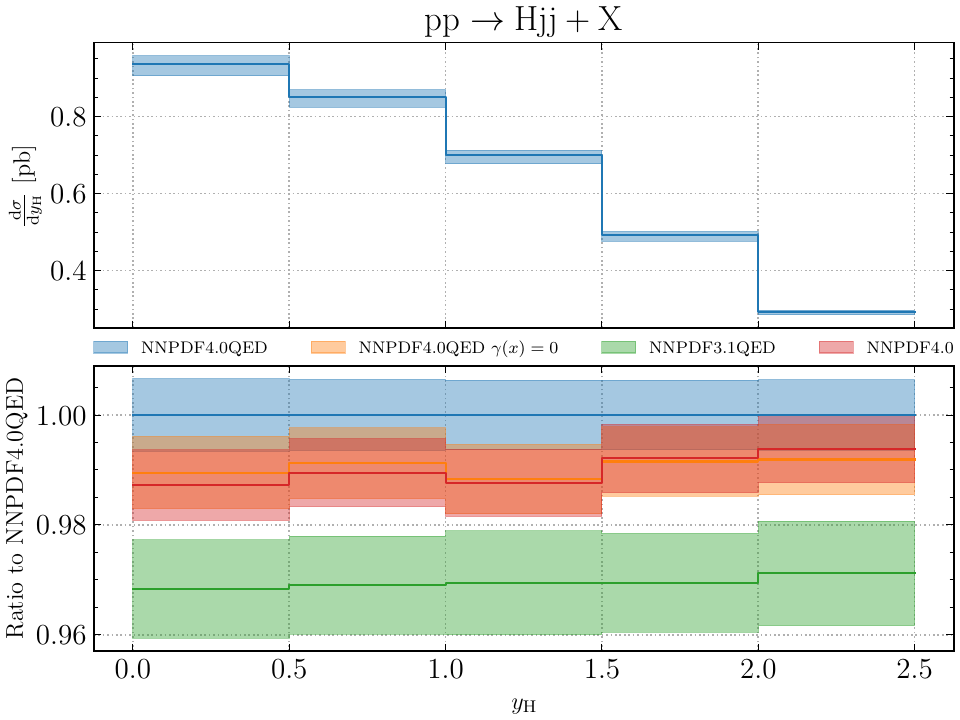}
  \includegraphics[width=.49\textwidth]{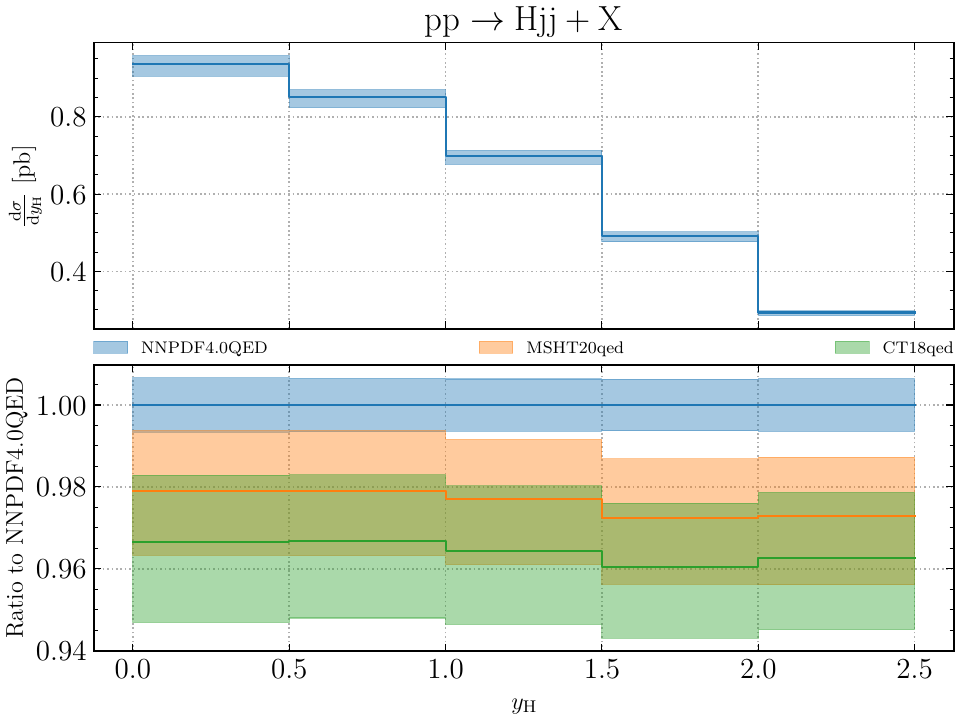}
  \caption{\small Same as Fig.~\ref{fig:NNPDF_DY_14TEV_40_PHENO}
    but for the rapidity distribution of the Higgs for production in
    gluon-gluon fusion (top panel), in association with a $W^+$ boson
    (middle panel) and in vector-boson fusion (bottom panel).
  }
  \label{fig:NNPDF_H_14TEV_40_PHENO} 
\end{figure}

\begin{figure}[!t]
  \centering
  \includegraphics[width=.49\textwidth]{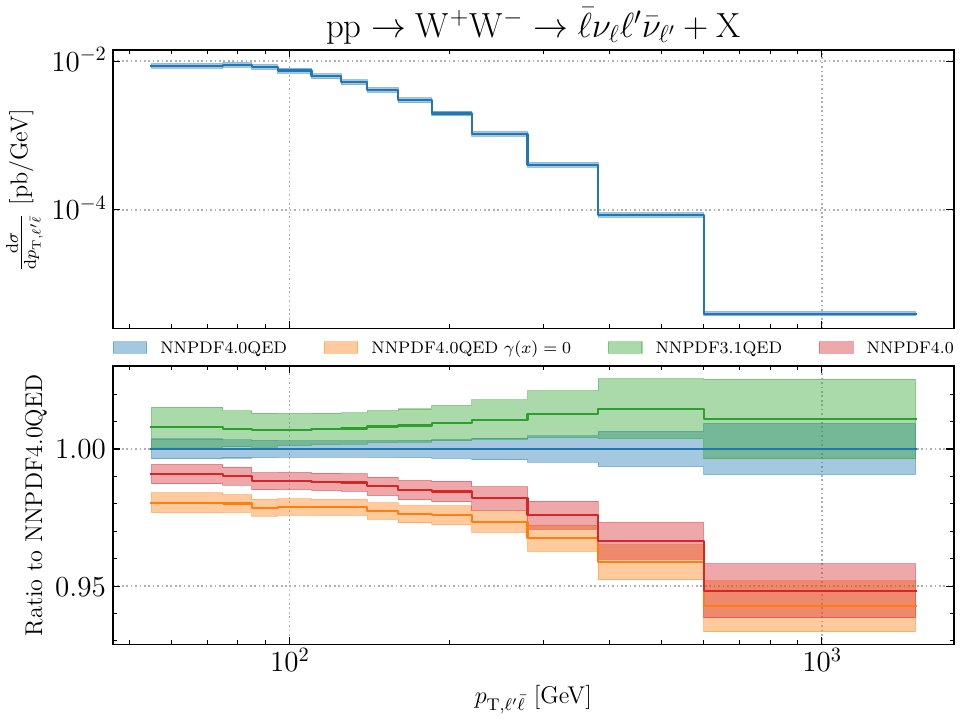}
  \includegraphics[width=.49\textwidth]{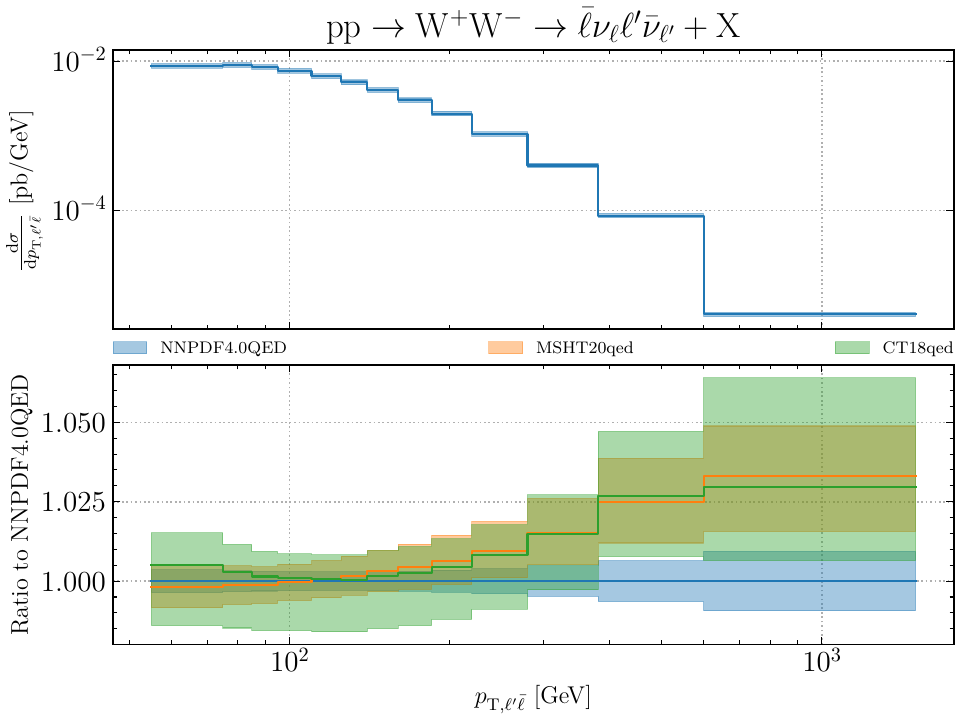}
  \includegraphics[width=.49\textwidth]{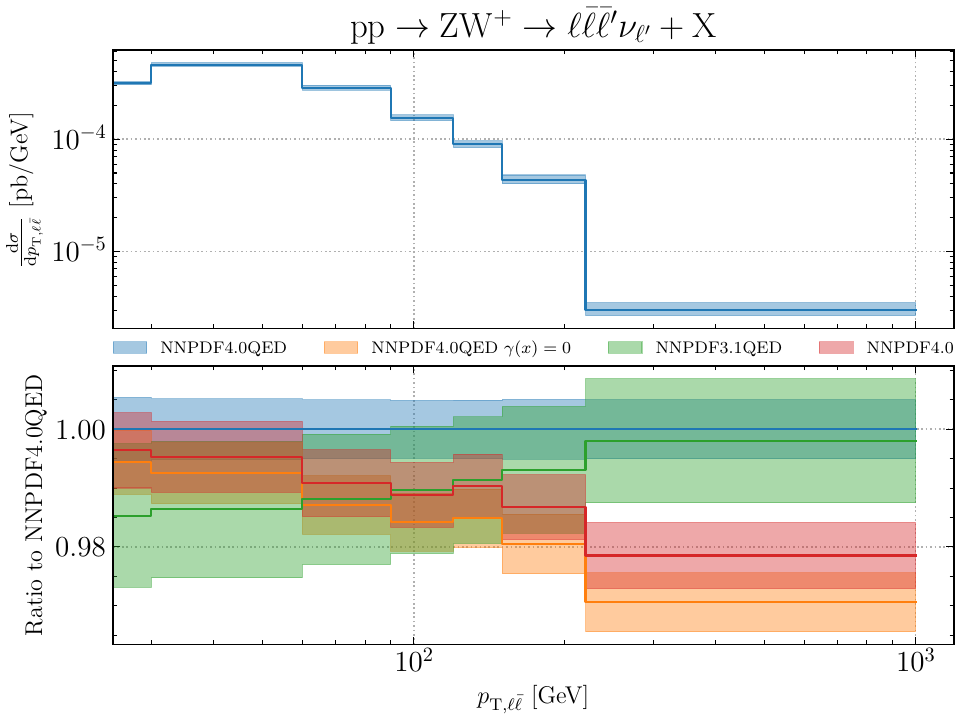}
  \includegraphics[width=.49\textwidth]{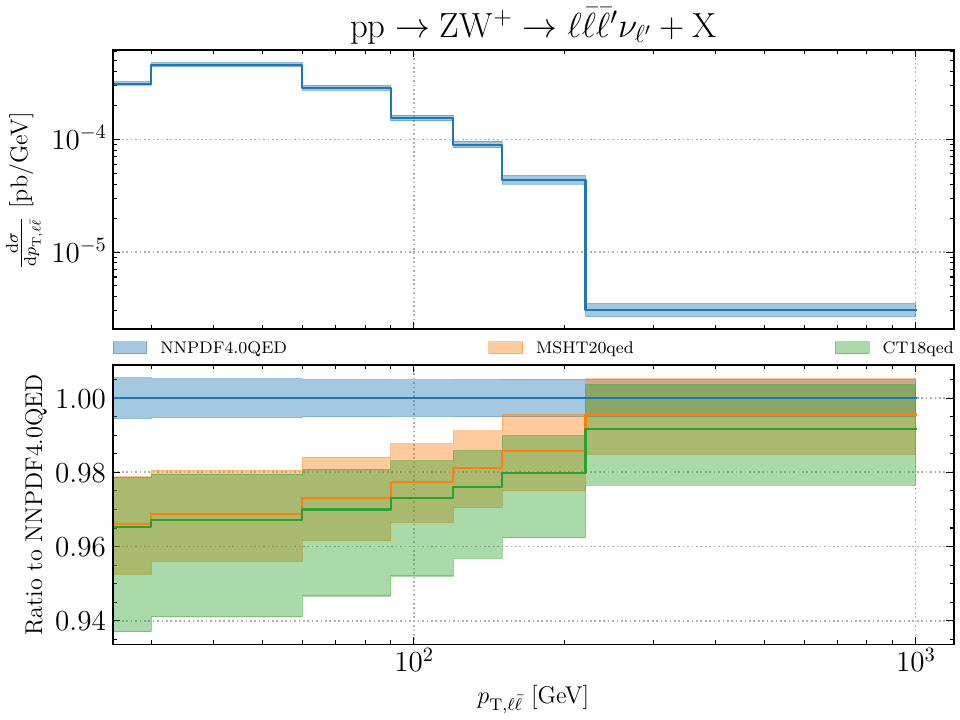}
  \includegraphics[width=.49\textwidth]{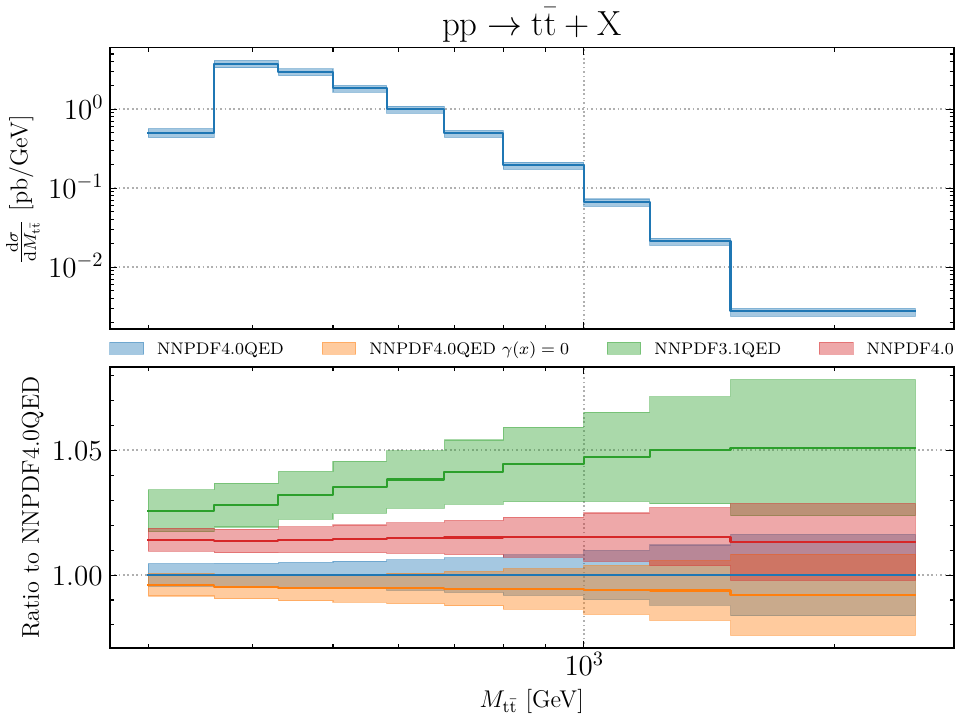}
  \includegraphics[width=.49\textwidth]{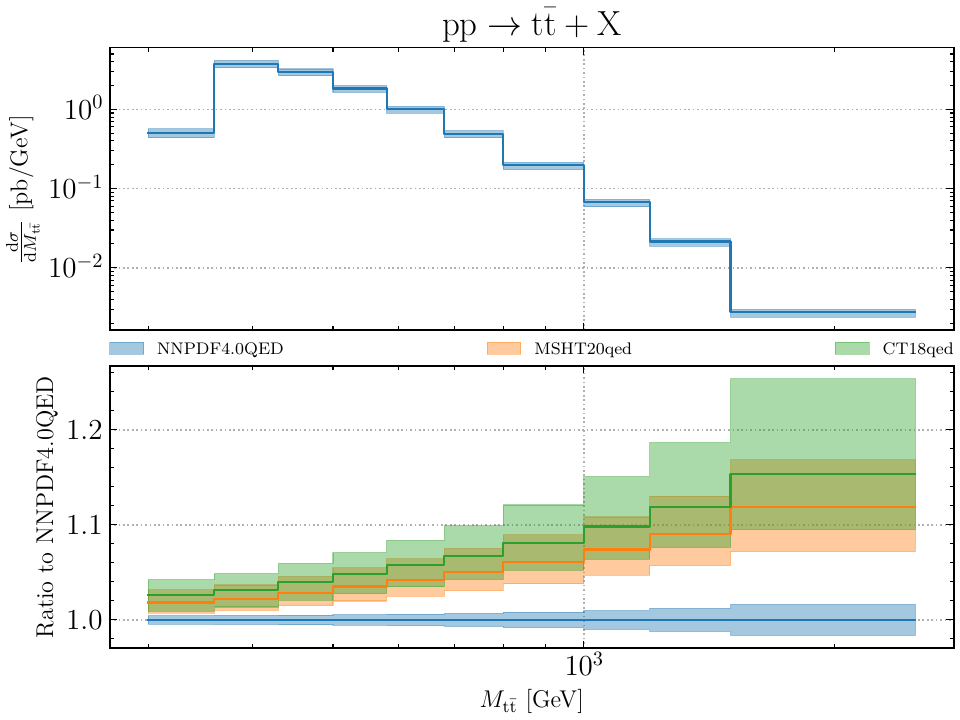}
  \caption{\small Same as Fig.~\ref{fig:NNPDF_DY_14TEV_40_PHENO}
    but for the dilepton transverse momentum distribution for weak boson pair
    production ($W^+W^-$ and $W^+Z$) and for the invariant mass distribution
    for top-quark pair production.
    No acceptance cuts on the decay productions of the $W, Z$ bosons
    and top quark have been imposed.
  }
  \label{fig:NNPDF_VV_14TEV_40_PHENO} 
\end{figure}

In Fig.~\ref{fig:NNPDF_DY_14TEV_40_PHENO} we show results  for
inclusive Drell--Yan production both in the neutral-current and
charged-current channels. For neutral current we show results for the
invariant mass distribution of the dilepton pair, while for
charged-current we display the rapidity distribution of the lepton. 
Predictions for the Higgs rapidity distribution 
are shown in Fig.~\ref{fig:NNPDF_H_14TEV_40_PHENO} for
gluon fusion, associated production with a $W^+$ boson, and  vector
boson fusion. Finally, in Fig.~\ref{fig:NNPDF_VV_14TEV_40_PHENO}
we show predictions for weak-boson pair production ($W^+W^-$ and $W^+Z$),
as a function of the dilepton transverse momentum, and for top-quark pair
production as a function of the invariant mass of the top quark pair.

For charged-current Drell--Yan, the QED corrections have essentially
no effect. In all the other cases
that we consider, the effects of the QED correction fall in one of two
categories. Either upon inclusion of QED effects
we see an enhancement of the cross-section, which
is only present when the photon-induced contribution is included,
while the NNPDF4.0QED result without photon-induced contribution is
very close to 
the pure QCD result. Or else we see a suppression of the
cross-section, but with the NNPDF4.0QED result with and without the
photon-induced contribution very close to each other. Of course, the
former case can be explained with the presence of a sizable positive
photon-induced contribution, while the latter case is explained by the
suppression of the gluon luminosity due to the transfer of momentum
fraction from  the gluon to the photon seen in Fig.~\ref{fig:lumis_1}.

The first situation --- enhancement due to the photon--induced
contribution --- is observed in neutral-current dilepton
production, where the enhancement increases with invariant mass and can reach up to 5\% at the TeV scale.
Note that the photon-induced contribution to the dilepton final state can proceed also through $t$- and $u$-channel leading-order diagrams.
The fact that the QED enhancement is absent at the $Z$ peak suggests that this non-resonant contribution provides the dominant part of the photon-induced contribution.
A similar
situation occurs in $W^+W^-$ and $ZW^+$, where the enhancement
increases  with $p_T$ and reaches 5\% in the former case and  2\% in
the latter case for transverse momenta in the TeV range. Finally, the
enhancement 
is also observed in associate Higgs production with $W^+$ and in
vector boson fusion. In the former case the enhancement is largest at forward
rapidity, where it reaches 4\%, and it decreases to 2\% for the
largest rapidity $y_H=2.5$. In the latter case the enhancement is very
moderate, around 1\%, and almost
independent of the rapidity (though slightly decreasing as
the rapidity increases).

The second situation --- suppression, independent of the
photon-induced contribution --- is found in processes that proceed
through gluon fusion. The effect is clearly seen in
Higgs production in
gluon fusion, where the suppression is weakly dependent on rapidity,
varying between 2\% at central rapidity and about 1\% at the largest
rapidity. A similar, but  more moderate effect, is also seen in top
pair production, where the suppression is of order 1\%, essentially
independent of the invariant mass of the top pair.

When comparing results obtained using different QED PDF sets, be they
NNPDF3.1QED, MSHT20QED or CT18QED, we observe  that 
differences are essentially driven by the difference in quark and
gluon PDFs. This is a direct consequence of the 
similarity of the photon PDF in all sets. Indeed, all comparisons are quite
similar to those shown in Sect.~9.3 of Ref.~\cite{Ball:2021leu},
where the same processes were studied in pure QCD.

All in all, we conclude that the inclusion of QED corrections is
important for precision phenomenology at the percent level, even in
cases in which the photon-induced contribution is negligible,
such as Higgs production in gluon fusion. Here neglecting the indirect effect
of including the photon PDF results in an overestimation of the peak
cross-section (and thus the total cross-section) by about 2\%, thus
biasing the prediction by an amount that is of the same order as the PDF
uncertainty, but not included in it.

%-------------------------------------------------------------------------------
%-------------------------------------------------------------------------------

%-------------------------------------------------------------------------------
%-------------------------------------------------------------------------------

% Summary and acknowledgements
\section{Summary and outlook}
\label{sec:summary}

We have presented a new determination of QED PDFs based on the
NNPDF4.0 set of parton distributions, using the methodology previously
adopted for the construction of the NNPDF3.1QED PDFs~\cite{Bertone:2017bme}.
This methodology implements the LuxQED
procedure~\cite{Manohar:2016nzj,Manohar:2017eqh}
to determine the photon PDF. Results are consistent with previous studies:
specifically, we find that QED effects have a small but non-negligible
impact mostly on the  gluon PDF, and that photon-initiated contributions are
most important for high-mass process such as neutral-current Drell--Yan
production. This PDF determination is based on a new NNPDF
pipeline for producing theory predictions~\cite{Barontini:2023vmr}. 
Thanks to the integration of QED evolution in this pipeline, and in particular thanks
to the use of the {\sc\small EKO} evolution code~\cite{Candido:2022tld},
the production of QED variants of NNPDF determinations is now essentially
automated, and will become the default in future releases. Indeed,
there is in general no reason to 
switch off the photon PDF, even if it is a small correction, nor to
neglect its effect on the momentum fraction carried by the gluon.

The NNPDF4.0QED PDF sets are made publicly available via the {\sc\small LHAPDF6}
interface, 
\begin{center}
  {\bf \url{http://lhapdf.hepforge.org/}~} .
\end{center}
All sets are delivered as sets of $N_{\rm rep}=100$ Monte Carlo replicas.
Specifically, we provide the NLO and NNLO global fits constructed with the
settings defined in Sect.~\ref{sec:theory} and denoted as
\begin{flushleft}
  \tt NNPDF40\_nlo\_as\_01180\_qed \\
  \tt NNPDF40\_nnlo\_as\_01180\_qed \\
\end{flushleft}
These sets are also made available via the NNPDF collaboration website
\begin{center}
  {\bf \url{https://nnpdf.mi.infn.it/nnpdf4-0-qed/}~} .
\end{center}
They should be considered the QED PDF counterparts of the published
NNPDF4.0 QCD-only PDF sets~\cite{Ball:2021leu}.

We also make available the set of 100 NNPDF4.0 (QCD only) replicas used to
produce the comparisons shown in Sect.~\ref{sec:results}. These
differ from the published NNPDF4.0 PDFs because they have been
produced using the new theory pipeline, and include some minor bug
corrections in the implementation of the NNPDF4.0 dataset. These are called:
\begin{flushleft}
  \tt NNPDF40\_nlo\_as\_01180\_qcd \\
  \tt NNPDF40\_nnlo\_as\_01180\_qcd .\\
\end{flushleft}
The equivalence of these replicas to the published NNPDF4.0 replicas
is demonstrated in Appendix~\ref{app:pineline}. They are made available
for completeness, on the NNPDF website only.

The NNPDF4.0QED determination is part of a family of developments
based on the NNPDF4.0 PDF set, and aimed at increasing its accuracy.
These will also include a determination of the theory
uncertainty on NNPDF4.0 PDFs based on the methodology of
Refs.~\cite{NNPDF:2019vjt,NNPDF:2019ubu}, and a first 
PDF determination based on NNPDF methodology at approximate
N$^3$LO~\cite{Hekhorn:2023gul}. 
All of these, as well as their combination,
will become part of the default NNPDF
methodology in future releases.

A natural development of QED PDFs is the full inclusion of electroweak
corrections in theory predictions used for PDF determination, which
enables a widening of both the set of processes and the
kinematic range that may be used for PDF determination, which are
currently constrained by the need of ensuring that electroweak  corrections are
small. This is especially important as electroweak effects become more relevant
in regions of phase space sensitive to the large-$x$ PDFs, which are in turn
relevant for new physics searches~\cite{Ball:2022qtp,Hammou:2023heg}. This development will be greatly facilitated
by the availability, within the new pipeline, of the {\sc\small PineAPPL}
interface to automated Monte Carlo generators, such as {\sc\small mg5\_aMC@NLO}.
All these developments will help achieve PDF determination with percent or
sub-percent accuracy.

\subsection*{Acknowledgments}

We thank Valerio Bertone for assistance with the benchmarking of QED evolution
with {\tt APFEL}, and Lucian Harland-Lang and Luca Buonocore for
discussions on the LuxQED procedure.
R.~D.~B, L.~D.~D., and R.~S. are supported by the U.K.
Science and Technology Facility Council (STFC) grant ST/T000600/1.
F.~H. is supported by the Academy of Finland
project 358090 and is funded as a part of the Center
of Excellence in Quark Matter of the Academy of Finland, project 346326.
E.~R.~N. is supported by the Italian
Ministry of University and Research (MUR) through the “Rita Levi-Montalcini”
Program. M. U. and Z. K. are supported by the European Research Council under
the European Union's Horizon 2020 research and innovation Programme (grant
agreement n.950246), and partially  by the STFC consolidated grant ST/L000385/1.
J.~R. is partially supported by NWO, the Dutch Research Council.
C.~S. is supported by the German Research Foundation (DFG) under reference
number DE 623/6-2.

% Appendices

\appendix

% NNPDF4.0 fits with new theory pipeline
\section{The new NNPDF theory pipeline}
\label{app:pineline}

As mentioned in Sect.~\ref{sec:results}, this work is based on a new
pipeline for the calculation of theoretical predictions. This new theory
pipeline is described in Ref.~\cite{Barontini:2023vmr};
it  supersedes and replaces the one
used for the NNPDF4.0 determination, which was based
on a combination of different pieces of code of various origin, 
specifically {\sc\small APFEL}~\cite{Bertone:2013vaa} for PDF evolution
and DIS structure function computation, and
{\sc\small APFELgrid}~\cite{Bertone:2016lga} for the generation of
interpolation grids of NLO partonic matrix elements. This last piece of code
made use of {\sc\small APPLgrid}~\cite{Carli:2010rw}
and {\sc\small FastNLO}~\cite{Wobisch:2011ij} interpolators.
The main benefit of the new pipeline is its unified, yet modular and
flexible, structure.

This theory pipeline has been extensively benchmarked for numerical
accuracy, including QCD evolution, the computation of structure
functions, the interpolation of grids, and the interfacing to
{\sc\small mg5\_aMC@NLO}. Using this new pipeline, interpolation grids for
a number of processes have been recomputed, as discussed below. Therefore
Tables~2.1--2.5 in Ref.~\cite{Ball:2021leu} should be updated
accordingly.
\begin{itemize}
\item All fully inclusive DIS processes (Table 2.1 in
  Ref.~\cite{Ball:2021leu}) have been recomputed using
  {\sc\small YADISM}~\cite{yadism,Candido:2023utz}. Specifically, in
  this code  heavy quark mass
  effects are accounted for using a new implementation of the FONLL
  prescription~\cite{num_fonll}.
\item All fixed-target Drell--Yan processes (see Table 2.3 in
  Ref.~\cite{Ball:2021leu}) have been recomputed using a modified
  version~\cite{Barontini:2023vmr} of {\sc\small VRAP}~\cite{anastasiou:2003ds}.
\item The following datasets of Table 2.4 and Table 2.5 in
  Ref.~\cite{Ball:2021leu} have been recomputed using
  {\sc\small mg5\_aMC@NLO} (all grids, in the {\sc \small PineAPPL} format,
  are available at \url{https://github.com/NNPDF/pineapplgrids}):
    \begin{itemize}
      \item ATLAS $W, Z$ 7 TeV ($\mathcal{L} = 4.6~\text{fb}^{-1}$)
      \item ATLAS low-mass DY 7 TeV
      \item ATLAS high-mass DY 7 TeV
      \item ATLAS $\sigma_{tt}^{\rm tot}$ 7, 8
      \item ATLAS $\sigma_{tt}^{\rm tot}$ 13 ($\mathcal{L} = 139~\text{fb}^{-1}$)
      \item ATLAS $t\bar t$ lepton+jets 8 TeV
      \item ATLAS $t\bar t$ dilepton 8 TeV
      \item CMS Drell--Yan 2D 7 TeV
      \item CMS $t\bar t$ 2D dilepton 8 TeV
      \item CMS $t\bar t$ lepton+jets 13 TeV
      \item CMS $t\bar t$ dilepton 13 TeV
      \item CMS $\sigma_{tt}^{\rm tot}$ 5.02 TeV
      \item CMS $\sigma_{tt}^{\rm tot}$ 7, 8 TeV
      \item CMS $\sigma_{tt}^{\rm tot}$ 13 TeV
      \item LHCb $Z$ 7 TeV ($\mathcal{L} = 940~\text{pb}^{-1}$)
      \item LHCb $Z \to ee$ 8 TeV ($\mathcal{L} = 2~\text{fb}^{-1}$)
      \item LHCb $W, Z \to \mu$ 7 TeV
      \item LHCb $W, Z \to \mu$ 8 TeV
    \end{itemize}
\end{itemize}
Grids for all remaining datasets have been converted to the {\sc\small PineAPPL}
format without recomputing them.
Of course, none of these changes except the new FONLL implementation
should make any difference to the
extent that the previous and new theory implementations are both
numerically accurate. The new FONLL implementation differs from the
previous one by subleading corrections and thus it does introduce NNLO
differences in the NLO fit and N$^3$LO differences in the NNLO
fit; these differences are confined to the charm and bottom mass
corrections to  deep-inelastic structure functions and thus only
affect a small number of datapoints, at NNLO at the sub-percent level.

Finally, in the process of transitioning to the new pipeline, a few bugs
were discovered in the  implementation of a few datapoints, such as
incorrect normalization, incorrect scale assignment or incorrect bin
size. These corrections in practice have a negligible impact, as they
involve a handful of points out of more than 4500.

%-----------------------------------------------------------------------
\begin{table}[!t]
  \centering
  \footnotesize
  \renewcommand{\arraystretch}{1.50}
\begin{tabularx}{\textwidth}{Xlllllll}
  \toprule
  &     \multicolumn{2}{c}{NLO QCD}
  &	\multicolumn{2}{c}{NNLO QCD} \\
  &     New & Published 
  &	New & Published \\
\midrule
$\chi^2$
& 1.26 & 1.24
& 1.17 & 1.16 \\
$\la E_{\rm tr}\ra_{\rm rep}$
& 2.41 $\pm$ 0.06
& 2.43 $\pm$ 0.08
& 2.28 $\pm$ 0.05
& 2.27 $\pm$ 0.07 \\
$\la E_{\rm val}\ra_{\rm rep}$
& 2.57 $\pm$ 0.10
& 2.62 $\pm$ 0.13
& 2.37 $\pm$ 0.11
& 2.35 $\pm$ 0.11 \\
$\la \chi^2\ra_{\rm rep}$
& 1.29 $\pm$ 0.02
& 1.27 $\pm$ 0.02
& 1.20 $\pm$ 0.02
& 1.18 $\pm$ 0.02 \\
$\la {\rm TL}\ra_{\rm rep}$
& 12900 $\pm$ 2000
& 13200 $\pm$ 2100
& 12400 $\pm$ 2600
& 13400 $\pm$ 2400 \\
$\phi$
& 0.156 $\pm$ 0.006
& 0.178 $\pm$ 0.007
& 0.153 $\pm$ 0.005
& 0.162 $\pm$ 0.005 \\
  \bottomrule
\end{tabularx}
\vspace{0.2cm}
\caption{\small Statistical indicators (defined as in
  Table~\ref{tab:chi2_qed_global} for a set of 100 NNPDF4.0 NNLO PDF
  replicas
    produced using the new theory pipeline compared to published NNPDF4.0
  NNLO 100 replica set. The  $\phi$ estimator, defined in
  Ref.~\cite{NNPDF:2014otw}, Eq.~(4.6), is also shown (see text). 
  \label{tab:chi2_newbaseline}
}
\end{table}
%-------------------------------------------------------------------------------
\begin{figure}[!t]
  \centering
  \includegraphics[width=.49\textwidth]{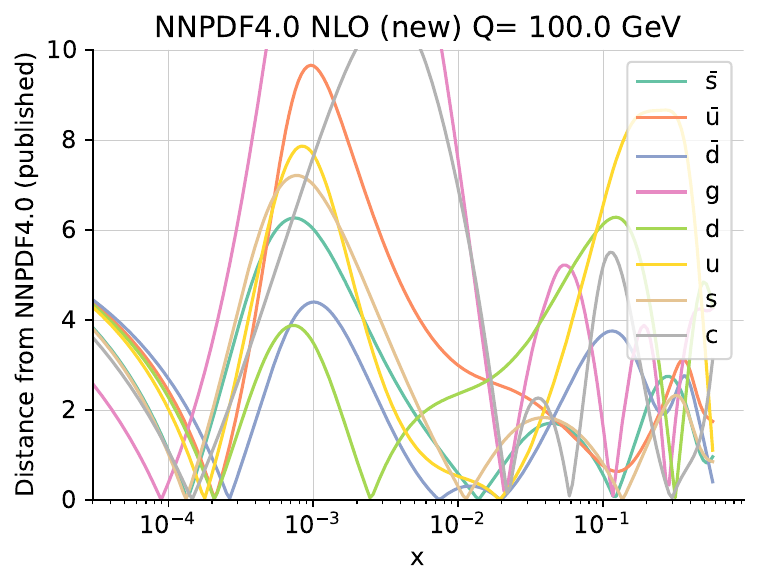}
  \includegraphics[width=.49\textwidth]{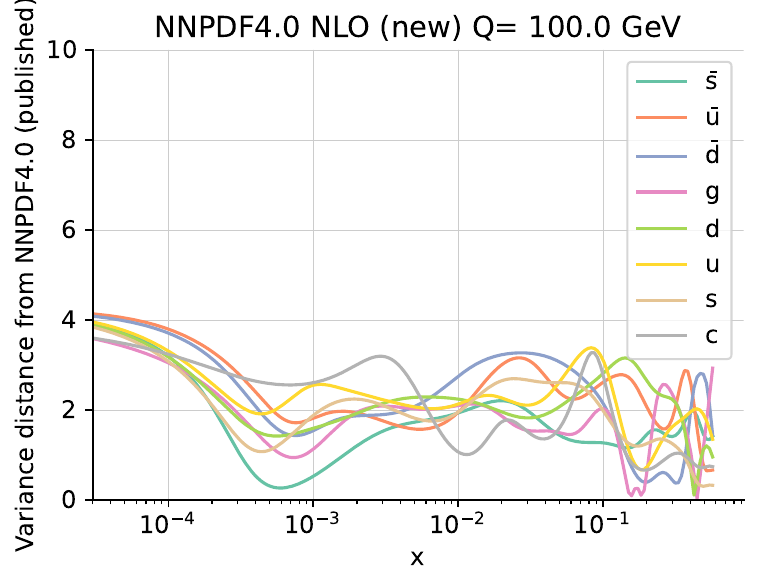}\\
  \includegraphics[width=.49\textwidth]{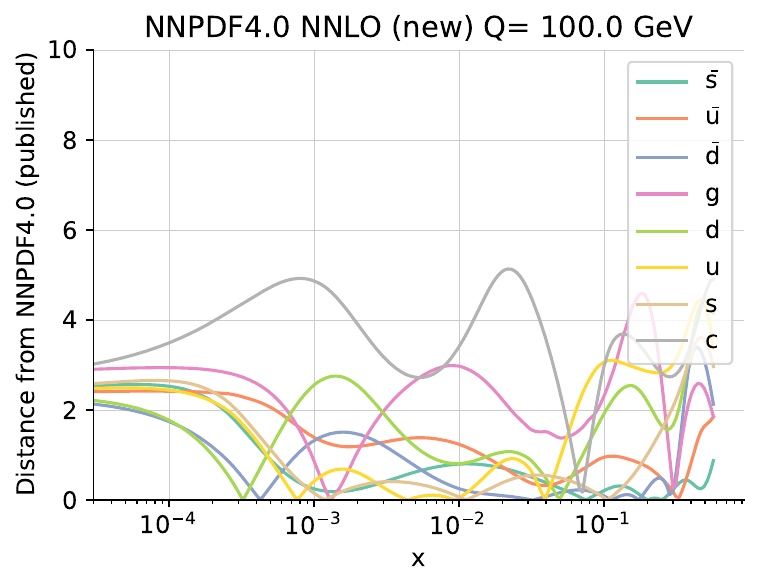}
  \includegraphics[width=.49\textwidth]{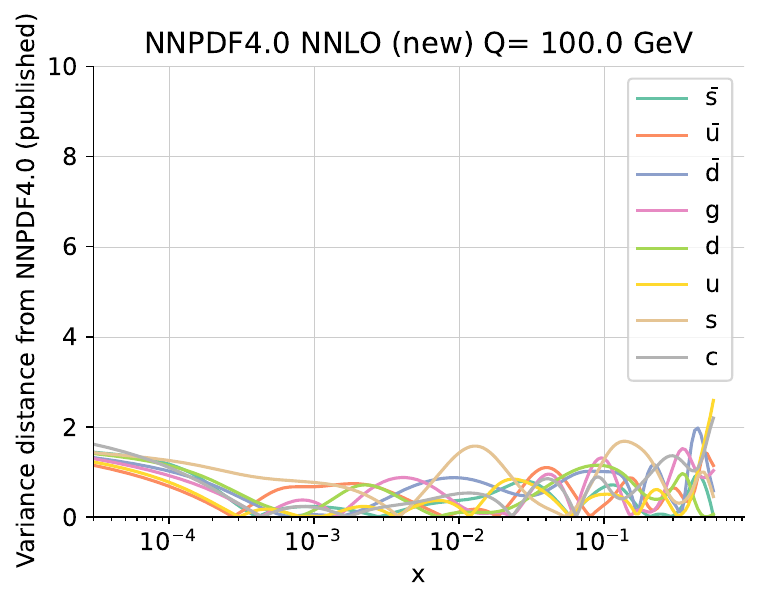}\\
  \caption{\small Distances at $Q=100$~GeV between the central values (left)
    and uncertainties (right) of the 100 NNPDF4.0 PDF replicas at NLO (top) and NNLO (bottom)
    whose statistical indicators are compared in Table~\ref{tab:chi2_newbaseline}.
}
  \label{fig:distances_newbaseline} 
\end{figure}
%------------------------------------------------------------------------------

%-------------------------------------------------------------------------------
\begin{figure}[!t]
  \centering
  \includegraphics[width=.49\textwidth]{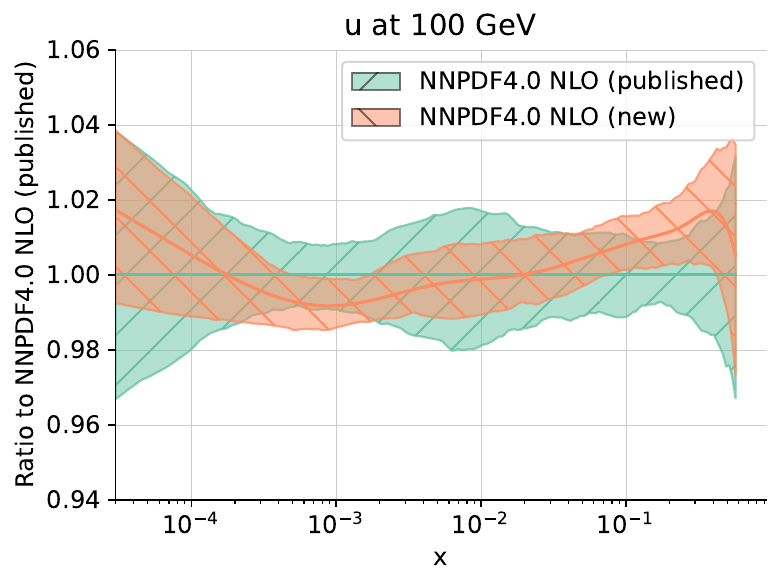}
  \includegraphics[width=.49\textwidth]{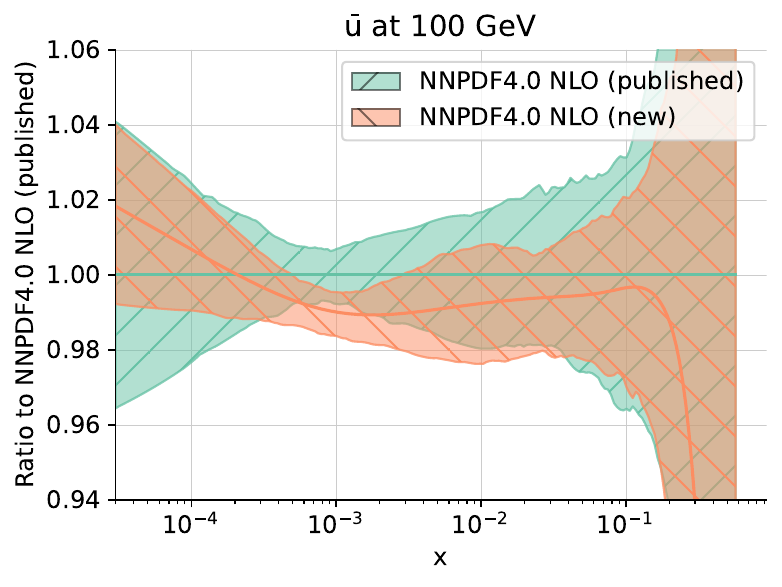}\\
  \includegraphics[width=.49\textwidth]{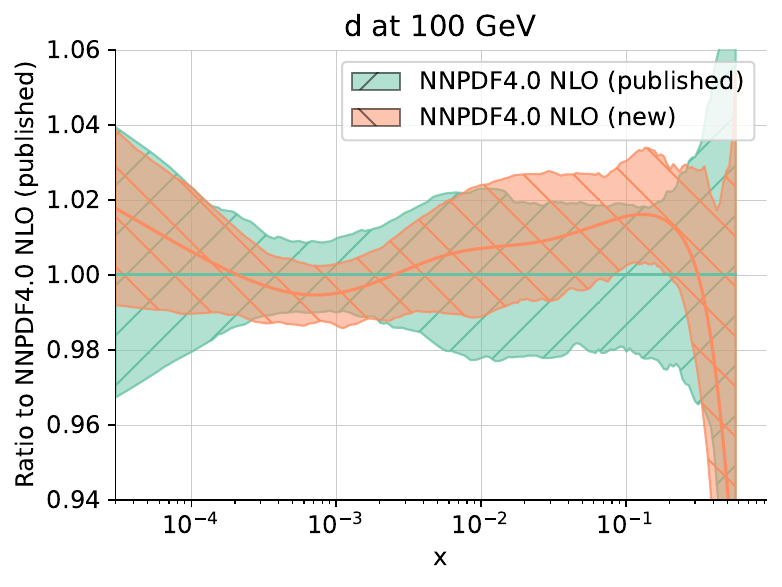}
  \includegraphics[width=.49\textwidth]{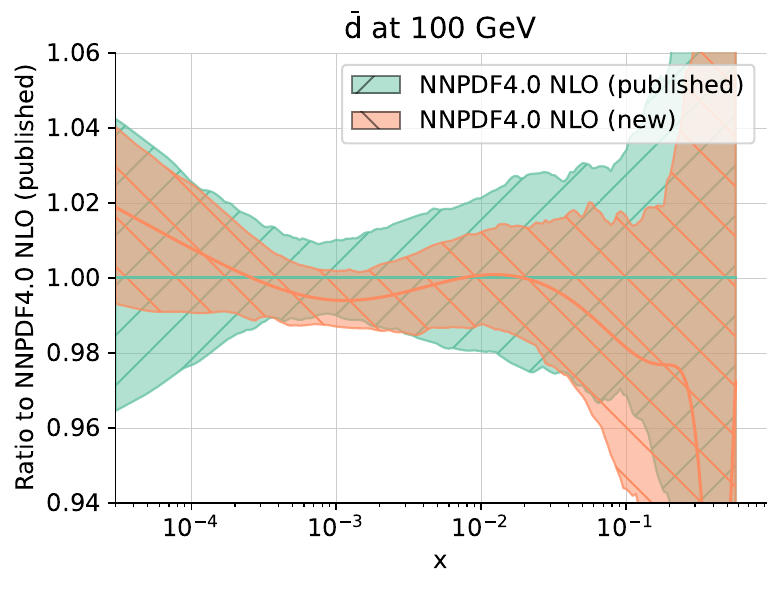}\\
  \includegraphics[width=.49\textwidth]{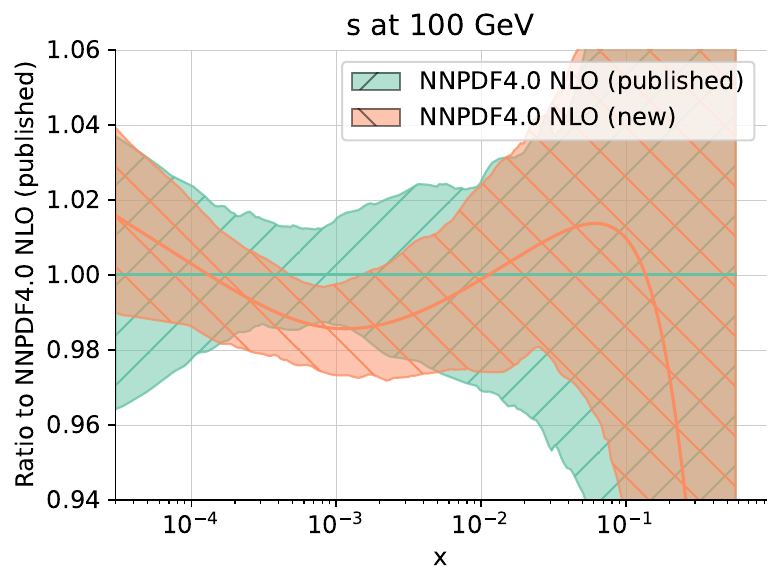}
  \includegraphics[width=.49\textwidth]{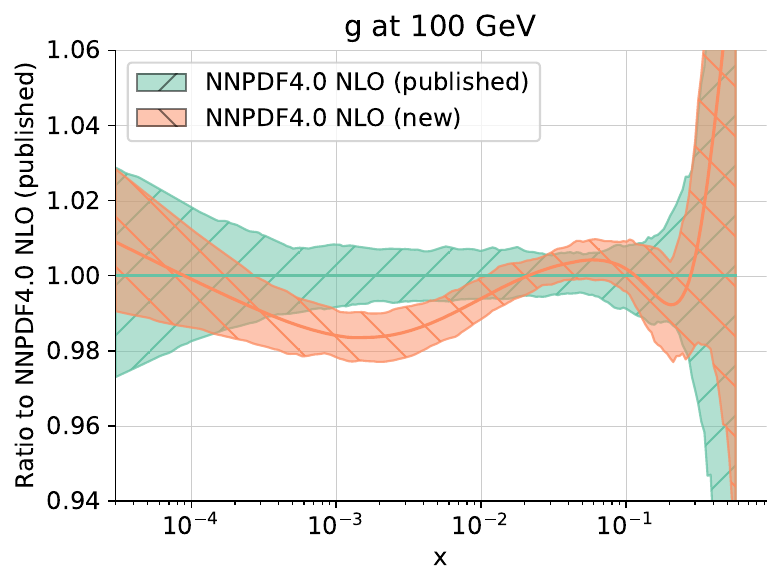}\\
  \caption{\small The NLO PDFs
    at $Q=100$ GeV from the 100 replica sets compared in
    Table~\ref{tab:chi2_newbaseline} and Fig.~\ref{fig:distances_newbaseline}.
    Results are shown normalized to the central value of the published set.
    Bands correspond to 1$\sigma$ uncertainties.
    From left to right and from top to bottom, we show the up, anti-up, down,
    anti-down, strange and gluon PDFs.
}
  \label{fig:NewBaseline-q100gev-ratios_nlo} 
\end{figure}
%----------------------------------------------------------------------------

%-------------------------------------------------------------------------------
\begin{figure}[!t]
  \centering
  \includegraphics[width=.49\textwidth]{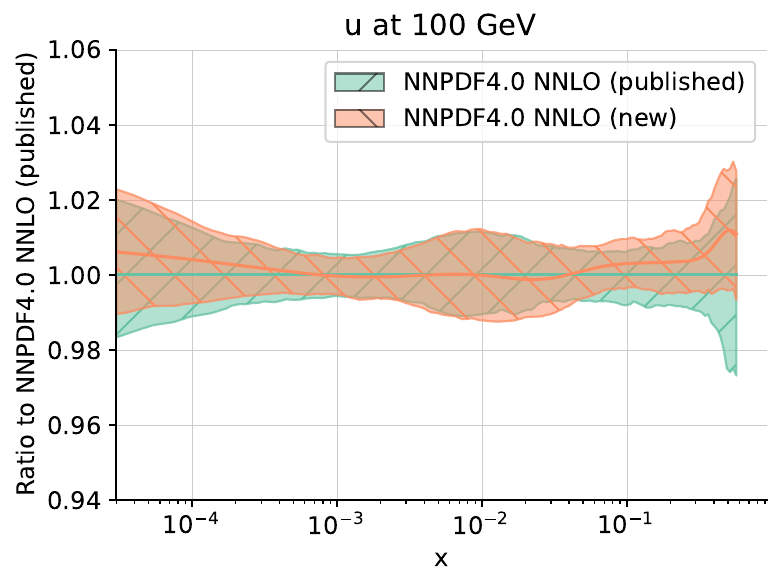}
  \includegraphics[width=.49\textwidth]{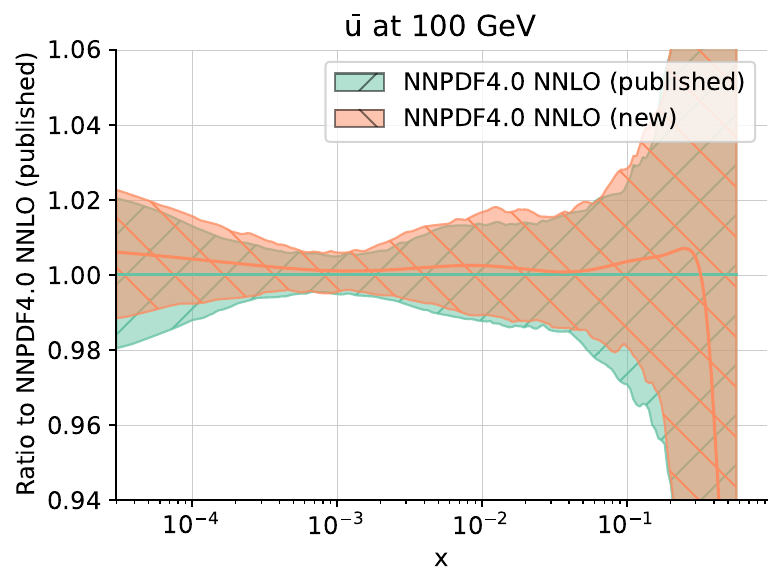}\\
  \includegraphics[width=.49\textwidth]{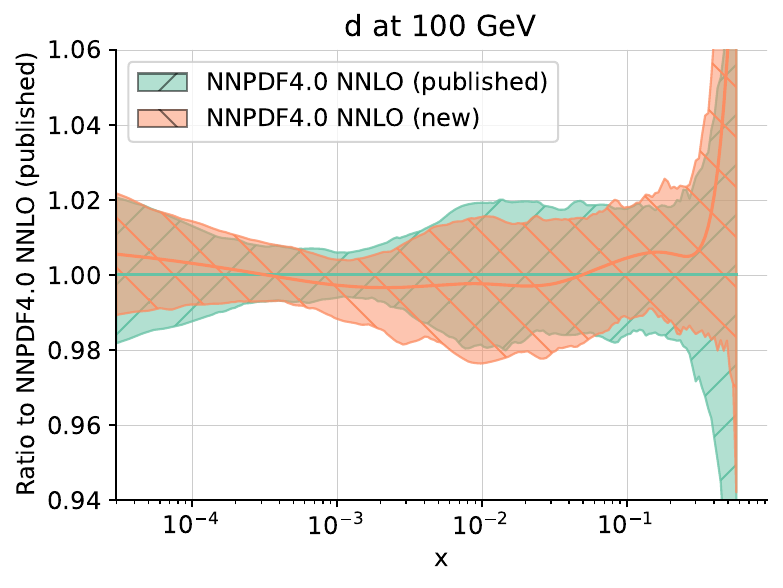}
  \includegraphics[width=.49\textwidth]{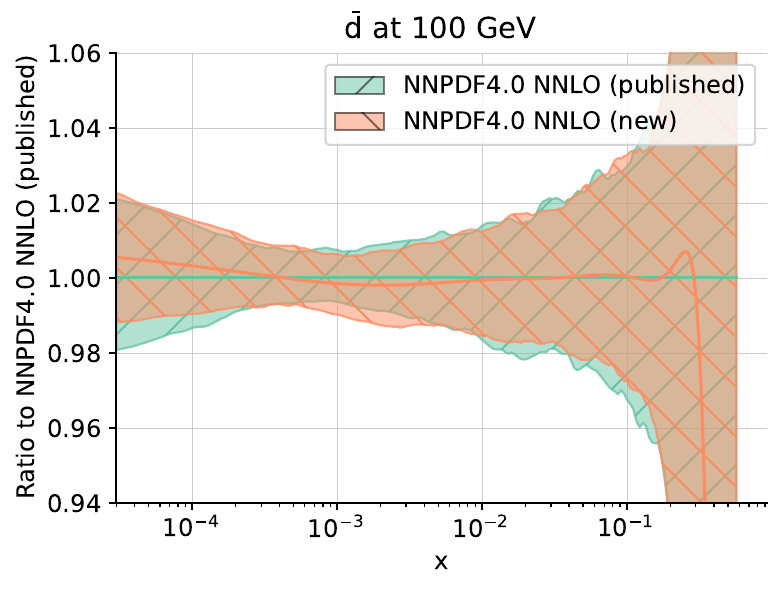}\\
  \includegraphics[width=.49\textwidth]{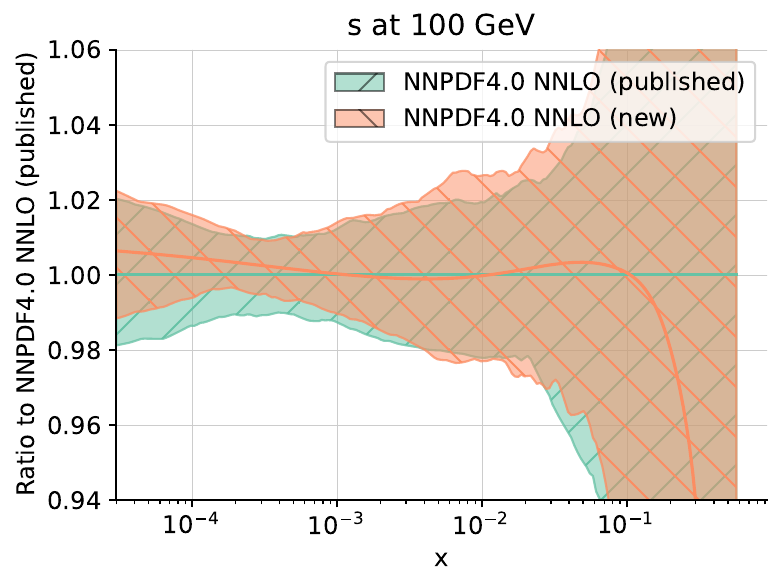}
  \includegraphics[width=.49\textwidth]{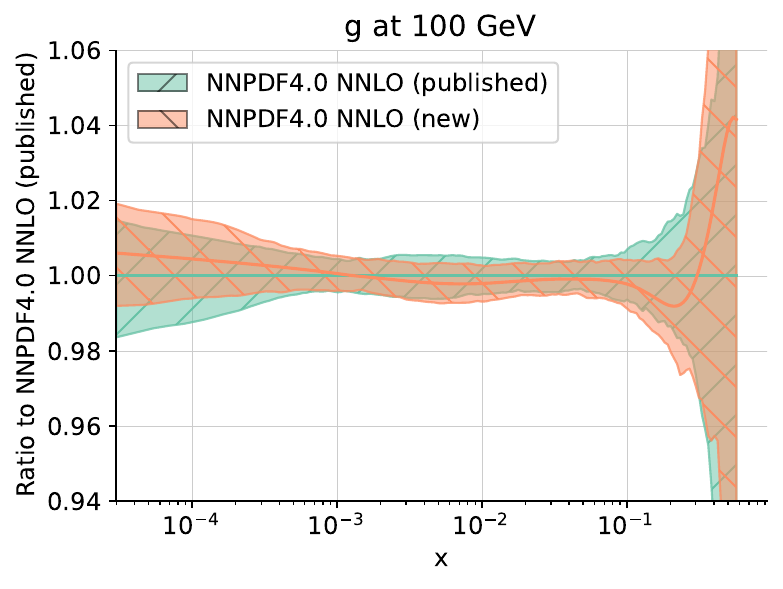}\\
  \caption{\small Same as Fig.~\ref{fig:NewBaseline-q100gev-ratios_nlo} at NNLO.}
  \label{fig:NewBaseline-q100gev-ratios_nnlo} 
\end{figure}
%----------------------------------------------------------------------------

We have generated variants of the published NNPDF4.0 NLO and NNLO fits using
the new theory pipeline. In order to illustrate the equivalence between these
fits and the NNPDF4.0 fits, obtained with the previous theory pipeline, we
compare each pair of fits. All fits are made of 100 replicas. The corresponding
statistical estimators are reported in Table~\ref{tab:chi2_newbaseline}.
In addition to the
fit quality estimators, we also show the $\phi$ estimator, defined in Eq.~(4.6)
of Ref.~\cite{NNPDF:2014otw}, which measures the ratio of the average
(correlated) PDF uncertainty on datapoints over the experimental uncertainty,
and is thus a fairly sensitive measure of fit quality.
All estimators are seen to coincide at NNLO (note that the statistical
uncertainty on the $\chi^2$ with the given number of datapoints is 0.02),
demonstrating explicitly the equivalence between the two replica sets. All
estimators are also very close at NLO: differences are compatible with the
statistical uncertainty on the $\chi^2$.

\begin{figure}[!t]
  \centering
  \includegraphics[width=.49\textwidth]{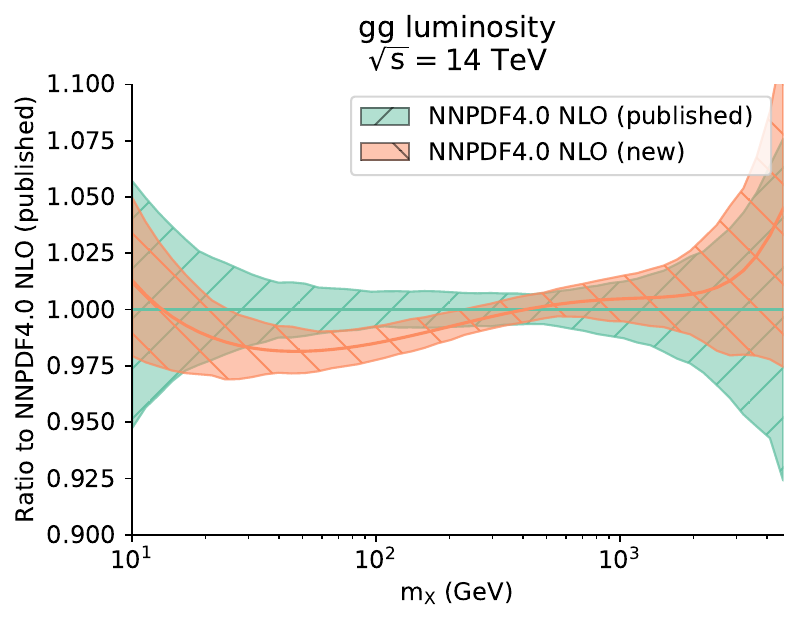}
  \includegraphics[width=.49\textwidth]{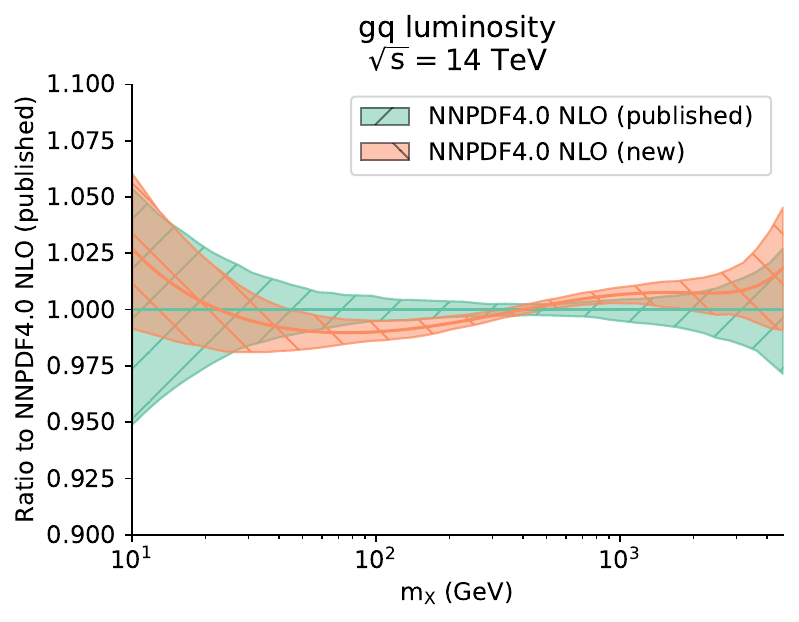}\\
  \includegraphics[width=.49\textwidth]{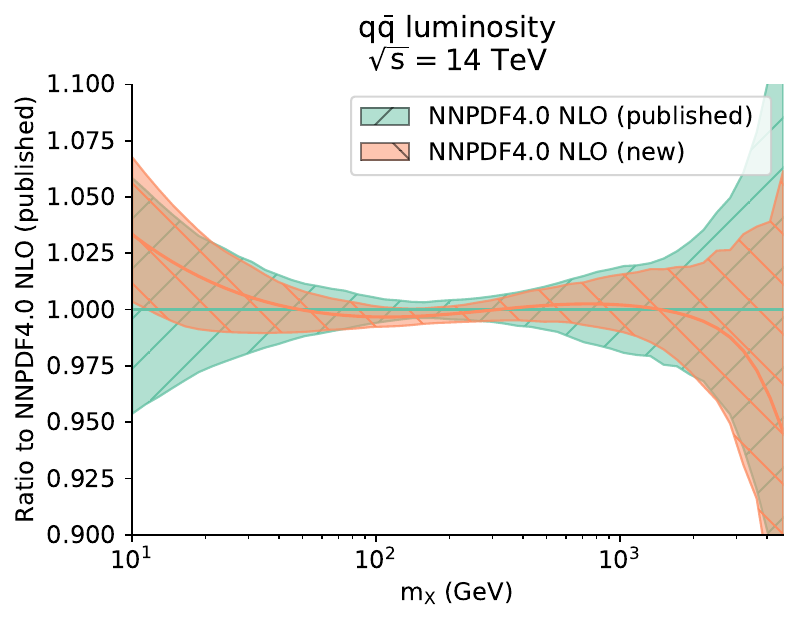}
  \includegraphics[width=.49\textwidth]{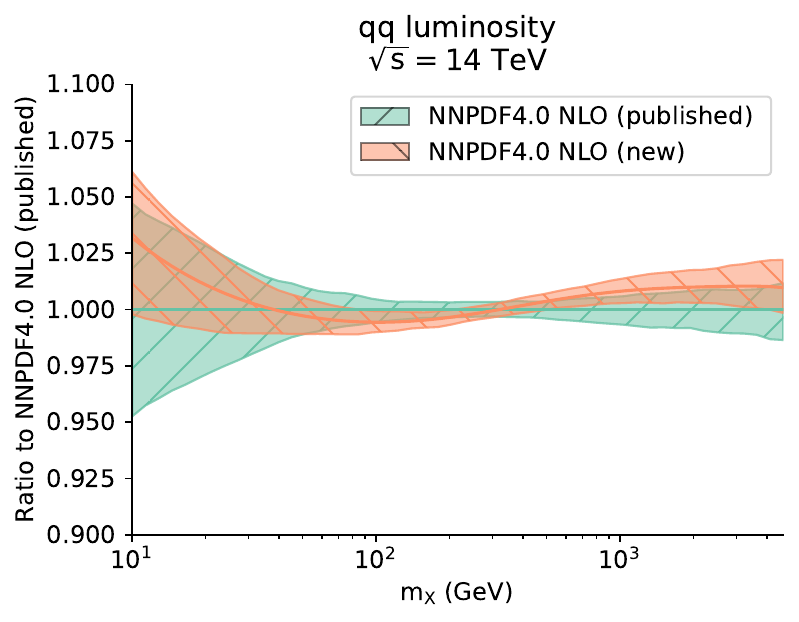}\\  
  \caption{\small Same as in Fig.~\ref{fig:NewBaseline-q100gev-ratios_nlo} for the
    partonic luminosities at the LHC with $\sqrt{s}=14$ TeV as a function of
    the invariant mass $m_X$. The $gg$ (left) and  $qg$ (right) luminosities
    are shown in the top row, the $q\bar{q}$ (left), and $qq$ (right)
    luminosities in the bottom row.}
  \label{fig:lumi-newbaseline_nlo} 
\end{figure}

\begin{figure}[!t]
  \centering
  \includegraphics[width=.49\textwidth]{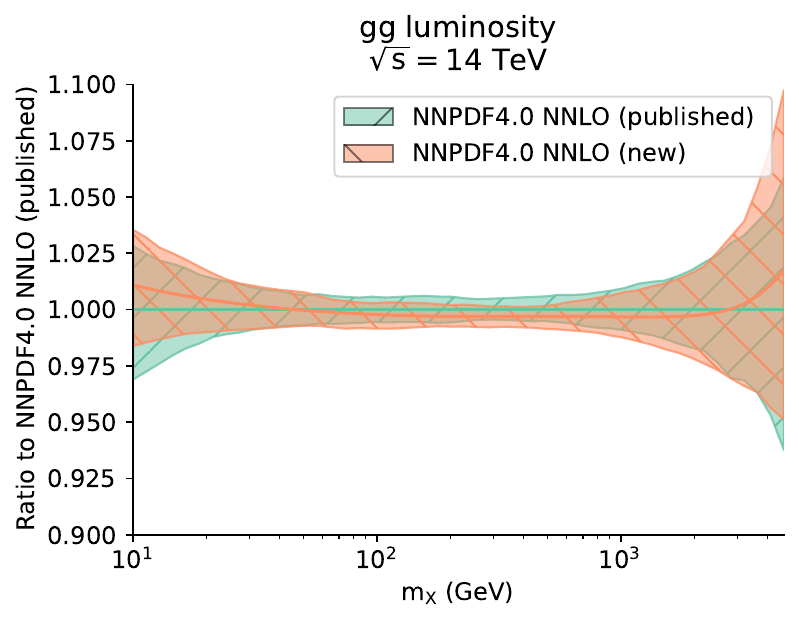}
  \includegraphics[width=.49\textwidth]{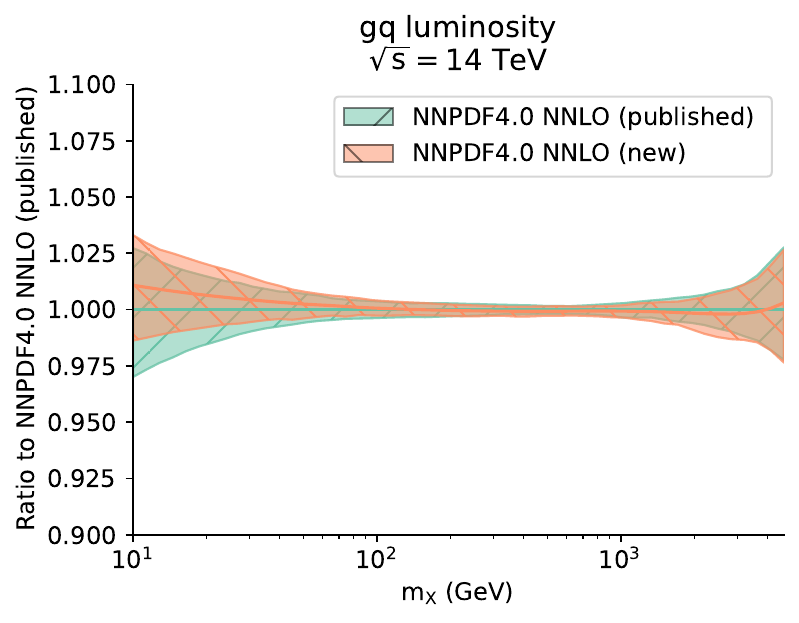}\\
  \includegraphics[width=.49\textwidth]{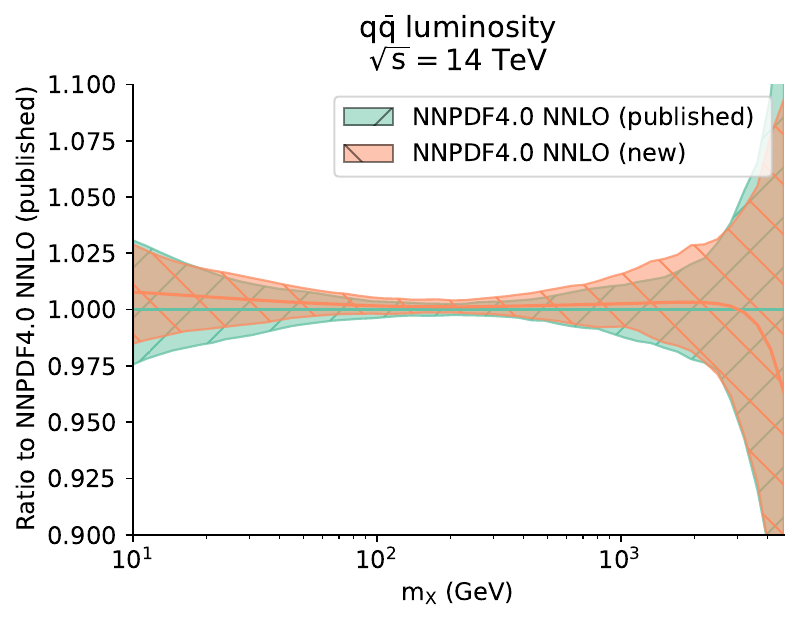}
  \includegraphics[width=.49\textwidth]{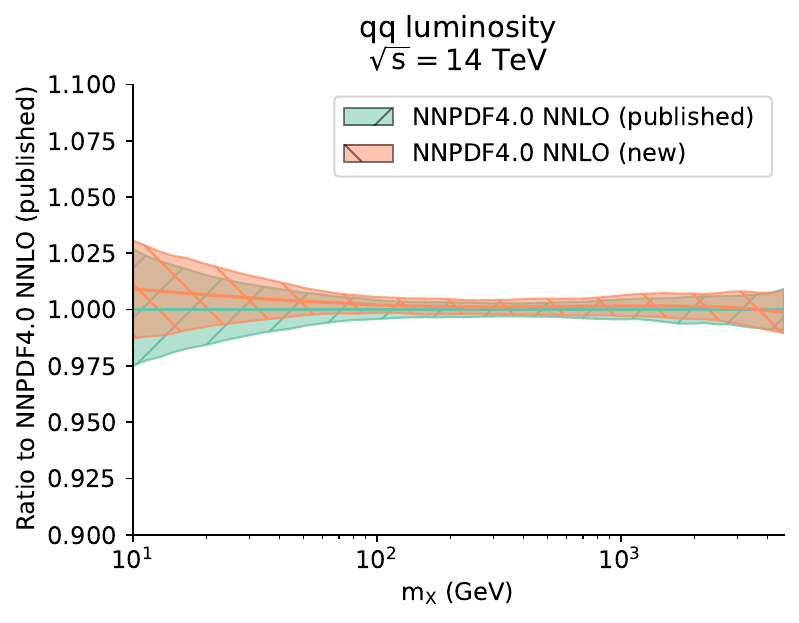}\\  
  \caption{\small Same as in Fig.~\ref{fig:lumi-newbaseline_nlo} at NNLO.}
  \label{fig:lumi-newbaseline_nnlo} 
\end{figure}

Parton distributions are compared in Fig.~\ref{fig:distances_newbaseline},
which displays the distances (defined in Ref.~\cite{Ball:2010de}) between
central values and uncertainties computed using the two replica sets at
$Q=100$ GeV. For a set of 100 replicas, $d\sim 1$ indicates statistical
equivalence, while $d\sim 10$ indicates one-sigma differences. At NNLO,
distances fluctuate about 1, indicating that the two replica
sets are drawn by the same underlying distribution. At NLO,
distances are somewhat larger, around the half-sigma level for most
PDFs and reaching 1$\sigma$ in some cases; this is mostly due to the
different FONLL implementation~\cite{num_fonll}.
The PDFs themselves are shown in
Figs.~\ref{fig:NewBaseline-q100gev-ratios_nlo}-\ref{fig:NewBaseline-q100gev-ratios_nnlo},
and the parton luminosities for the LHC at $\sqrt{s}=14$~TeV are shown in
Figs.~\ref{fig:lumi-newbaseline_nlo}-\ref{fig:lumi-newbaseline_nnlo},
respectively at NLO and NNLO.
The complete agreement between the two pairs of NNLO replica sets, and
the small differences at NLO, are manifest.

%-------------------------------------------------------------------------------
%----------------------------------------------------------------------------

%-------------------------------------------------------------------------------
%----------------------------------------------------------------------------

% EXA vs. TRN
\section{Solution of evolution equations}
\label{app:exatrn}

We summarize results and benchmarks concerning the solution to
evolution equations. First, we comparatively review 
solutions to evolution equations, for which a different choice is made
in this work in comparison to the previous NNPDF3.1QED and NNPDF4.0
PDF determinations: namely the use of an {exact} (EXA) solution instead
of the {truncated} (TRN) solution~\cite{Candido:2022tld}.
We then study the impact of this different choice on PDFs.
Having established the equivalence of (pure QCD) PDFs obtained
using either solution, we perform an inversion test of the {exact}
QED$\times$QCD solution implemented in {\sc\small EKO}, which we adopt,
in order to verify its accuracy. We finally benchmark this {\sc\small EKO}
QED$\times$QCD solution against the implementation in the {\sc\small APFEL}
code, which was used in the previous NNPDF3.1QED PDF determination.

\subsection{Exact vs.\ expanded solutions: formal aspects}

Perturbative solutions of QCD evolution equations of the form
\begin{equation}
  \label{eq:DGLAP}
  \mu^2 \frac{d f_i}{d \mu^2}
  =
  - \left( a_s\gamma_{ij}^{(1,0)} + a_s^2\gamma_{ij}^{(2,0)} + \dots \right) f_j\,,
\end{equation}
are based on the observation that if the equation is solved exactly at
leading order, with the leading-order term in the beta function
Eq.~(\ref{RGE:as}), then the solution $f_{_{ \rm LO}}(Q^2)$ in
terms of a boundary condition $f_{_{ \rm LO}}(Q_0^2)$ is a pure
leading log (LL) function, {\it i.e.}\ it is a function of $a_sL$ only, where
$L=\ln(Q^2/Q_0^2)$,
rather than depending on $a_s$ and $L$ separately. However, if the
next-to leading order contributions $\gamma_{ij}^{(2,0)}$ and
$\beta_1$ are also included, then the solution is
next-to-leading log (NLL) accurate, but the exact solution
\cref{eq:basiceko,eq:exeko} includes terms to all
logarithmic orders. A solution that only includes NLL
terms, namely with the structure 
\begin{equation}
  \label{eq:fnlo}
  f = f_{_{ \rm LO}} + a_s f_{_{ \rm NLO}}, \, 
\end{equation}
where both $f_{_{ \rm LO}}$ and $f_{_{ \rm NLO}}$ are pure LL
functions, can be constructed by linearization of the exact solution,
namely by expanding the exact solution and by truncating the expansion.
Various intermediate options are also possible. The argument can be repeated
at any logarithmic accuracy. Of course all these solutions are
equivalent up to subleading logarithmic terms.

A truncated solution to the pure QCD evolution equation
of the form Eq.~(\ref{eq:fnlo}) can be determined
in closed form to all orders in Mellin space~\cite{pegasus}, by
diagonalizing order by order the anomalous dimension matrix. This
corresponds to the solution referred to as {truncated} in
Ref.~\cite{Candido:2022tld}. However,
this strategy fails for combined QED$\times$QCD evolution, because the
QED and QCD anomalous dimension matrices do not commute, and thus
cannot be diagonalized simultaneously. Because $a_s$ and $\aem$ depend
on scale in different ways, this implies that the anomalous dimension
matrices $\gamma_{ij}$ evaluated at different scales do not commute:
\begin{equation}
  \label{eq:nocomm}
  \left[\gamma\left(N,a_s(\mu^2),
    \aem(\mu^2)\right),
    \gamma\left(N,a_s({\mu'}^2),
    \aem({\mu'}^2)\right)\right]\not=0\quad{\rm if}\quad \mu\not=\mu'.
\end{equation}
This means that the LO solution, constructed only including the LO
contributions to the anomalous dimensions, takes the form of a
path-ordered exponential
\begin{equation}
  \label{eq:DGLAP:lo:sol}
  f_{_{\rm LO}}(Q^2)
  =
  \mathcal P \exp\left(-\int_{Q_0^2}^{Q^2} \frac{\d\mu^2}{\mu^2}\,
  \left( a_s(\mu^2)\gamma^{(1,0)}
  + \aem(\mu^2)\gamma^{(0,1)}  \right)\right)_{_{\rm LO}}(Q_0^2)
\end{equation}
that cannot be written in closed form. Note that the commutator terms
are not subleading: for example, the quadratic term is proportional to
$a_s\aem L^2$, so it is a LL contribution.

The problem persists to all orders, and requires a truncated
solution to be also given as a path-ordered exponential. For instance,
including NLO QCD and LO QED contributions, the evolution equation is
\begin{equation}
  \label{eq:DGLAP:nlo}
  \mu^2 \frac{d f_i}{d \mu^2}
  =
  - \left( a_s\gamma_{ij}^{(1,0)}
  + \aem\gamma_{ij}^{(0,1)}
  + a_s^2\gamma_{ij}^{(2,0)}\right) f_j \, .
\end{equation}
The perturbative NLO QCD solution then has the form
Eq.~(\ref{eq:fnlo}), where in the general case $f_{_{ \rm LO}}$  and
$f_{_{ \rm NLO}}$ are LO accurate in QCD and QED, but may also include
subleading contributions, {\it i.e.}\
$f_{_{ \rm N^kLO}}=f_{_{\rm N^kLO}}(a_sL,\aem L)[1+\mathcal{O}(a_s,\aem)]$.
Substituting the perturbative expansion Eq.~(\ref{eq:fnlo}) in the NLO  
Eq.~(\ref{eq:DGLAP:nlo}), and using the LO solution,
\cref{eq:DGLAP:lo:sol}, leads to        
\begin{equation}
  \label{eq:fnlo:diffeq}
  \mu^2 \frac{d f_{_{ \rm NLO},i}}{d \mu^2}
  =
  - a_s \left( \gamma_{ij}^{(1,0)}
  - \beta_{_{\rm QCD}}^{(2,0)} \delta_{ij} \right)f_{_{ \rm NLO},j}
  - a_s \gamma_{ij}^{(2,0)} f_{_{ \rm LO},j}
  - \aem \gamma_{ij}^{(0,1)} f_{_{ \rm NLO},j}\, .
\end{equation}
Again, due to the non-commutativity of $\gamma_{ij}^{(1,0)}$ and
$\gamma_{ij}^{(0,1)}$, the solution for $f_{_{ \rm NLO}}$ is 
a path-ordered exponential and cannot be given in closed analytic
form. This continues to be the case  at higher orders.

A truncated solution can be constructed numerically, by starting with
the exact path-ordered solution
Eqs.~(\ref{eq:basiceko}-\ref{eq:exeko}), and then expanding
numerically. Such an approach, however, involves approximating
higher-order derivatives by finite differences~\cite{Bertone:2013vaa},
which may lead to numerical instabilities. This method was
adopted in  {\sc\small APFEL}:   {\sc\small APFEL} is an
$x$-space evolution 
code, so the analytic truncated solution of Ref.~\cite{pegasus} cannot
be used, and the expansion is performed numerically anyway, even in the case
of pure QCD. On the other hand, in the {\sc\small EKO} evolution code
the pure QCD {truncated} solution is implemented through the analytic
solution of Ref.~\cite{pegasus}, while the numerical path-ordered solution only needs to be 
used  for the {exact} solution
\cref{eq:basiceko,eq:exeko}. Therefore, in order to ensure greater
accuracy, we  adopt the exact solution
for the QED$\times$QCD case. The impact of this choice is benchmarked in the
next subsection.

\subsection{Exact vs.\ expanded solution: impact on PDFs}

We wish to  quantify the impact of the choice of solution: {exact},
Eqs.~(\ref{eq:basiceko}-\ref{eq:exeko}), vs.\ {truncated}, based on an
expansion of the form of Eq.~(\ref{eq:fnlo}). To this goal, we have determined
a set of NNPDF4.0 NNLO (pure QCD) PDF replicas, but now using the
{exact} solution, instead of the  {truncated} solution used
for the  published NNPDF4.0 PDF sets. We have used the new theory pipeline.

%-----------------------------------------------------------------------
\begin{table}[!t]
  \centering
  \footnotesize
  \renewcommand{\arraystretch}{1.50}
\begin{tabularx}{\textwidth}{Xlllll}
  \toprule
  &	\multicolumn{2}{c}{NNPDF4.0 NNLO QCD} \\
&	 Exact Solution &	 Truncated Solution \\
\midrule
$\chi^2$
&  1.17  & 1.17   \\
$\la E_{\rm tr}\ra_{\rm rep}$
&  2.26 $\pm$  0.06
&  2.28  $\pm$ 0.05   \\
$\la E_{\rm val}\ra_{\rm rep}$
&  2.34 $\pm$ 0.10
&  2.37 $\pm$ 0.11   \\
$\la \chi^2\ra_{\rm rep}$
&  1.19  $\pm$ 0.01
&  1.20	 $\pm$ 0.02    \\
$\la {\rm TL}\ra_{\rm rep}$
&  12100  $\pm$   2500
&  12400   $\pm$  2600  \\
$\phi $
&  0.147  $\pm$ 0.005
&  0.153  $\pm$ 0.005     \\
  \bottomrule
\end{tabularx}
\vspace{0.2cm}
\caption{\small Same as Table~\ref{tab:chi2_newbaseline}, now comparing
  a set of 100 NNPDF4.0 replicas obtained using the default {truncated}
  solution of evolution equations with a set obtained using the {exact}
  solution (pure NNLO QCD theory). Both PDF sets are produced using the
  new theory pipeline.}
\label{tab:chi2_exact_vs_truncated_baseline}
\end{table}
%--------------------------------------------------------------------------

%-------------------------------------------------------------------------------
\begin{figure}[!t]
  \centering
  \includegraphics[width=.49\textwidth]{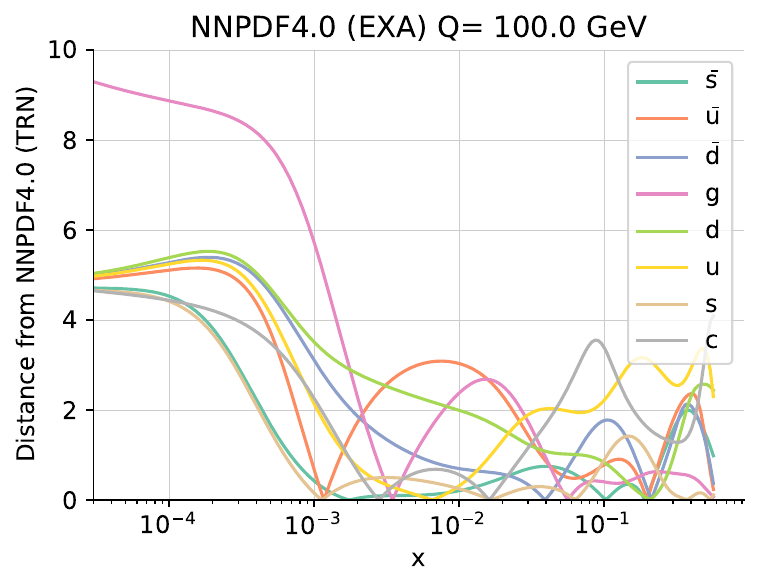}
  \includegraphics[width=.49\textwidth]{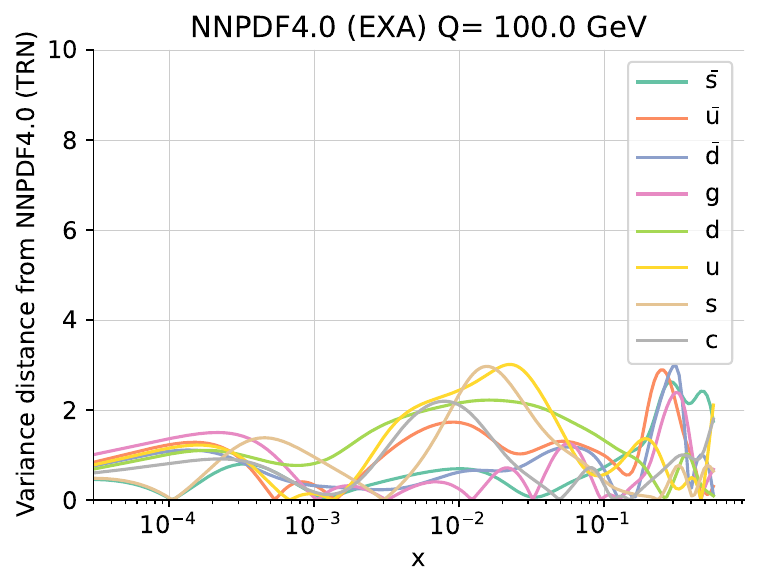}
  \caption{\small Statistical distances between the central values (left)
    and uncertainties (right) of the 100 NNPDF4.0 PDF replicas whose
    statistical indicators are compared in
    Table~\ref{tab:chi2_exact_vs_truncated_baseline}.}
  \label{fig:distance_EXA_vs_TRN} 
\end{figure}
%-------------------------------------------------------------------------------

%-------------------------------------------------------------------------------
\begin{figure}[!t]
  \centering
  \includegraphics[width=.49\textwidth]{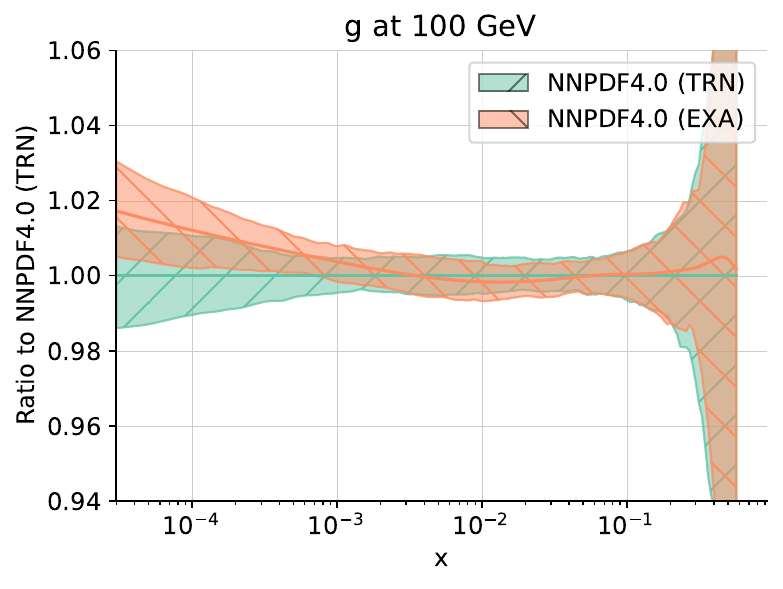}
  \includegraphics[width=.49\textwidth]{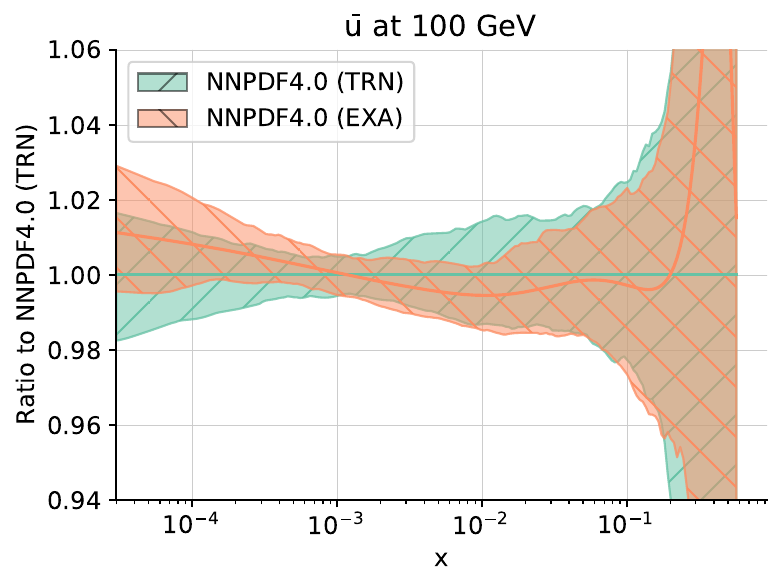}
  \caption{\small The gluon (left) and anti-up quark (right) PDFs
    at $Q=100$ GeV from the PDF sets compared in
    Table~\ref{tab:chi2_exact_vs_truncated_baseline} and
    Fig.~\ref{fig:distance_EXA_vs_TRN}.
    Results are shown normalized to the central value of the PDFs
    obtained with {truncated} evolution (published NNPDF4.0).
  }
  \label{fig:upbar_EXA_vs_TRN_100gev} 
\end{figure}
%-------------------------------------------------------------------------------

The  statistical indicators for these two sets of replicas are compared
in Table~\ref{tab:chi2_exact_vs_truncated_baseline}, and are seen to
be indistinguishable.
The corresponding PDFs are compared in \cref{fig:distance_EXA_vs_TRN},
where we display the distances (defined in Ref.~\cite{Ball:2010de}) between
the central values and uncertainties of all PDFs in either set, and in
\cref{fig:upbar_EXA_vs_TRN_100gev}, where we specifically compare the
gluon and antiup PDFs at $Q=100$ GeV.
It is clear that for $x\gsim 10^{-3}$ the PDFs are statistically
indistinguishable. At small $x$ they start differing at the
half-$\sigma$ level, with differences at most reaching the one-$\sigma$ level
for the gluon.

This is consistent with the fact that the {exact}
and {truncated} solutions differ by higher-order perturbative
corrections that go beyond the NNLO accuracy of the
computation. Indeed, it is well known that at small $x$ the
perturbative convergence starts deteriorating because of high-energy
logarithms that need resummation in order for PDF determination to be
accurate~\cite{Ball:2017otu}. The qualitative behavior of the
PDFs shown in \cref{fig:upbar_EXA_vs_TRN_100gev} agrees with this
explanation: the {exact} solution exponentiates a set of subleading
small-$x$ logarithms which are linearized in the {truncated} solution.
It follows that in the small-$x$ region, where there is no data, these
contributions lead to a stronger rise of the gluon,
which then feeds back onto the quark-antiquark sea.

\subsection{Exact solution: invertibility}

The unexpanded EKO \cref{eq:exeko} manifestly satisfies the exact inversion property
\begin{equation}
  \label{eq:figureB3}
  E(Q^2 \leftarrow Q_0^2)E(Q_0^2 \leftarrow Q^2)=\mathbb{1} \, .
\end{equation}
Hence, checking that evolving a PDF back and forth between two scales
gives back the starting PDF provides a stringent test of the
implementation of evolution equations and its accuracy.

\begin{figure}[!t]
  \centering
  \includegraphics[width=.49\textwidth]{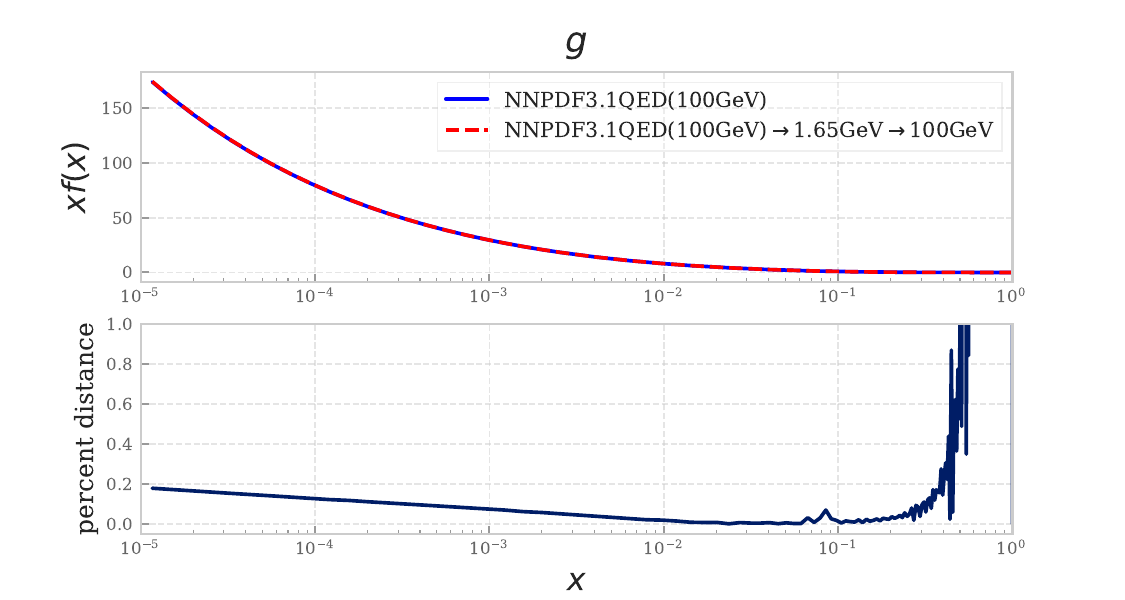}
  \includegraphics[width=.49\textwidth]{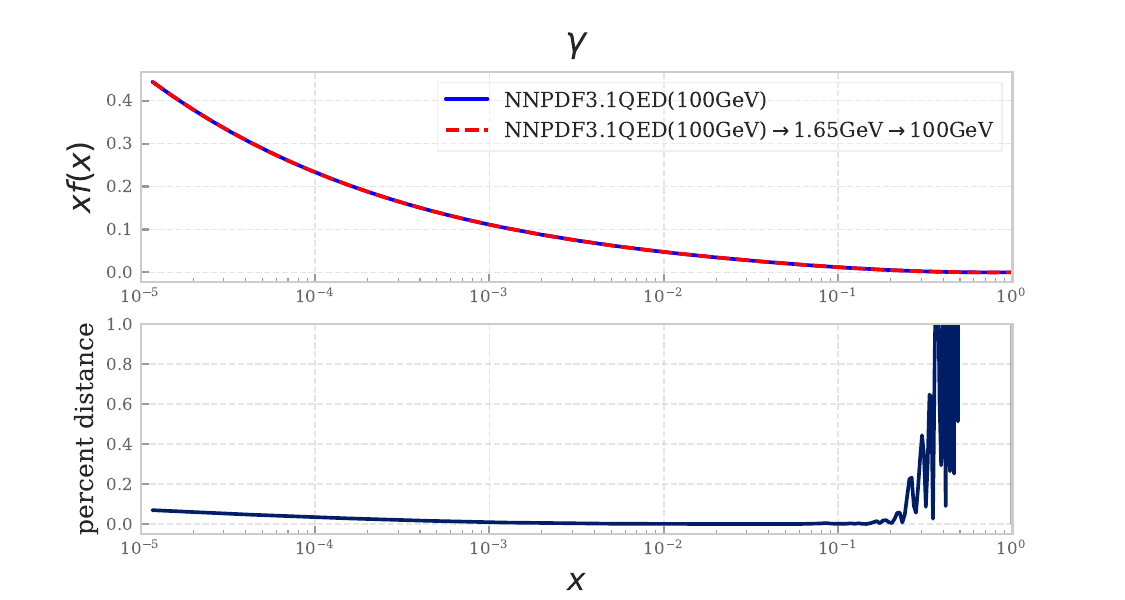}
  \includegraphics[width=.49\textwidth]{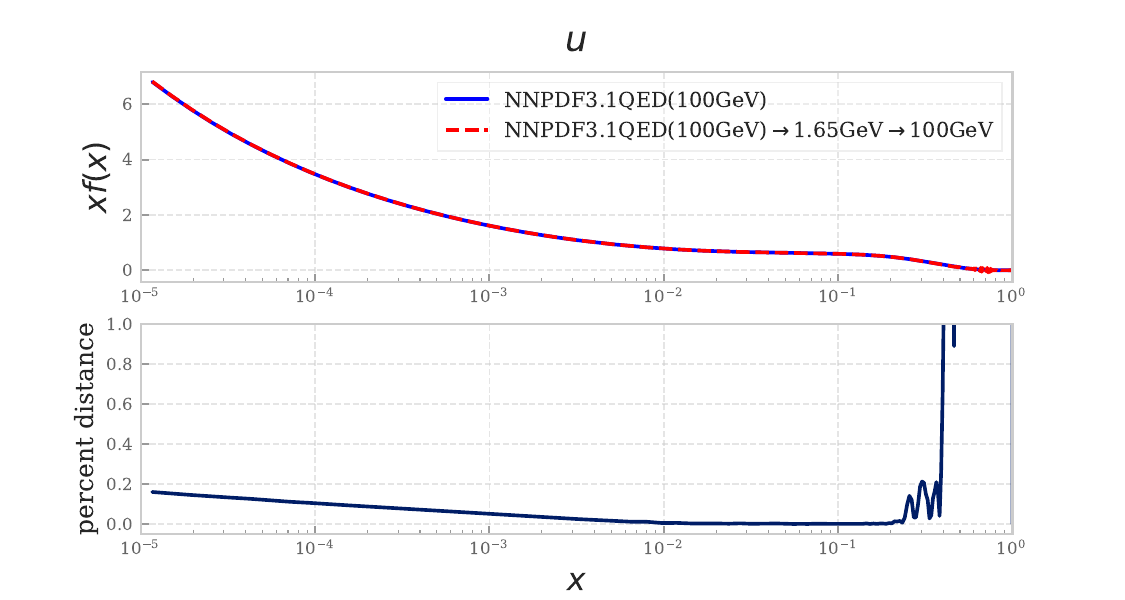}
  \includegraphics[width=.49\textwidth]{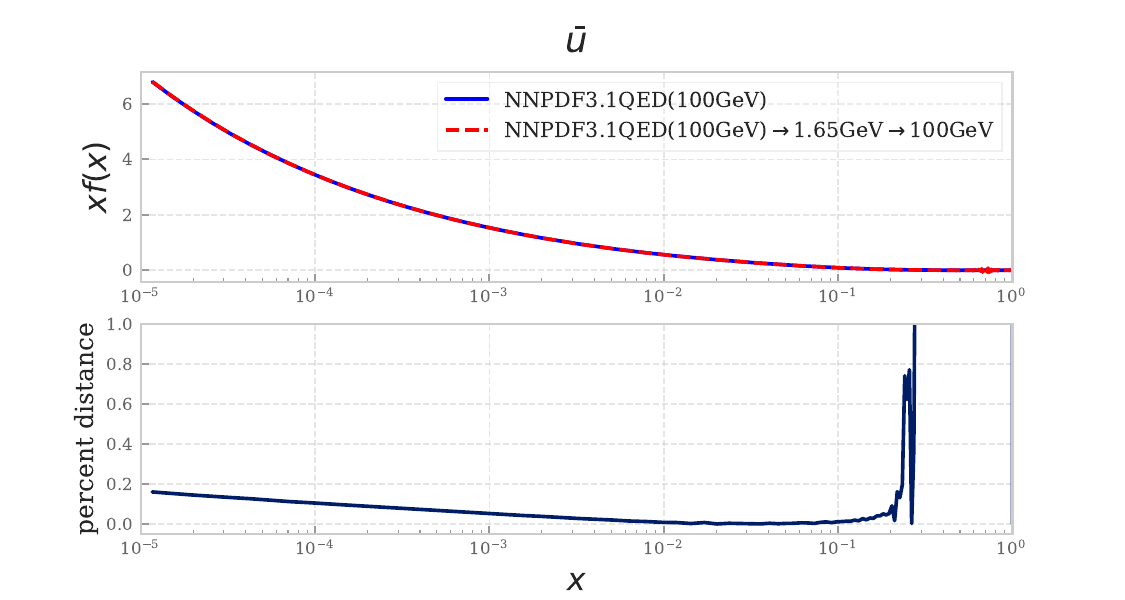}
  \includegraphics[width=.49\textwidth]{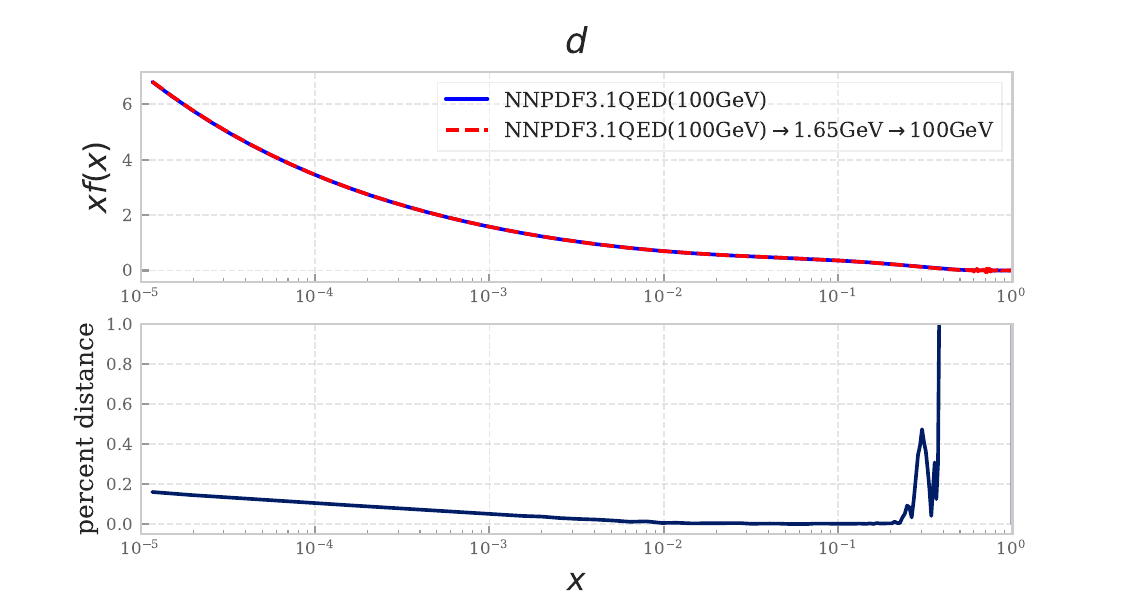}
  \includegraphics[width=.49\textwidth]{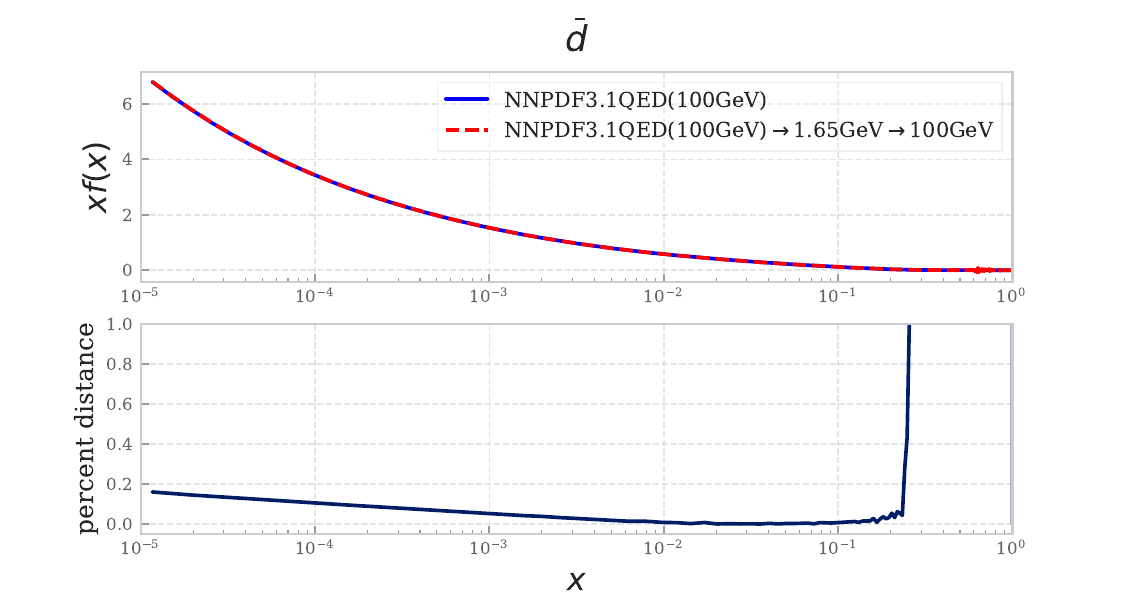}
  \caption{\small Comparison between the 
    NNPDF3.1QED PDFs at $Q_a=100$~GeV, with the results of evolution
    of the same PDFs to $Q_b=1.65$ GeV followed by evolution back to
    $Q_a=100$~GeV. We show, from top to bottom, the gluon, up and down (left)
    and the photon, anti-up and anti-down (right). In each case we show both the
    pair of PDFs, and their percentage relative difference.}
  \label{fig:closure_trn_vs_exact} 
\end{figure}

We have performed this check, by starting with the NNPDF3.1QED NNLO
PDFs at $Q_a=100$~GeV, then evolving down to $Q_b=1.65$ GeV and back to
$Q_a=100$~GeV. Note that this evolution crosses back and forth the
bottom quark threshold, so this also checks the accuracy of the
inversion of matching conditions between the 
$N_f=4$ and $N_f=5$ flavor schemes.
Results are displayed in \cref{fig:closure_trn_vs_exact}, where we compare the
initial and final PDFs. Differences are  at most at the permille level,
except at very large $x$ where they can reach the percentage level,
but PDFs are becoming rapidly very small. We conclude that {exact}
evolution as implemented in the  {\sc\small EKO} code has (at
least) this permille accuracy.

%-------------------------------------------------------------------------------
%-------------------------------------------------------------------------------

\subsection{Benchmarking of QCD$\times$QED evolution: {\sc\small EKO} vs.\ {\sc\small APFEL} }
\label{app:benchmark}

We finally compare the implementation of  QCD$\times$QED evolution in
the  {\sc\small EKO} code used in this paper, with the  {\sc\small APFEL}
code used for the NNPDF3.1QED PDF determination. The comparison
is performed by taking as input the NNPDF3.1QED NNLO PDF set at the
initial parametrization scale of $Q_0=1.65$ GeV,  evolving to
$Q=100$~GeV and determining the percentage difference for all PDFs.

\begin{figure}[!t]
  \centering
  \includegraphics[width=.49\textwidth]{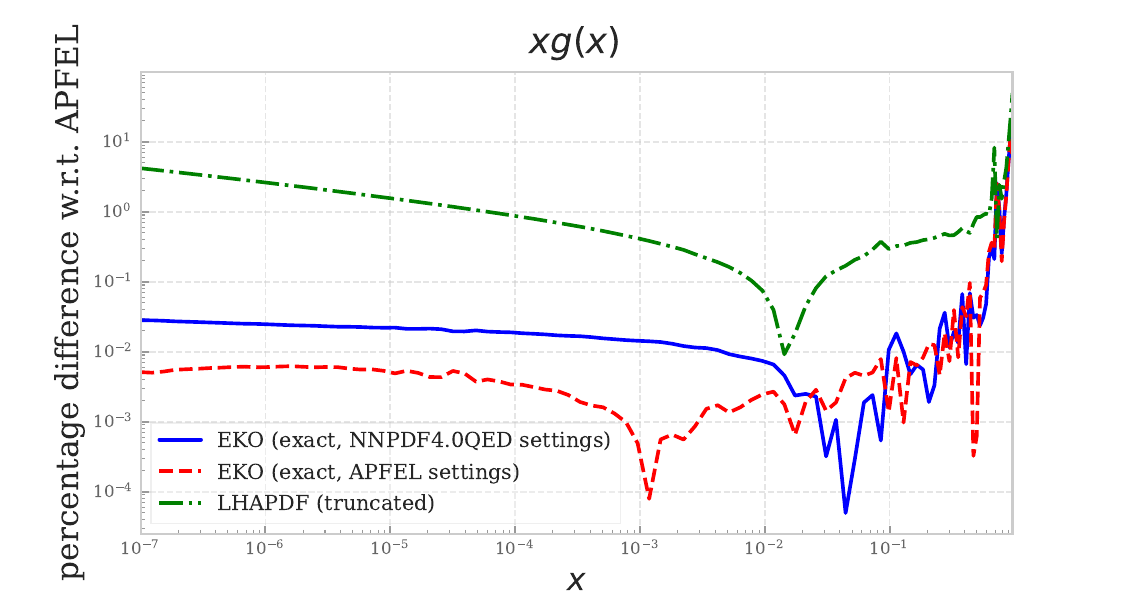}
  \includegraphics[width=.49\textwidth]{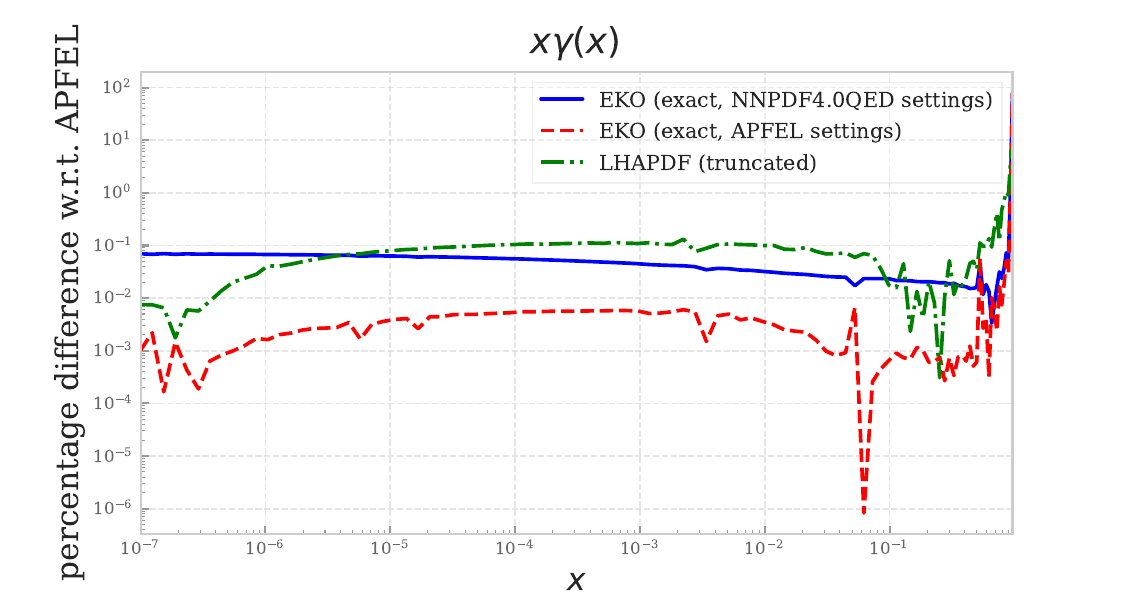}
  \includegraphics[width=.49\textwidth]{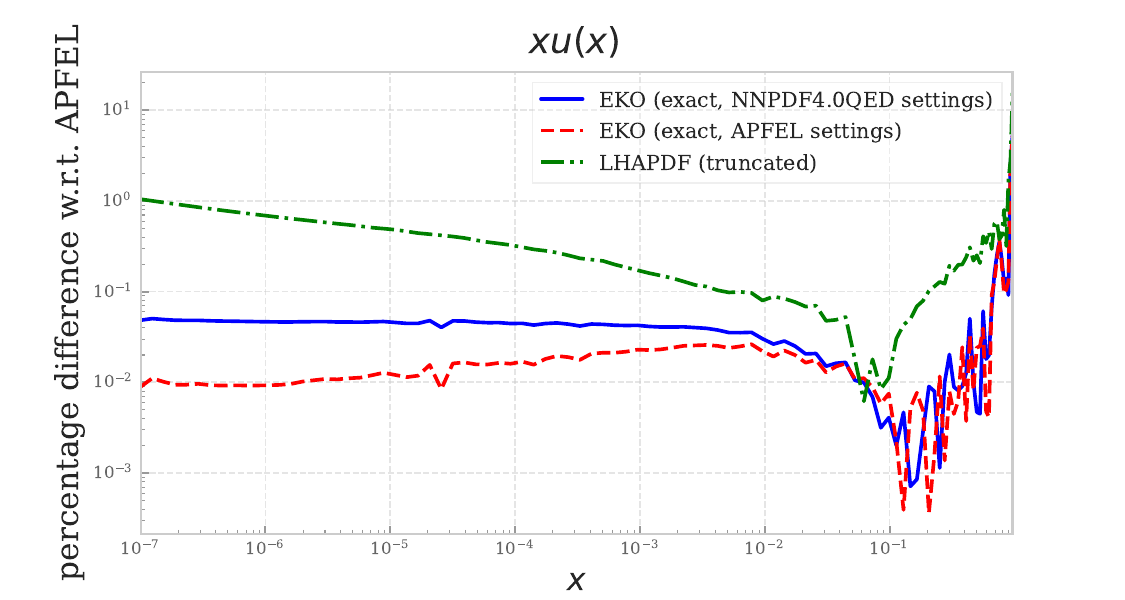}
  \includegraphics[width=.49\textwidth]{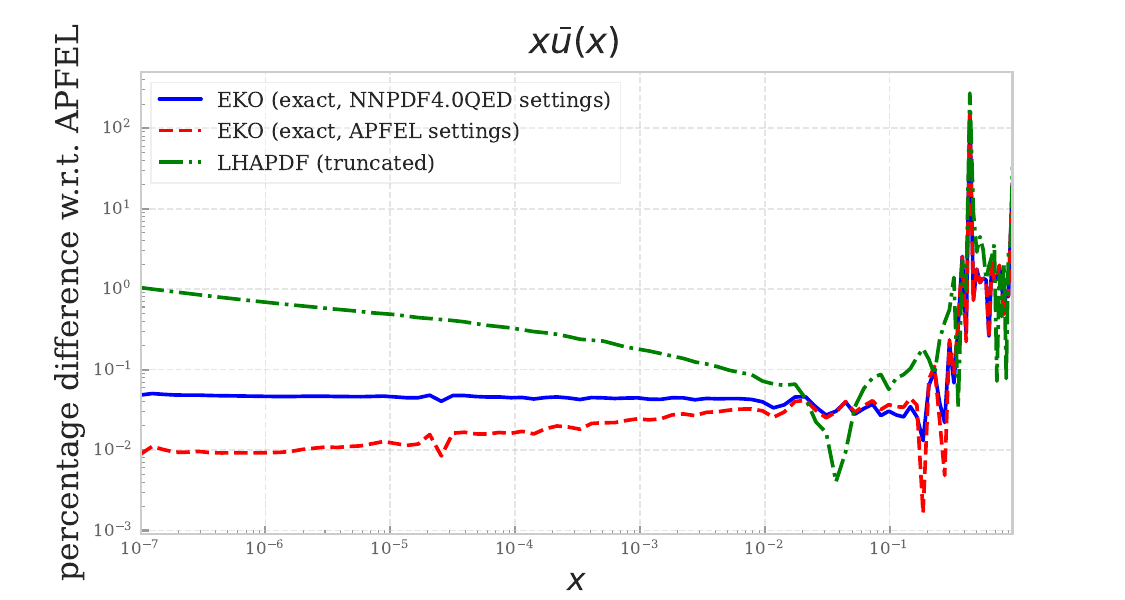}
  \includegraphics[width=.49\textwidth]{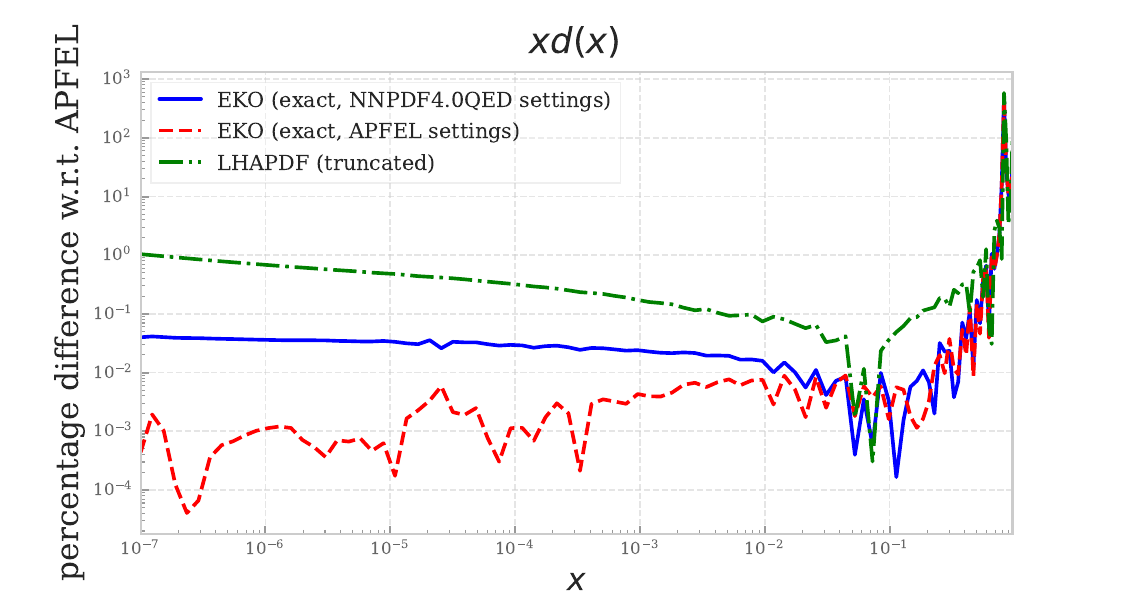}
  \includegraphics[width=.49\textwidth]{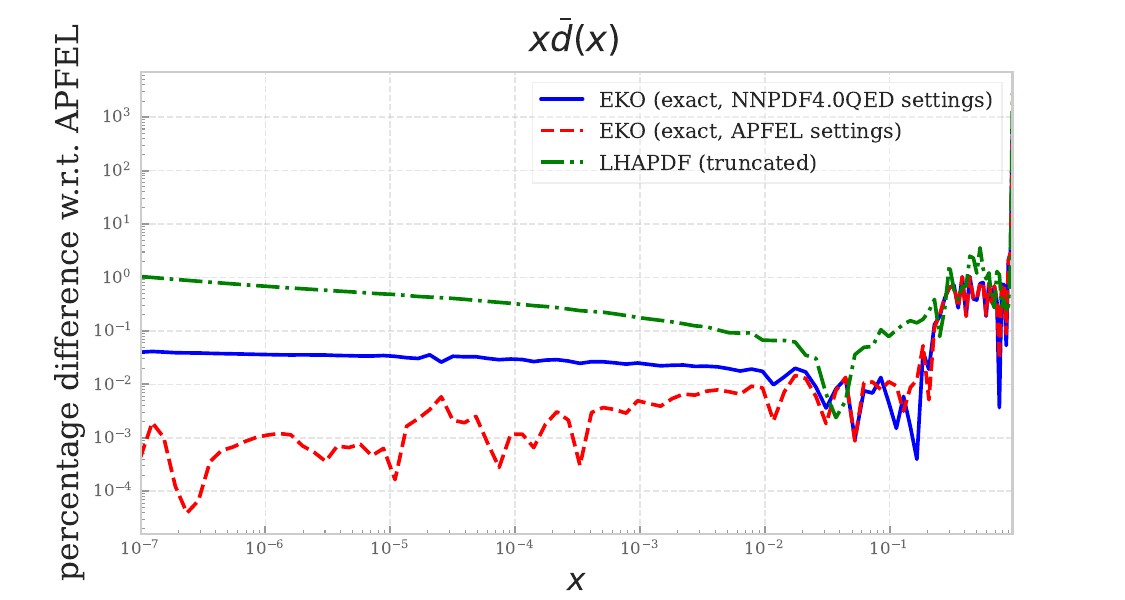}
  \includegraphics[width=.49\textwidth]{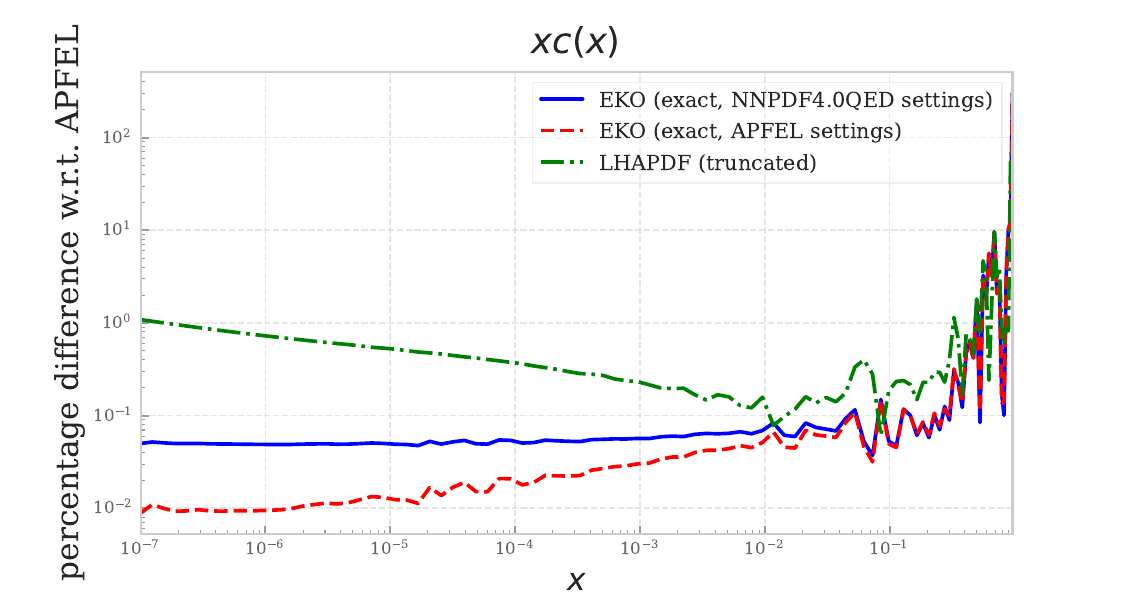}
  \includegraphics[width=.49\textwidth]{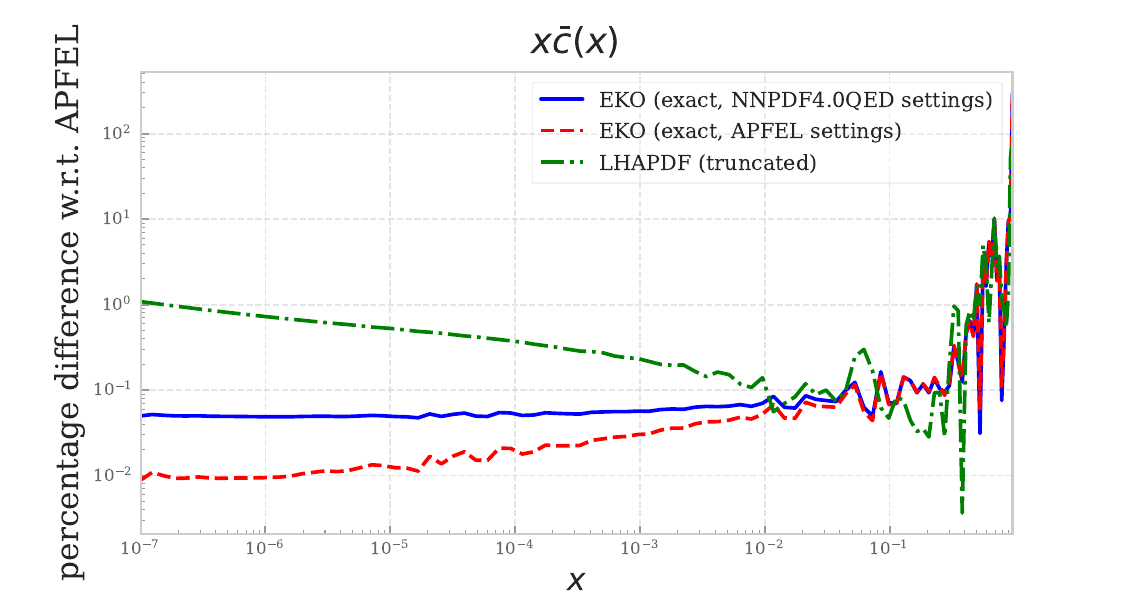}
  \caption{\small Percentage difference between pairs of PDFs at
    $Q=100$~GeV obtained evolving NNPDF3.1QED PDFs from
    $Q_0=1.65$~GeV with different implementation of the
    QED$\times$QCD evolution. From top to bottom the
    gluon, up, down and charm (left), photon, antiup, antidown and
    anticharm (right) are shown. The three curves compare:
    {\sc\small APFEL} {exact} vs.\ {truncated} evolution
    (green, dot-dashed); {\sc\small APFEL} vs.\ {\sc\small EKO}
    {exact} with in each case default settings for the running
    of the couplings (see text) (blue, solid); {\sc\small APFEL}
    vs.\ {\sc\small EKO} {exact} both with {\sc\small APFEL}
    settings for the running of the coupling (red, dashed). Note the logarithmic
    scale on the $y$ axis; note also that the range  on the $y$ axis
    for gluon
    plot differs from that of all other PDFs}
  \label{fig:EKO_vs_APFEL} 
\end{figure}

Results are shown in Fig.~\ref{fig:EKO_vs_APFEL} for  three pairwise
comparison. First, we compare {\sc\small APFEL} {exact} vs.\ {truncated}
evolution (green curve); the APFEL {truncated} result is the
published NNPDF3.1 PDF set, as given by public LHAPDF grids. Then
we compare {\sc\small EKO} vs.\ {\sc\small APFEL} evolution
using the {exact} solution in both cases, and the default
settings of either code as respectively used in this work and for
the NNPDF3.1QED PDF set (blue curve). These settings differ in the running of
the couplings: in the {\sc\small APFEL} settings the coefficients
$\beta_{_{\rm QCD}}^{(2,1)}$, $\beta_{_{\rm QED}}^{(0,3)}$ and
$\beta_{_{\rm QED}}^{(1,2)}$ in Eqs.~\eqref{RGE:as} and
\eqref{RGE:aem} are neglected, {\it i.e.}, the two equations are
decoupled and $\aem$ runs at  leading order. Finally, we compare
{\sc\small EKO} vs. {\sc\small APFEL} evolution using the same
solution and the same settings (red curve), namely {exact} solution and
running of the couplings as in  {\sc\small APFEL}. Note that the scale
on the $y$ axis is logarithmic, and that it shows percentage
difference (so $10^{-3}$ denotes a relative difference of $10^{-5}$).
   
%------------------------------------------------------------------------
%------------------------------------------------------------------------
  
The percentage differences between {\sc\small EKO} and {\sc\small APFEL}
with common settings are always below $10^{-2}$, for all PDFs except
at very large $x$ and 
for charm, where they are about a factor 10 larger. This sets the
accuracy of the evolution codes that are being compared.
The impact of the different running of the couplings is moderate: it
increases the percentage difference by about a factor 3, and then
only for the photon for all $x$ values, for other PDFs only at small
$x\lsim 10^{-2}$. Even so, the difference between {\sc\small EKO} and
{\sc\small APFEL} with their respective settings is at the
sub-permille level. The difference between {exact}  and {truncated}
evolution is at the permille level for $x\gsim 0.003$,
and it can grow up to a few percent
for the gluon and a few permille for all other PDFs at very small
$x\sim 10^{-6}$, in agreement with what already discussed when
comparing  {exact}  and {truncated} evolution at the level of PDFs in
Fig.~\ref{fig:upbar_EXA_vs_TRN_100gev}.

% Bibliography
\bibliography{nnpdf40qed}

\end{document}